\newcounter{myctr}

\newcommand\modelabbrv{HM-LDM}
\newcommand\modelname{Hybrid-Membership Latent Distance Model}
\documentclass{ws-acs}
\usepackage{url}
\usepackage{hyperref}

\usepackage{cite}
\usepackage{amsmath,amssymb,amsfonts}
\usepackage{algorithmic}
\usepackage{booktabs}  
\usepackage{xcolor}
\usepackage{subcaption}
\usepackage{epsfig}
\usepackage[outdir=figs_shm_ldm/]{epstopdf}

\begin{document}

\makeatletter
\def\@biblabel#1{[#1]}
\makeatother


%
\catchline{}{}{}{}{}
%

\title{A Hybrid Membership Latent Distance Model for Unsigned and Signed Integer Weighted Networks}

\author{\footnotesize Nikolaos Nakis}

\address{Department of Applied Mathematics and Computer Science, Technical University of Denmark, Anker Engelunds Vej 101\\
Kongens Lyngby, 2800, Denmark\\
nnak@dtu.dk}

\author{Abdulkadir Çelikkanat}

\address{Department of Applied Mathematics and Computer Science, Technical University of Denmark, Anker Engelunds Vej 101\\
Kongens Lyngby, 2800, Denmark\\
abce@dtu.dk}

\author{Morten Mørup}

\address{Department of Applied Mathematics and Computer Science, Technical University of Denmark, Anker Engelunds Vej 101\\
Kongens Lyngby, 2800, Denmark\\
mmor@dtu.dk}

\maketitle

\begin{history}
\end{history}

\begin{abstract}
Graph representation learning (GRL) has become a prominent tool for furthering the understanding of complex networks providing tools for network embedding, link prediction, and node classification. In this paper, we propose the Hybrid Membership-Latent Distance Model (\textsc{HM-LDM}) by exploring how a Latent Distance Model (LDM) can be constrained to a latent simplex. By controlling the edge lengths of the corners of the simplex, the volume of the latent space can be systematically controlled. Thereby communities are revealed as the space becomes more constrained, with hard memberships being recovered as the simplex volume goes to zero. We further explore a recent likelihood formulation for signed networks utilizing the Skellam distribution to account for signed weighted networks and extend the \textsc{HM-LDM} to the signed Hybrid Membership-Latent Distance Model (\textsc{sHM-LDM}). Importantly, the induced likelihood function explicitly attracts nodes with positive links and deters nodes from having negative interactions. We demonstrate the utility of \textsc{HM-LDM} and \textsc{sHM-LDM} on several real networks. We find that the procedures successfully identify prominent distinct structures, as well as how nodes relate to the extracted aspects providing favorable performances in terms of link prediction when compared to prominent baselines. Furthermore, the learned soft memberships enable easily interpretable network visualizations highlighting distinct patterns. 
\end{abstract}


\section{Introduction}\label{sec:introduction}
In various scientific disciplines, including but not limited to physics, sociology, science-of-science, and biology, networks naturally arise to describe different interactions. These contain spin glasses, friendship interactions, scholarly collaborations, protein-to-protein interactions, structural and functional brain connectivity, and many more \cite{newman}. In order to study these networks and understand their underlying structures, scientists turn to graph analysis tools. The most prominent way for analyzing networks lies in Graph Representation Learning (GRL) \cite{GRL-survey-ieeebigdata20}, which includes approaches capable of performing downstream tasks such as link prediction, node classification, network reconstruction, and community detection with superior performance when compared to prior classical methods. Contrary to GRL, traditional algorithms are characterized by limited flexibility and capacity since they utilize node and graph-level statistics requiring careful design of heuristics and usually high time complexity \cite{GRL_HAM}. The main goal of GRL is to find a mapping, through a learning process, projecting a network into a low-dimensional (usually Euclidean) latent space where node similarity in the graph is translated to node similarity in the latent space, i.e., by positioning related nodes close in proximity in the latent space \cite{survey_hamilton_leskovec}. 

Early GRL approaches capitalized on Natural Language Processing (NLP) where they performed random walks to generate node sequences that correspond to sentences in terms of the NLP terminology \cite{deepwalk-perozzi14,node2vec-kdd16,expon_fam_emb,netmf-wsdm18,line}. The core idea lies in simulating random walks over graphs and optimizing the co-occurrence probability for node pairs based on their obtained distance through the simulated walks. Relatively recent pioneering works \cite{graphsage_hamilton} have extended GRL to the deep learning theory, giving rise to Graph Neural Networks (GNN). Essentially, GNNs perform iterative message-passing extending convolution operations to graphs. One of their limitations is usually the need for node features or else meta-data to avoid the over-smoothing pitfall hampering performance \cite{oversmoothing_gnn} when the GNN model defines deep architectures. Another major category of approaches for GRL relies on matrix decomposition tools \cite{netmf-wsdm18,netsmf-www2019}. Such models learn representations based on the decomposition of a target matrix, which can be constructed to convey first and high-order nodal proximity information \cite{HOPE-kdd16,netmf-wsdm18}. Despite Non-negative Matrix Factorization (NMF) being a prevalent technique for unsupervised signal decomposition and approximation of multivariate non-negative data, few GRL methods utilize such a decomposition. NMF methods have attracted considerable interest since they can extract interpretable part-based representations by revealing the latent factors of the imposed decomposition, which aids in structure retrieval \cite{lee99}.

NMF has been utilized in the context of network analysis and GRL \cite{nmf1,nmf2,nmf3,nmf4}, enabling efficient, unsupervised, and overlapping community detection. This has been explored in various studies, including a mixed-membership stochastic block model (MM-SBM) \cite{JMLR:v9:airoldi08a} defined based on a symmetric-NMF decomposition \cite{nmf4}. This method allows for part-based community assignments for networks while providing uniqueness guarantees. To obtain the propensity of nodes belonging to different communities, standard least-squares NMF optimization was replaced with a Poisson likelihood optimization \cite{nmf1}. Another study used a Poisson distribution to infer mixed memberships for overlapping community detection \cite{nmf2}. These studies involve the generation of mixed-membership vectors for part-based representations. These vectors are then projected onto a space generated by an NMF method, which captures abstract representations of node similarities, positions, and metric properties. Another popular application for NMF is the hyperspectral unmixing \cite{HU} via variational minimum volume regularization \cite{hart2015inferring,zhuang2019regularization}. A well-known approach is the Minimum Volume Constrained-Nonnegative Matrix Factorization (MVC-NMF) \cite{MVCNMF}, which tries to approximate the hyperspectral data matrix with minimum error while including a volume constraint on the simplex matrix. MVC-NMF uses an alternating minimization procedure alternating over a quadratic programming problem and a nonconvex programming problem.

The Latent Space Models (\textsc{LSM}s) are also one of the most powerful ways to learn low-dimensional latent representations \cite{nakis2022hierarchical,çelikkanat2022piecewisevelocity}. These methods employ generalized linear models for constructing latent node embeddings which express important network characteristics. More specifically, the \textsc{LDM} \cite{exp1} utilizes the Euclidean norm for positioning similar nodes closer in the latent space, which comes as a direct consequence of the triangular inequality, naturally representing transitivity ({\it``a friend of a friend is a friend''}) and homophily ({\it a tendency where similar nodes are more likely to connect to each other than dissimilar ones}) properties. An immediate result of obeying the triangular inequality is that the \textsc{LDM} successfully models high-order interactions, as present in complex systems \cite{high_order1,high_order2}. The \textsc{LDM} can be generalized through the Eigenmodel \cite{hoff2007modeling} that can account for stochastic equivalence ({\it``groups of nodes defined by shared intra- and inter-group relationships''}) akin to the SBM \cite{JMLR:v9:airoldi08a} and the mixed membership SBM \cite{JMLR:v9:airoldi08a}. Furthermore, \textsc{LDM}s have been endowed with a clustering model imposing a Gaussian Mixture Model as prior forming the latent position clustering model \cite{handcock2007model,ryan2017bayesian}.

Archetypal Analysis (AA) \cite{cutler1994a} has become a popular tool for extracting polytopes in tabular data. AA was originally defined as an unsupervised learning approach where input data are expressed as linear mixtures (convex combinations) of archetypes/distinct aspects being present in the data \cite{5589222}. AA has been recently extended to the context of network analysis and the modeling of signed networks \cite{slim}, characterizing polarization and conflict over graphs.

Unlike traditional networks modeling only positive and neutral links between entities, signed networks can capture more complex relations, such as cooperative and antagonistic approaches. They are instrumental in modeling more realistic and richer representations of real social structures. Hence, the analysis of the signed networks can reveal significant insights into understanding how the network structure is actually formed. The proverb {\it ``The enemy of my enemy is my friend''} is a very known example demonstrating that driving forces leading individuals to form connections are not merely positive inclinations. The \textit{balance theory} \cite{balance_theory} explains these motives by proposing that individuals have an inner desire to provide balance and consistency in their relationships. Inspired by the theory, \textsc{POLE} \cite{pole} proposes a novel network embedding method for signed networks based on generating random walks. It assigns a sign for each random walk by incorporating the balance theory. \textsc{SIDE} \cite{side} also utilizes fixed-length random walks to extract the node representations of signed networks, but it employs a different optimization strategy. \textsc{SIGAT} \cite{sigat}, and \textsc{SDGNN} \cite{SDGNN} propose approaches leveraging the successful graph neural network architectures for signed networks. The \textsc{SLF} approach \cite{slf} relies on extracting multiple latent factors to model four relationship types: positive, negative, non-link, and neutral. Most recently, \textsc{SLDM} \cite{slim} combined the latent space models and archetypal analysis to learn node embeddings reflecting the different aspects of networks, such as polarized groups or overlapping community structures.

This paper serves as an extension to the {\it HM-LDM: A Hybrid-Membership Latent Distance Model} paper as appeared in \cite{hmldm}. The main contributions of the paper and its extended version can be summarized as:

\begin{itemize}
    \item We introduce a novel method for unsupervised representation learning on graphs called \modelname \ (\textsc{\modelabbrv}), which combines the strengths of LDM and NMF. The \textsc{\modelabbrv} approach reconciles part-based network representations with low-dimensional latent spaces that satisfy similarity properties like homophily and transitivity. These properties play a critical role in GRL because they enable a straightforward interpretation of network structure. Moreover, the proposed method captures the latent community structure of the networks using a simple continuous optimization procedure based on the log-likelihood of the network. Unlike most existing methods that impose hard constraints on community memberships, the assignment of community memberships in our hybrid model can be controlled and altered using the simplex volume as defined by the latent node representations. We extensively evaluate the proposed method's performance in link prediction and community discovery tasks across various network types and demonstrate its superiority over existing methods.
   \item We hereby, extend the framework to the analysis of signed networks via the use of the Skellam distribution forming the \textsc{s}igned \textsc{H}ybrid-\textsc{M}embership \textsc{L}atent \textsc{D}istance \textsc{M}odel (\textsc{sHM-LDM}) inspired by recent advances in GRL \cite{slim}. The model characterizes and uncovers distinct aspects of signed networks by constraining the latent space to the $D$-simplex. We show that the \textsc{sHM-LDM} relates to archetypal analysis for relational data \cite{slim} as a minimal volume approach and as a special case when polytopes are constrained to simplexes. We benchmark the performance of our model against prominent signed network representation learning approaches and across four real signed networks, as well as two real bipartite networks.
\end{itemize}

\noindent\textbf{Source code:} \href{https://github.com/Nicknakis/HM-LDM}{\textit{https://github.com/Nicknakis/HM-LDM}}.

\section{Problem statement and proposed method}\label{sec:method}
Let $\mathcal{G}=(\mathcal{V},\mathcal{E})$ be an undirected graph where $\mathcal{V}$ shows the vertex set and $\mathcal{E} \subseteq \mathcal{V}\times \mathcal{V}$ the edge set. We use $\mathbf{Y}_{N \times N}=\left(y_{i,j}\right)$ to denote the adjacency matrix of the graph where $y_{i,j} =0$ if the pair $(i,j) \not\in \mathcal{E}$ otherwise it is non-zero value for all $ 1\leq i< j\leq N := |\mathcal{V}|$. It is worth noting that we will also consider signed weighted networks in the paper, so the edge weight or the entries of the adjacency matrix can take any positive or negative integer value ($y_{ij} \in \mathbb{Z}$). In the latter case, we will further denote $\mathcal{E}^{+}$ as the positive edge set, and
$\mathcal{E}^{-}$ as the negative edge set. The detailed list of the symbols used throughout the manuscript and their corresponding definitions can be found in Table \ref{tab:table_of_symbols}.

\begin{table}[!b]
\centering
\caption{Table of symbols}
\label{tab:table_of_symbols}
\resizebox{\textwidth}{!}{
\begin{tabular}{ll}
\toprule
\textbf{Symbol} & \textbf{Description} \\ \midrule
$\mathcal{G}$ & Graph \\
$\mathcal{V}$ & Vertex set \\
$\mathcal{E}$ & Edge set \\
$\mathcal{E}^{+}$ & Positive edge set \\
$\mathcal{E}^{-}$ & Negative edge set \\
$N$ & Number of nodes \\
$D$ & Dimension size \\
$\gamma_i,\beta_i,\psi_i$ & Bias terms of node $i$ \\
$\mathbf{w}_i$ & Latent embedding for node $i$ \\
$\lambda_{ij}$ & Poisson rate (intensity) of node pair $(i,j)$\\
$\lambda^+_{ij}$ & Positive interaction Poisson rate (intensity) of node pair $(i,j)$ of the Skellam distribution\\
$\lambda^-_{ij}$ & Negative interaction Poisson rate (intensity) of node pair $(i,j)$ of the Skellam distribution\\
$\mathcal{I}_{|y|}$ & Modified Bessel function of the first kind and order $|y|$\\
$\delta$ & Simplex side length with $\delta \in \mathbb{R}_+$ \\
$p$ & Power of the $\ell_2$ norm with $p \in\{1,2\}$\\
$\Delta^{D}$ & The standard $D-$simplex\\
$\bm{\Lambda}$ & Eigenmodel non-negative relational matrix\\
$\mathbf{A}$ & The matrix containing the archetypes (extreme points of the convex hull) with $\mathbf{A} \in \mathbb{R}^{(D+1)\times (D+1)}$\\
\bottomrule
\end{tabular}%
 }
\end{table}

Our main goal is to learn a representation, $\mathbf{w}_i \in \mathbb{R}^{D}$, for each node $i \in \mathcal{V}$ in a lower dimensional space ($D \ll N$) such that similar nodes in the network should have close embeddings. More specifically, we concentrate on mapping the nodes into the unit $D$-simplex, $\Delta^{D} \subset \mathbb{R}_{+}^{D+1}$, which is defined by
\begin{align*}
   \Delta^{D}= \left\{ (x_0,\ldots, x_{D})\in\mathbb{R}^{D+1}\Bigg| \sum_{d=0}^{D}x_d \!=\! 1, \ x_d \geq 0, \ \forall d \in \{0,\ldots, D\} \right\}.
\end{align*}
Consequently, for unsigned networks, the inferred node representations carry information about latent community memberships. While in the case of signed networks, node embeddings define memberships over distinct aspects and profiles being present in the network. Importantly, in contrast with other GRL approaches, in this study, we seek and construct identifiable solutions which can only be achieved up to a permutation invariance, as reported in Def. \ref{def:identifiabilty}. Identifiability guarantees are also extended to the modeling of signed networks providing embedding spaces that can easily be interpreted. 

In the following part, we will first introduce the Hybrid-Membership Latent Distance Model (\textsc{HM-LDM}) focused on unsigned networks, and later we will generalize the framework to the analysis of signed networks forming the \textsc{s}igned \textsc{H}ybrid-Membership \textsc{L}atent \textsc{D}istance Model (\textsc{sHM-LDM}).
  

\begin{definition}[\textbf{Identifiability}]\label{def:identifiabilty}
Let $\mathbf{W}$ be an optimal embedding matrix whose rows indicate the corresponding node representations. We call $\mathbf{W}$ an \textit{identifiable solution up to a permutation} if there is a matrix $\mathbf{P}$ satisfying $\widetilde{\mathbf{W}}=\mathbf{W}\mathbf{P}$ for some optimal solution $\widetilde{\mathbf{W}}$, then $\mathbf{P}$ must be a permutation matrix.
\end{definition}

\subsection{The Hybrid-Membership Latent Distance Model}
For a given unsigned network $\mathcal{G}=(\mathcal{V},\mathcal{E})$, we suppose that the random variables representing the links for a pair of nodes $i$ and $j$ independently follow a Poisson distribution when conditioned on the latent representations $\{\bm{W},\boldsymbol{\gamma}\}$, as introduced later. In this section, we consider unweighted networks, so the entries of the adjacency matrix, $\mathbf{Y}=(y_{ij})\in\{0,1\}^{|\mathcal{V}|\times|\mathcal{V}|}$ are binary values, and we can write the log-likelihood function as follows:
\begin{equation}
    \label{eq:prob_adj}
    \log P(\mathbf{Y}|\bm{W},\boldsymbol{\gamma})=\!\!\sum_{\substack{i<j \\ y_{ij}=1}}\!\log(\lambda_{ij}(\mathbf{w}_i,\mathbf{w}_j,\gamma_i,\gamma_j))\;-\;\sum_{\substack{i< j }}\Big(\lambda_{ij}(\mathbf{w}_i,\mathbf{w}_j,\gamma_i,\gamma_j)+\log(y_{ij}!)\Big) \:.
\end{equation}
Similar to the work in \cite{doi:10.1198/016214504000001015}, we here employ the Poisson regression approach for unweighted networks since it successfully generalizes to the modeling of binary networks \cite{nmf2}.

We utilize the rates of the distributions to learn the representations of nodes in the latent space by defining the Poisson rate $\lambda_{ij}$ as follows:
  \begin{equation}
     \label{eq:nmf_rate}
     \log \lambda_{ij}=\Big(\gamma_i+\gamma_j-\delta^p\cdot||\mathbf{w}_i -\mathbf{w}_j||_2^p\Big),
 \end{equation}
 where $\mathbf{w_i} \in [0,1]^{D+1}$ are the latent embeddings constrained to the $D-$simplex, i.e. $\sum_{d=1}^{D+1} w_{id}=1$, $\delta \in \mathbb{R}_+$ is the non-negative value controlling the simplex volume, and $\gamma_i \in \mathbb{R}$ a bias term of node $i\in\mathcal{V}$ accounting for node-specific effects \cite{doi:10.1198/016214504000001015,KRIVITSKY2009204} such as degree heterogeneity. Lastly, $p$ is the power of the $\ell_2$ norm with $p \in\{1,2\}$ controlling the model specification. Specifically, power $p$ adjusts the influence of the embedding distances in the rate functions. In other words, in Eq. \ref{eq:nmf_rate} we constrain the latent space to the $D-$simplex, and the simplex's edge lengths ($1$-faces) are scaled by the non-negative constant $\delta$, controlling the simplex side length and thus the simplex volume. In the rest of the paper, we will call this proposed method by \modelname \ (\textsc{\modelabbrv}).
 
 It can be seen that a non-negative Eigenmodel with bias terms (i.e. $\tilde{\gamma}_i+\tilde{\gamma}_j +(\mathbf{\tilde{w}}_i\bm{\Lambda}\mathbf{\tilde{w}}_j^{\top})$) corresponds to Eq. \eqref{eq:nmf_rate} for $p=2$ if $\bm{\Lambda}$ is chosen as a diagonal matrix with constant entries $2\delta^2$, and if the bias terms are reparameterized as $\tilde{\gamma}_i=\gamma_i-\delta^2\cdot||\mathbf{w}_i||^2_2$ since expression $\tilde{\gamma}_i+\tilde{\gamma}_j +(\mathbf{\tilde{w}}_i\bm{\Lambda}\mathbf{\tilde{w}}_j^{\top})$ turns into:
\begin{align*}
 (\gamma_i-\delta^2||\mathbf{w}_i||^2_2) + (\gamma_j-\delta^2||\mathbf{w}_j||^2_2) + (2\delta^2\mathbf{w}_i\mathbf{w}_j^\top) = \gamma_i + \gamma_j - \delta^2\| \mathbf{w}_i-\mathbf{w}_j \|^2.
\end{align*}
Therefore, the squared Euclidean distance incorporates the conventional LDM to the non-negativity-constrained Eigenmodel. Although the squared Euclidean distance is not a metric, it still embodies the homophily property, resulting in an interpretable latent space. Despite not exactly satisfying the triangle inequality, it preserves the relative ordering of pairwise Euclidean distances. That's why it is highly preferred in many applications since it is a strictly convex smooth function. By using the well-known cosine formula, we can write:
\begin{align}
||\mathbf{w}_i-\mathbf{w}_j||_2^2 &= ||\mathbf{w}_i-\mathbf{w}_k||_2^2+|| \mathbf{w}_k-\mathbf{w}_j ||_2^2-2||\mathbf{w}_i-\mathbf{w}_k||_2||\mathbf{w}_k-\mathbf{w}_j||_2\cos(\theta),\nonumber
\end{align}
\noindent where $\theta \in (-\pi/2, \pi/2)$ represents the angle between $\mathbf{w}_i-\mathbf{w}_k$ and $\mathbf{w}_k-\mathbf{w}_j$. Note that the third term also approaches to $0$ for $\theta \rightarrow \pi/2$. For the case where $\theta \in [\pi/ 2, 3\pi/ 2]$, it satisfies the triangle inequality: $||\mathbf{w}_i - \mathbf{w}_j ||_2^2 \leq || \mathbf{w}_i - \mathbf{w}_k ||_2^2  + || \mathbf{w}_k - \mathbf{w}_j ||_2^2$.

\begin{figure}
  \centering
    \subfloat[Translation invariances.]{{
  \includegraphics[width=0.21\columnwidth]{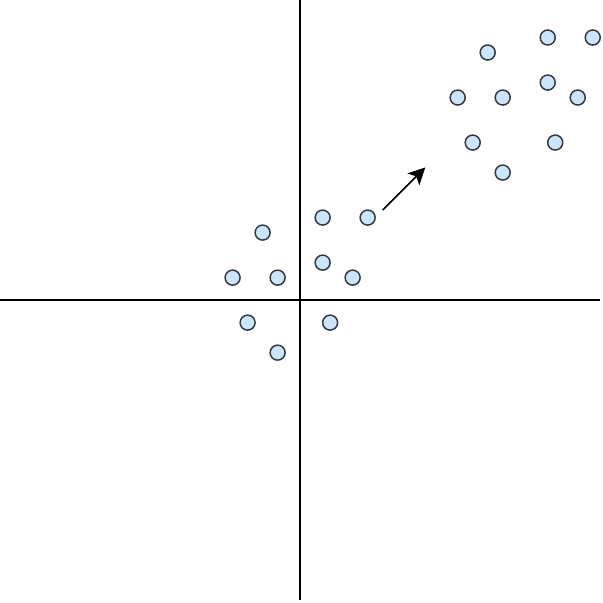} }}%
  \hfill
   \subfloat[Rotation invariances.]{{
  \includegraphics[width=0.25\columnwidth]{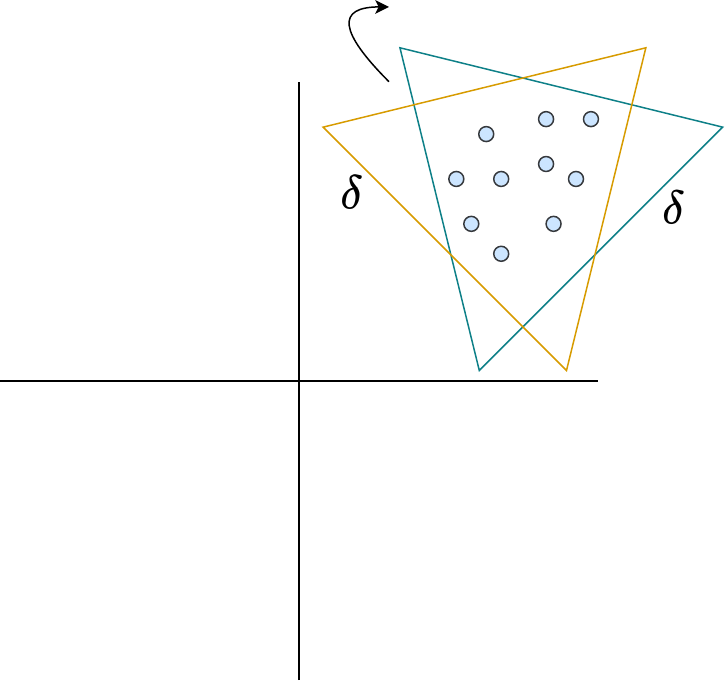} }}%
  \hfill
  \subfloat[Decreased simplex volume ensuring identifiability.]{{\includegraphics[width=0.34 \columnwidth]{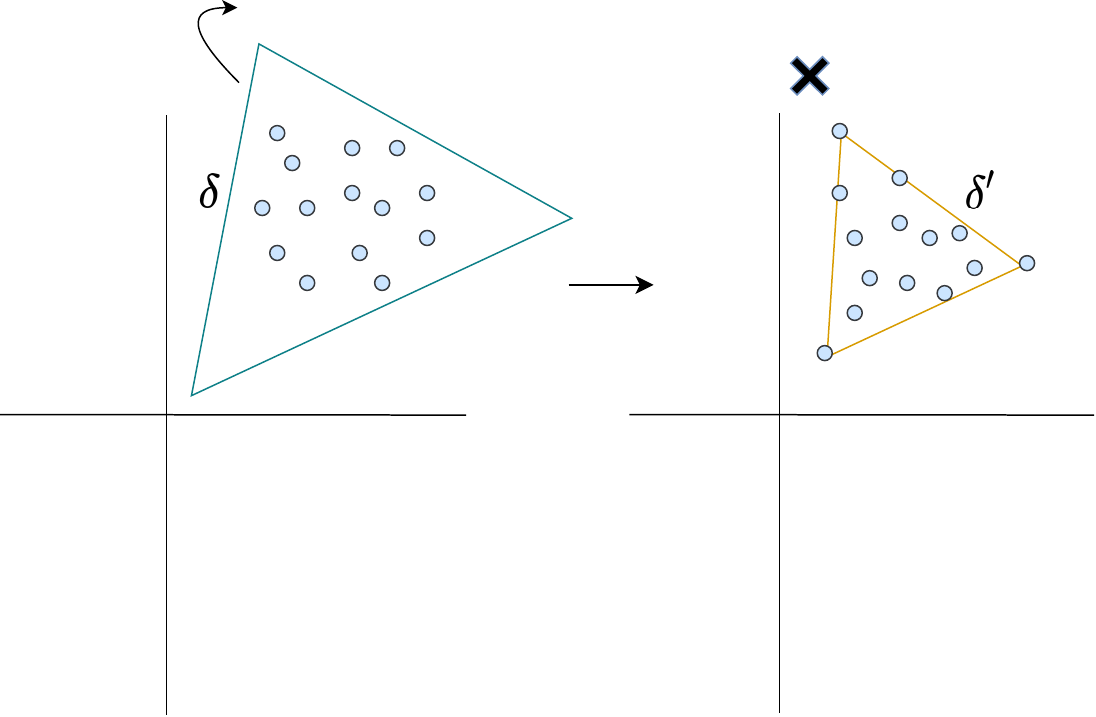} }}%
  \caption{A $2$-dimensional latent space with the $2$-simplex given as the green and yellow triangles, the blue points denote embedding positions of the $\textsc{LDM}$ and $\delta$ is the simplex size.}\label{fig:invariances}
\end{figure}

Since we learn the node representations in a $D-$simplex space, each entry of an embedding vector, in fact, points out a latent community membership, so the node representations also provide information regarding the community structure of the network. Note that we can translate the learned embeddings to the non-negative orthant without any loss in performance or in expressive capability since the translation is invariant to the distance metric, as shown in Fig \ref{fig:invariances} (a). A rotation operation also does not affect the pairwise distance among the embedding vectors but the node representations must be positioned inside a ball lying in a $D-$simplex otherwise, the embeddings cannot be rotated (see Fig \ref{fig:invariances} (b)). 

However, as we mentioned before, the embedding vectors also define the nodes' community memberships. Therefore, a rotation operation alters the community assignments while leaving the distance matrix invariant. As a result, the latent representations cannot be used to express community information in this case. It is worth noticing that we can have \textit{identifiable} node representations if the corners of the simplex include at least one node because it makes the rotation operation inapplicable. In this regard, this condition can be satisfied by the distance scaling parameter (i.e., $\delta \in \mathbb{R}^+$) introduced in Eq. \ref{eq:nmf_rate}. Since we know that $\| x \|_1^p/ \sqrt{D^p} \leq \| x \|_2^p \leq \| x \|_1^p$ for $p\in\{1,2\}$, shrinking the volume of the simplex sufficiently (equivalently decreasing the $\delta$ value) forces nodes to populate around the corners of the simplex. The node embeddings move towards the corners of the simplex to balance the change in the term $\delta^p\| \mathbf{w}_i - \mathbf{w}_j \|_2^p$ since we have $\| \mathbf{w}_i \|=1$ for all $i\in\mathcal{V}$.

We will name a node \textit{champion} if it is located in one of the corners of the simplex. In other words, its latent representation must be a standard binary unit vector in a $D-$simplex space. 
The champion nodes play a crucial role in achieving identifiability since the learned representations become identifiable (up to a permutation matrix) if every corner of the simplex is occupied by at least one champion node (please see the definition below). In this case, any random rotation will no longer leave the solution invariant, as illustrated in Fig \ref{fig:invariances} (c). Hence, the scaling parameter, $\delta$, determines the model's type of memberships and expressive capabilities. Large values of $\delta$ make the solution rotationally invariant. On the other hand, small values of $\delta$ result in identifiable solutions and hard cluster assignments, where nodes are exclusively assigned to the corners of the simplex. Moreover, certain regimes of $\delta$ values can provide identifiable solutions with similar performance to \textsc{LDM}.

\begin{definition}[\textbf{Community champion}]\label{def:champion}
A node for a latent community is called \textit{champion} if it belongs to the community (simplex corner) while forming a binary unit vector.
\end{definition}

We can approach model identifiability for the $p=2$ model specification from a different perspective using the Non-negative Matrix Factorization (NMF) theory. We achieve this by re-parameterizing Eq. \eqref{eq:nmf_rate} with $\tilde{\gamma}_i+\tilde{\gamma}_j+2\delta^2\cdot(\mathbf{w}_i\mathbf{w}_j^{\top})$ as previously discussed. In this formulation, the product $\mathbf{W}\mathbf{W}^{\top}$ defines a symmetric NMF problem that is uniquely factorized (up to permutation invariance) and identifiable when $\mathbf{W}$ is full-rank, and at least one node resides solely in each corner of the simplex, ensuring separability condition \cite{nmf4,nmf5}. Under this NMF formulation, the product $\mathbf{w}_i\mathbf{w}_j^{\top} \in [0,1]$ reaches its upper bound only if nodes $i$ and $j$ reside in the same corner of the simplex. When $\delta$ is small, the model favors hard latent community assignments of nodes since nodes with similar features achieve high values only when they belong to the same latent community (simplex corner). On the other hand, when nodes head towards the corners of the simplex for large values of $\delta$, the second term of the log-likelihood function in Eq. \eqref{eq:prob_adj} changes exponentially. Hence, assigning dissimilar nodes to the same community severely penalizes the likelihood. For this reason, a high value of $\delta$ is beneficial for mixed-membership allocations.

\subsection{The Signed Hybrid-Membership Latent Distance Model}
Recent advances in GRL \cite{slim}, extended \textsc{LDM}s to the study of signed networks while characterizing network polarization via the use of Archetypal Analysis (AA) \cite{cutler1994a,5589222} and the Skellam distribution \cite{skellam}. The Skellam distribution is the difference of two independent Poisson-distributed random variables ($y=N_1 - N_2\in\mathbb{Z}$) with respect to the rates $\lambda^{+}$ and $\lambda^{-}$: 
\begin{align}\label{eq:likelihood_sk}
P(y|\lambda^{+},\lambda^{-}) = e^{-(\lambda^{+}+\lambda^{-})}\left(\frac{\lambda^{+}}{\lambda^{-}}\right)^{y/2}\mathcal{I}_{|y|}\left(2\sqrt{\lambda^{+}\lambda^{-}}\right),
\end{align}
where $N_1 \sim Pois(\lambda^{+})$ and $N_2 \sim Pois(\lambda^{-})$, and $\mathcal{I}_{|y|}$ is the modified Bessel function of the first kind and order $|y|$.

Whereas in \cite{slim} the network representations were constrained to the convex hull as defined by the inferred representations, it is discussed that other approaches to model pure/ideal forms have been Minimal Volume (MV) approaches as defined by
\begin{eqnarray}
\mathbf{Z}\approx \mathbf{AW}\quad \text{s.t. } vol(\mathbf{A})= v \text{ and } \boldsymbol{w}_j\in \Delta^{D},
\end{eqnarray}
where $\mathbf{A} \in \mathbb{R}^{(D+1)\times (D+1)}$ is the matrix describing the archetypes (extreme points of the convex hull) of the latent space, and $vol(\mathbf{A})$ is the volume of matrix $\mathbf{A}$ which can be expressed through the determinant as $|det(\mathbf{A})|$, when $\mathbf{A}$ is a square matrix \cite{hart2015inferring,zhuang2019regularization}. Extraction of distinct aspects/profiles through MV does not require the presence of ``pure'' observations defining the convex-hull or else the extracted polytope/simplex. As the volume decreases, observations are ``forced'' to populate the corners of the polytope, yielding archetypal characterization when the reconstruction of data is defined through convex combinations of these corners.

 The main disadvantage of MV procedures is the need for careful regularization tuning to define volumes ensuring identifiability as well as maintaining enough capacity to express the data with a small reconstruction error \cite{zhuang2019regularization}. In addition, analytical and tractable computation of the volume of polytopes requires calculating the sum of determinants for all simplexes used to construct the inferred polytope \cite{bueler2000exact}. This is computationally expensive (especially in high dimensions) and sometimes unstable when $\mathbf{A}$ comes close to singular.
 
 In this paper, we constrain the columns of matrix $\mathbf{A}$ to the $D$-simplex with length $\delta$. Thus, by controlling the volume of $\mathbf{A}$, we essentially define a constrained-to-simplexes MV approach. Calculating the volume for the $D-$simplex with length $\delta$ is straightforward and computationally efficient. Rather than including regularization over the volume of $\mathbf{A}$ in the loss function during inference, we deterministically control the simplex length $\delta$ which is given as an input to the model and gradually decreased until uniqueness guarantees are obtained. Volume minimization can be obtained trivially by decreasing $\delta$. Such a procedure gives us explicit control over the model capacity by fixing the volume which is harder to be obtained with classical MV approaches where the volume expression is inserted in the loss function.
 
 Essentially, by defining $\mathbf{A}$ as $\mathbf{A}=\delta\cdot \mathbf{I}$, with $\mathbf{I}$ being the $(D+1)\times (D+1)$ identity matrix, we obtain as a special case of archetypal analysis under a constrained MV formulation. In addition, if every corner of the introduced simplex is populated by at least one node champion we obtain unique representations defining hybrid memberships.

 We now introduce the \textsc{s}igned \textsc{H}ybrid-\textsc{M}embership \textsc{L}atent \textsc{D}istance \textsc{M}odel (\textsc{sHM-LDM}).
 The \textsc{sHM-LDM} is able to analyse signed networks, and similar to \cite{slim} it introduces two Skellam rate parameters for Eq. \eqref{eq:likelihood_sk} as:
\begin{align}
\lambda_{ij}^{+} &= \exp\big(\beta_{i} + \beta_{j} - \delta^p||\mathbf{w}_i-\mathbf{w}_j||_2^p\big)\label{eq:rate1},
\\
\lambda_{ij}^{-} &= \exp\big(\psi_{i} + \psi_{j} + \delta^p||\mathbf{w}_i-\mathbf{w}_j||_2^p\big),
\label{eq:rate2}
\end{align}
where again $\mathbf{w_i} \in [0,1]^{D+1}$ and $\sum_{d=1}^{D+1} w_{id}=1$, $\delta \in \mathbb{R}_+$ and $\beta_i,\psi_j \in \mathbb{R}$ denote the node-specific random-effects. As explained in \cite{slim}, $\beta_i,\beta_j$ represent the ``social'' effects/reach of a node and the tendency to form (as a receiver and as a sender, respectively) positive interactions, expressing positive degree heterogeneity (indicated by $+$ as a superscript of $\lambda$). In contrast, $\psi_i,\psi_j$ provides the ``anti-social'' effect/reach of a node to form negative connections and thus models negative degree heterogeneity (indicated by $-$ as a superscript of $\lambda$). The norm degree $p \in \{1,2\}$ controls the power of the $\ell^2$-norm, and thus the model specification, as in the unsigned case. 

As in \cite{slim}, we define a maximum-a-posteriori (MAP) estimation, utilizing the Skellam likelihood over the adjacency matrix $\mathbf{Y}$ of the network $\mathcal{G}=(\mathcal{V}, \mathcal{E})$. We conditionally assume an independent likelihood given the unobserved latent positions and random effects. The corresponding loss function excluding constant terms is:
\begin{equation}\label{eq:loss_sk}
    L = \sum_{i<j}\Bigg( \lambda_{ij}^{+}+\lambda_{ij}^{-} -\frac{y_{ij}}{2}\log\left( \frac{\lambda_{ij}^{+} }{\lambda_{ij}^{-}}\right)\Bigg)  - \sum_{i<j}\log I_{|y_{ij}|}\Big (2\sqrt{\lambda_{ij}^{+}\lambda_{ij}^{-}}\Big )+ \frac{\rho}{2}\Big(||\bm{\beta}||_F^2+||\bm{\psi}||_F^2\Big),
\end{equation}
where $\mathcal{I}_{|y|}$ is the modified Bessel function of the first kind and order $|y|$, $||\cdot||_F$ denotes the Frobenius norm. In addition, $\rho$ is the regularization strength where $\rho=1$ is assumed throughout this paper yielding a normal prior with zero mean and unit variance for the random effects. For the latent positions, we assume a uniform Dirichlet distribution as a prior which only adds a constant term in Eq. \ref{eq:loss_sk} and thus is excluded.

Choosing the case where $p=2$, meaning that the \textsc{sHM-LDM} utilizes the squared Euclidean norm, we are able once more to relate the model to an Eigenmodel by creating the following reparameterizations. For the rate responsible for positive interactions $\{\lambda_{ij}^+\}$ as: $ \tilde{\beta}_i+\tilde{\beta}_j+(\mathbf{\tilde{w}}_i\bm{\Lambda}\mathbf{\tilde{w}}_j^{\top})$ where $\bm{\Lambda}$ is a diagonal matrix having non-negative elements, i.e. $\tilde{\beta}_i=\beta_i-\delta^2\cdot||\mathbf{w}_i||^2_2$, $\tilde{\beta}_j=\beta_j-\delta^2\cdot||\mathbf{w}_j||^2_2$ and $\tilde{\mathbf{w}}_i\bm{\Lambda}\tilde{\mathbf{w}}_j^\top=2\delta^2\cdot \mathbf{w}_i\mathbf{w}_j^\top$. Similarly, for the rate responsible for expressing animosity $\{\lambda_{ij}^-\}$ as: $ \tilde{\psi}_i+\tilde{\psi}_j+(\mathbf{\tilde{w}}_i\bm{\Lambda}\mathbf{\tilde{w}}_j^{\top})$ where $\bm{\Lambda}$ is a diagonal matrix having non-positive elements, i.e. $\tilde{\psi}_i=\psi_i-\delta^2\cdot||\mathbf{w}_i||^2_2$, $\tilde{\psi}_j=\psi_j-\delta^2\cdot||\mathbf{w}_j||^2_2$ and $\tilde{\mathbf{w}}_i\bm{\Lambda}\tilde{\mathbf{w}}_j^\top=2\delta^2\cdot \mathbf{w}_i\mathbf{w}_j^\top$. We witness that homophily in the case of \textsc{sHM-LDM} is expressed through a non-negative Eigenmodel (as in the unsigned case) while animosity/heterophily is expressed through a non-positive Eigenmodel able to express stochastic equivalence \cite{hoff2007modeling}. These two formulations admit the same embedding matrix $\mathbf{W}$ which balances the expression of ``opposing'' forces (homophily and animosity) in the latent space. Lastly, for $p=2$ both expressions admit to an NMF operation, obtaining an identifiable and unique factorization (up to permutation invariance) when $\mathbf{W}$ is full-rank and at least one node resides solely in each simplex corner \cite{nmf4,nmf5} as in the case of \textsc{HM-LDM} for unsigned networks.

\section{Experimental evaluation}\label{sec:experimental_evaluations}

We continue by assessing the effectiveness and efficacy of the suggested techniques. We start with the case of unsigned networks, including both latent and ground-truth community structures, and test \textsc{HM-LDM} based on its capability to detect identifiable latent structures as well as to perform link prediction. Additionally, for the networks with known community structures, we assess how the model can successfully infer the ground-truth community labels. We then continue with the case of signed networks for evaluating the performance of \textsc{sHM-LDM} in its ability to perform signed link prediction and discovery of distinct profiles.

For both the training of \textsc{HM-LDM} and \textsc{sHM-LDM}, we make use of the Adam optimizer \cite{kingma2017adam}, minimizing for the two models the log-likelihood function of Eq. \eqref{eq:prob_adj} and the MAP expression of Eq. \eqref{eq:loss_sk}, respectively. The learning rate is set as $lr \in [0.01,0.1]$. The node-specific random effects vectors for all models are randomly initialized and then tuned separately (for $1000$ iterations) by detaching initially the gradients from the latent representations $\mathbf{W}$. The latent embeddings matrix $\mathbf{W}$ is initialized based on the eigenvalues obtained by the spectral decomposition of the normalized Laplacian matrix of the network as expressed for unsigned \cite{10.5555/2980539.2980649,868688} and signed \cite{norm_lapl} networks. 

\begin{table}[!t]
\centering
\caption{Network statistics; $|\mathcal{V}|$: \# Nodes, $ |\mathcal{E}|$: \# Edges, $|\mathcal{K}|$: \# Communities.}
\label{tab:network_statistics}
\resizebox{1\textwidth}{!}{%
 \begin{tabular}{rcccccccc}\toprule
 & \textsl{AstroPh}\cite{snapnets}  & \textsl{GrQc}\cite{snapnets} & \textsl{Facebook}\cite{snapnets} & \textsl{HepTh}\cite{snapnets}& \textsl{Hamilton}\cite{fb_nets}  & \textsl{Amherst}\cite{fb_nets} & \textsl{Rochester}\cite{fb_nets} & \textsl{Mich}\cite{fb_nets} \\\midrule
$|\mathcal{V}|$ & 17,903 & 5,242 & 4,039 & 8,638 & 2,118 & 2,021 & 4,145 & 2,933 \\
$|\mathcal{E}|$ & 197,031 & 14,496 & 88,234 & 24,827 & 87,486 & 87,496 & 145,305 & 54,903 \\
$|\mathcal{K}|$& - & - & - & - & 15 & 15 & 19 & 13  \\\bottomrule
\end{tabular}%
}
\end{table}

\subsection{Unsigned Network Experiments}

We consider eight unsigned networks of various sizes and structures. We hereby supply the reader with additional information for the considered networks. The four networks with unknown community labels include (i) \textsl{AstroPh}, (ii) \textsl{GrQc}, and (iii) \textsl{HepTh} \cite{astroph_grqc_hepth} are collaboration networks based on papers submitted to the astrophysics, general relativity and quantum cosmology, and high energy physics categories of the e-print ArXiv, respectively. An edge between a pair of nodes (representing authors) is created if they have co-authored a paper. (iv) \textsl{Facebook} \cite{facebook} is a social network based on data obtained by a survey on a Facebook application. The additional four networks with ground-truth community labels include (v) \textsl{Hamilton}, (vi) \textsl{Amherst}, (vii) \textsl{Rochester}, and (viii) \textsl{Mich} which are all Facebook networks describing online friendships/connections of four American universities with the class year serving as the ground truth community \cite{fb_nets}. Network statistics are summarized by Table \ref{tab:network_statistics}. We treat the above networks as unweighted and undirected.

For the experiments, we consider eleven various prominent graph representation learning methods to evaluate the performance of our proposed approach. These are: (i) \textsc{DeepWalk} \cite{deepwalk-perozzi14}, (ii) \textsc{Node2Vec} \cite{node2vec-kdd16} which are two random-walk based methods. (iii) \textsc{LINE} \cite{line} learning node embeddings vectors by optimizing the first- and second-order proximity information. (iv) \textsc{NetMF} \cite{netmf-wsdm18} that factorizes the pointwise mutual information matrix of node co-occurrences obtained by random walks. (v) \textsc{NetSMF} \cite{netsmf-www2019} the scalable extension of the \textsc{NetMF} method \cite{netmf-wsdm18}. (vi) \textsc{LouvainNE} \cite{louvainNE-wsdm20} obtaining node representations by aggregating hierarchical embeddings of extracted network sub-graphs. (vii) \textsc{ProNE} \cite{prone-ijai19} which finds representation based on a sparse matrix factorization and spectral propagation operations. We also consider four NMF-based embedding approaches able to convey information about community memberships. These include (viii) \textsl{NNSED} utilizing an encoder-decoder approach for community detection. (ix) \textsl{MNMF} unifying NMF representation learning with modularity-based community detection. (x) \textsl{BigClam} defining a model-based community detection algorithm able to detect overlapping community structures. (xi) \textsl{SymmNMF} decomposing a pairwise similarity measure matrix between nodes of the network admitting graph clustering properties.

\textbf{Link prediction:}
To conduct the link prediction experiments, we adopt a commonly used approach \cite{deepwalk-perozzi14, node2vec-kdd16}, where we eliminate half of the network edges while ensuring that the remaining network stays connected. The removed edges, together with the equivalent number of node pairs (that were not part of the original network edges), create the negative instances for the test set. The models learn network embeddings based on the remaining network.

For the link prediction experiments, we use the four networks with unknown community structures and compare the performance in Table \ref{tab:auc_roc}, in terms of the Area Under Curve-Receiver Operating Characteristic (AUC-ROC) metric. We benchmark \textsc{HM-LDM} against other notable GRL and NMF models while considering the performance across various dimensions. All baselines are fine-tuned, and feature vectors for dyads are generated using binary operators (average, Hadamard, weighted-L1, weighted-L2) \cite{node2vec-kdd16}. For the baselines, we further train a logistic regression model with $L_2$ regularization and based on the constructed feature vectors make link predictions. Specifically, we choose the optimal hyperparameters and binary operator for each baseline model, based on which operator and hyperparameters return the highest AUC-ROC score.

For our frameworks, we follow a different approach leading to an unbiased estimation of link prediction performance. More specifically, we report results based on the first $\delta$ value (as we decrease the volume) that makes the solution identifiable, meaning the $\delta$ value where at least one community champion resides in a simplex corner. Importantly, there exist additional values for $\delta$ which define identifiable solutions as well as increased performance with respect to the reported one but are disregarded so the evaluation stays unbiased. In addition, predictions for \textsc{HM-LDM} are based directly on the Poisson rates $\lambda_{ij}$ defined for test set pairs $\{i,j\}$ with AUC-ROC scores as reported in Table \ref{tab:auc_roc}. This comes as an advantage of \textsc{HM-LDM} since it defines a likelihood function over the network connections and thus has no need for post-processing steps (such as training a logistic regression model) to make predictions. The true dimensions for \textsc{\modelabbrv} are $D+1$ but reported as $D$ since this is the true number of model parameters, for a fair comparison with the baselines. Results for our method are reported based on the average performance over five independent runs of the model (error bars were found to be in the scale of $10^{-3}$ and thus not presented). 

Upon contrasting our findings with the non-NMF models, we found that our \textsc{\modelabbrv} (either $p=1$ or $p=2$) outperforms these baselines and, in most cases, by a significant margin, producing favorable results. We notice a considerable difference in performance when comparing \textsc{\modelabbrv} with other part-based representation models, indicating the existence of identifiable regimes based on $\delta$ values where we can successfully obtain community memberships while simultaneously demonstrating the link prediction abilities of unconstrained \textsc{LDM}. (AUC Precision-Recall scores are similar to the AUC-ROC scores and thus not presented)

\begin{table*}[!t]
\centering
\caption{Area Under Curve (AUC-ROC) scores for varying representation sizes.}
\label{tab:auc_roc}
\resizebox{0.85\textwidth}{!}{%
\begin{tabular}{rcccccccccccc}\toprule
\multicolumn{1}{l}{} & \multicolumn{3}{c}{\textsl{AstroPh}} & \multicolumn{3}{c}{\textsl{GrQc}} & \multicolumn{3}{c}{\textsl{Facebook}}& \multicolumn{3}{c}{\textsl{HepTh}}\\\cmidrule(rl){2-4}\cmidrule(rl){5-7}\cmidrule(rl){8-10}\cmidrule(rl){11-13}
\multicolumn{1}{r}{Dimension ($D$)} & $8$ & $16$ & $32$ & $8$ & $16$ & $32$ & $8$ & $16$ & $32$& $8$ & $16$ & $32$ \\\cmidrule(rl){1-1}\cmidrule(rl){2-2}\cmidrule(rl){3-3}\cmidrule(rl){4-4}\cmidrule(rl){5-5}\cmidrule(rl){6-6}\cmidrule(rl){7-7}\cmidrule(rl){8-8}\cmidrule(rl){9-9}\cmidrule(rl){10-10}\cmidrule(rl){11-11}\cmidrule(rl){12-12}\cmidrule(rl){13-13}
\textsc{DeepWalk}\cite{deepwalk-perozzi14}    &.945 	&.950	&.952  & .919	&.916 &	.929 &   .986 &	.986	& .984 &.874 &.867 & .873 \\
\textsc{Node2Vec}\cite{node2vec-kdd16}    &.950 	&\underline{.962}	&\underline{.957} & .897	&.913	&.930    & \underline{.988}    & \underline{.988} & \underline{.987} &.881 &.882 &.881   \\
\textsc{LINE}  \cite{line}      &.909 	&.938	&.947 & .920 &.925	&.919 & .981 & .987 & .983 &.873 &.886 &.882 \\
\textsc{NetMF} \cite{netmf-wsdm18}     &.813 	&.823	&.839 & .860 & .866&.877	 &  .935 & .963   & .971  &.792&.806 &.821\\
\textsc{NetSMF} \cite{netsmf-www2019}     & .891	&.901	&.919 & .837&.858	&.886 &.975  &.981  &.985 & .809&.822  &.836  \\
\textsc{LouvainNE}\cite{louvainNE-wsdm20}   & .813	&.811	&.819 & .868	&.875	&.873 & .958 &.961 &.963 &.874 &.867 &.873 \\
\textsc{ProNE}\cite{prone-ijai19}  & .907	&.929	&.947 & .885	&.911	&.921 & .971 & .982 & .987   &.827 &.846 &.859  \\\midrule
\textsc{NNSED}\cite{NNSED}   &.861 	&.882	&.891   & .792 	&.808 	&.828    &.908   &.927   &.935    & .756   & .779   &.796  \\
\textsc{MNMF}\cite{MNMF}   & .893	&.925	&.943   & .911 	&.928 	&.937    &.965   &.978   &.982    & .857   &.880    &.891  \\
\textsc{BigClam}\cite{nmf3}  &.500 	&.723	&.810   & .752 	&.769 	&.780    & .744  &.722   &.647    & .776   &.700    &.748  \\
\textsc{SymmNMF}\cite{SymmNMF}  &.767 	&.779	&.800   & .729 	&.772 	&.835    & .933  &.942   &.951    & .696   &.727    &.766  \\\midrule
\textsc{HM-LDM($p=1$)}  & \underline{.956}	&.952	&.952   &\textbf{.944} 	&\textbf{.948}	&\textbf{.951}   & .982  & .979 & .974   &\textbf{.916}   & \textbf{.921}  &\textbf{.924}
\\
\textsc{\modelabbrv ($p=2$)}      &\textbf{.972}   &\textbf{.973}   & \textbf{.963}  &\underline{.940} & \underline{.942} & \underline{.946} & \textbf{.992}     & \textbf{.993}      & \textbf{.993}      &\underline{.908} &\underline{.910} &\underline{.911}
\\\bottomrule    
\end{tabular}%
 }
\end{table*}
 
\begin{figure}[!b]
  \centering
  \subfloat[\textsl{AstroPh}]{{\includegraphics[width=0.23\columnwidth]{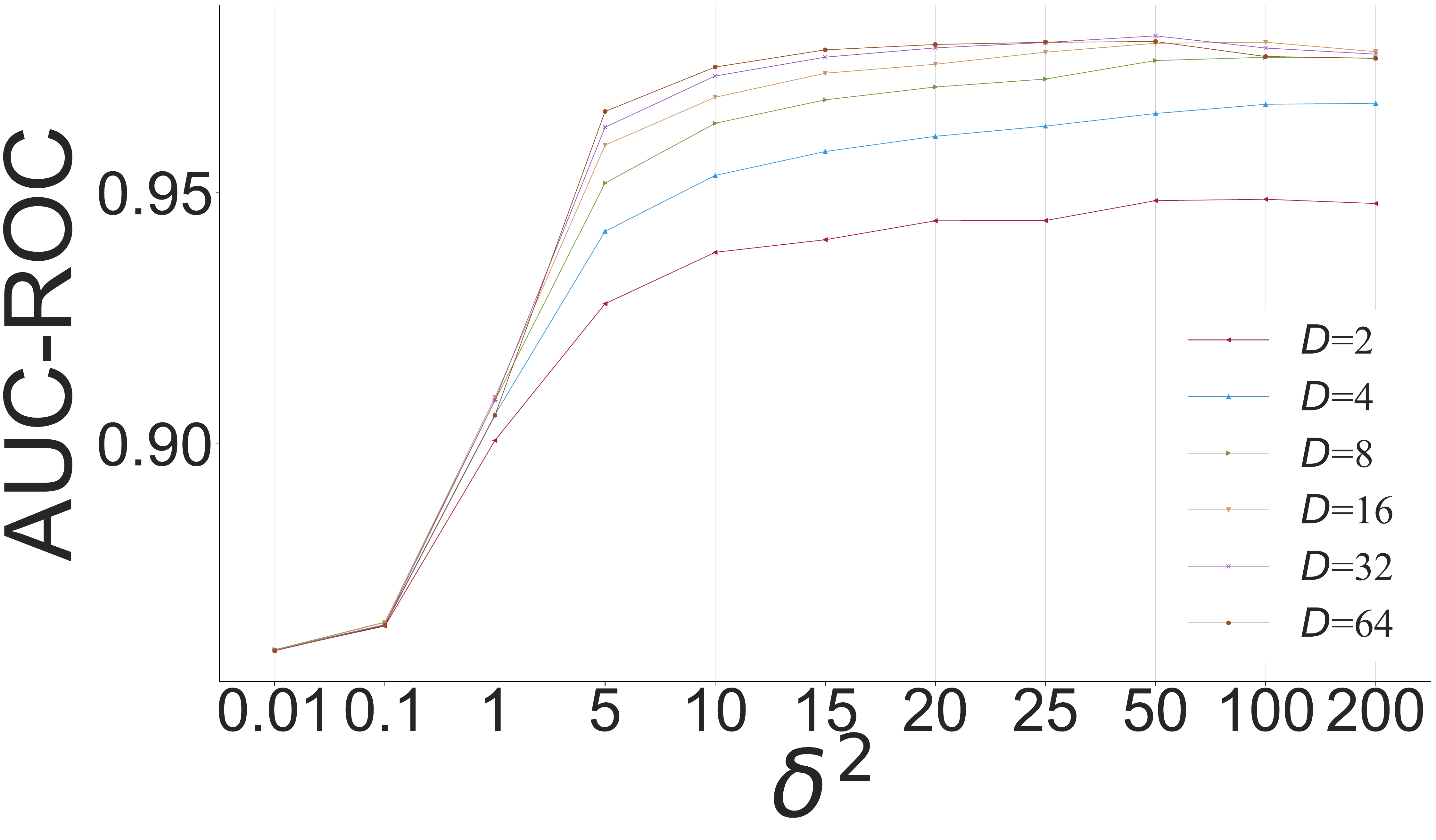} }}%
\hfill 
  \subfloat[\textsl{Facebook}]{{ \includegraphics[width=0.23\columnwidth]{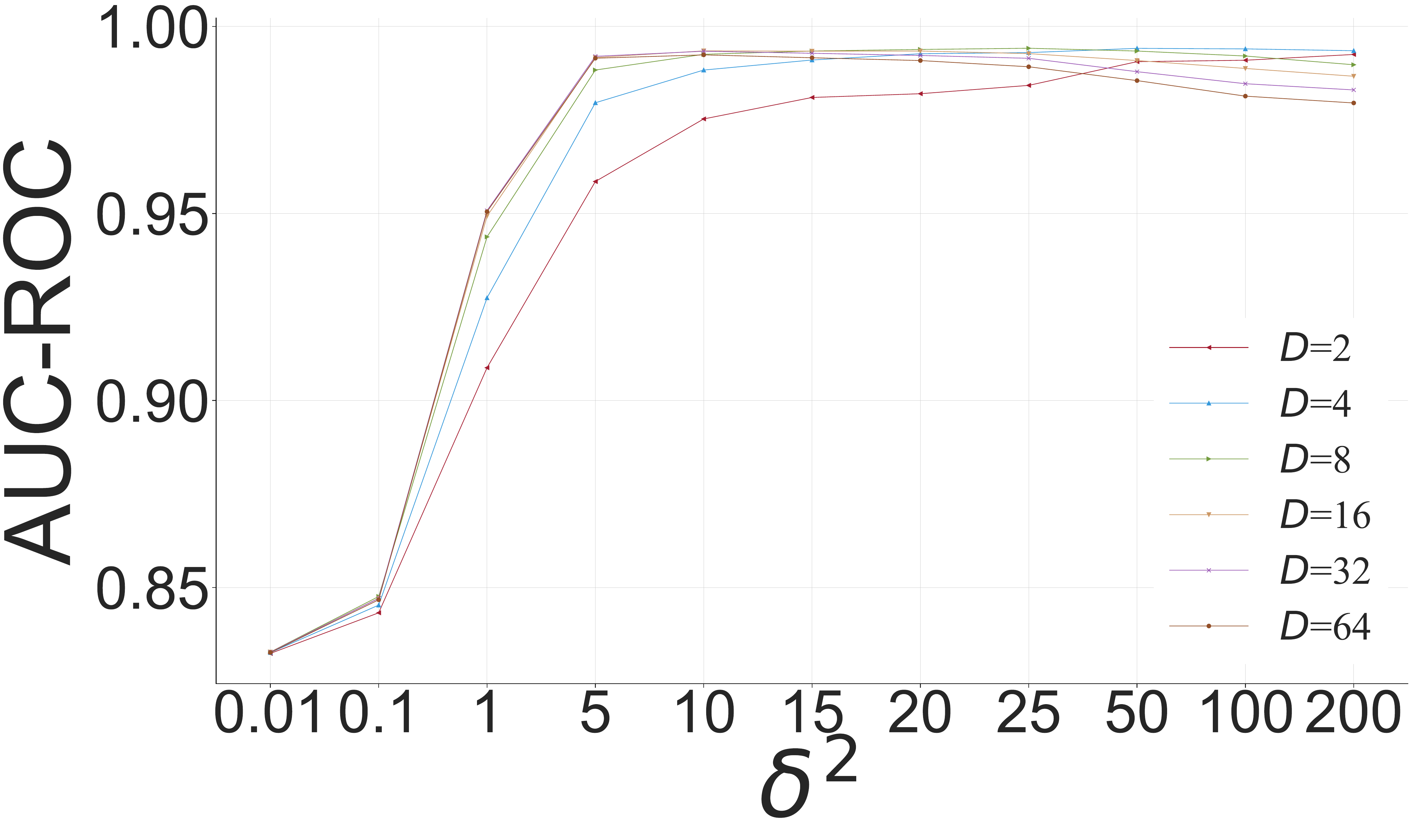} }}%
\hfill 
  \subfloat[\textsl{GrQc}]{{ \includegraphics[width=0.23\columnwidth]{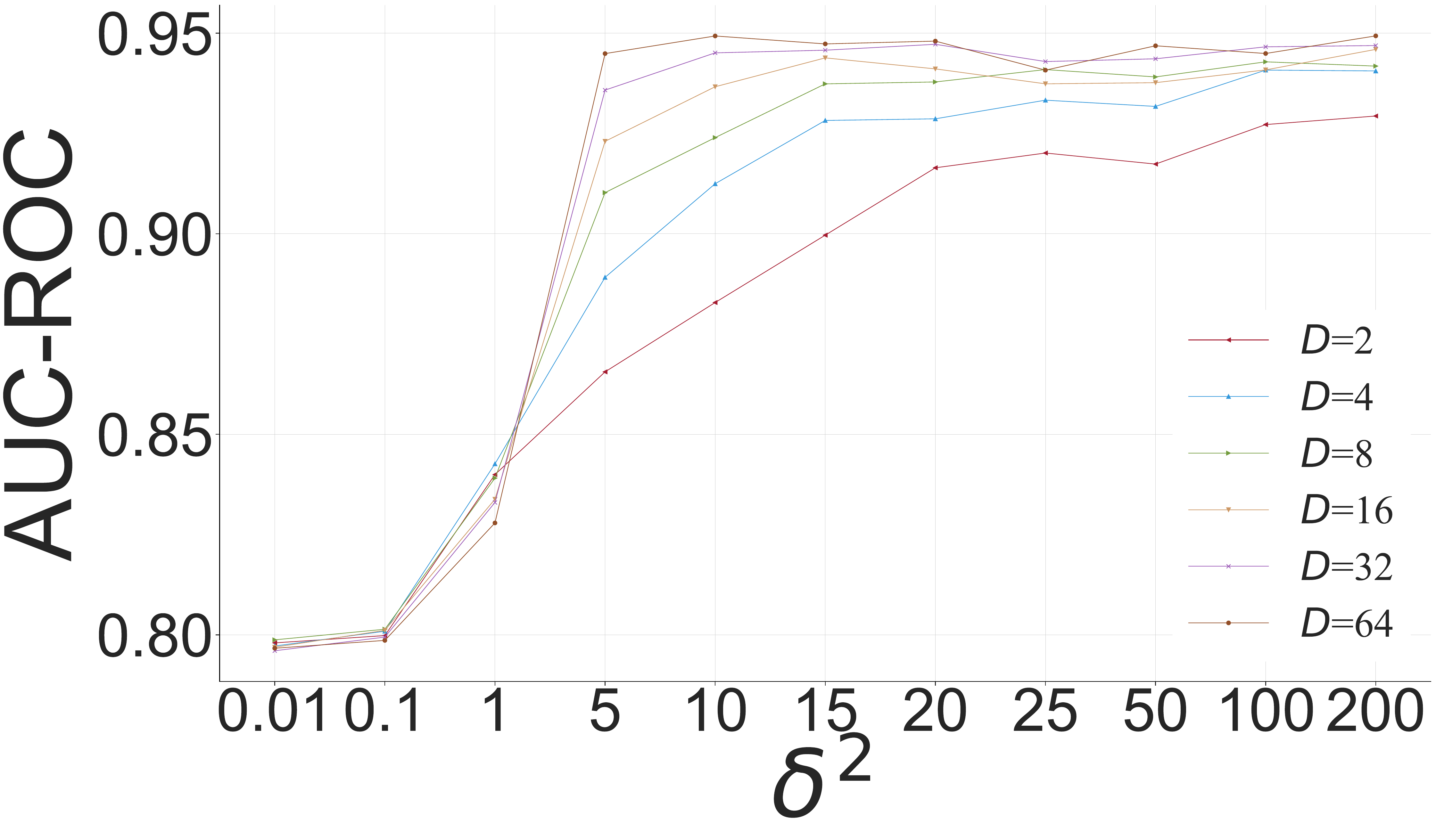} }}%
\hfill
  \subfloat[\textsl{HepTh}]{{ \includegraphics[width=0.23\columnwidth]{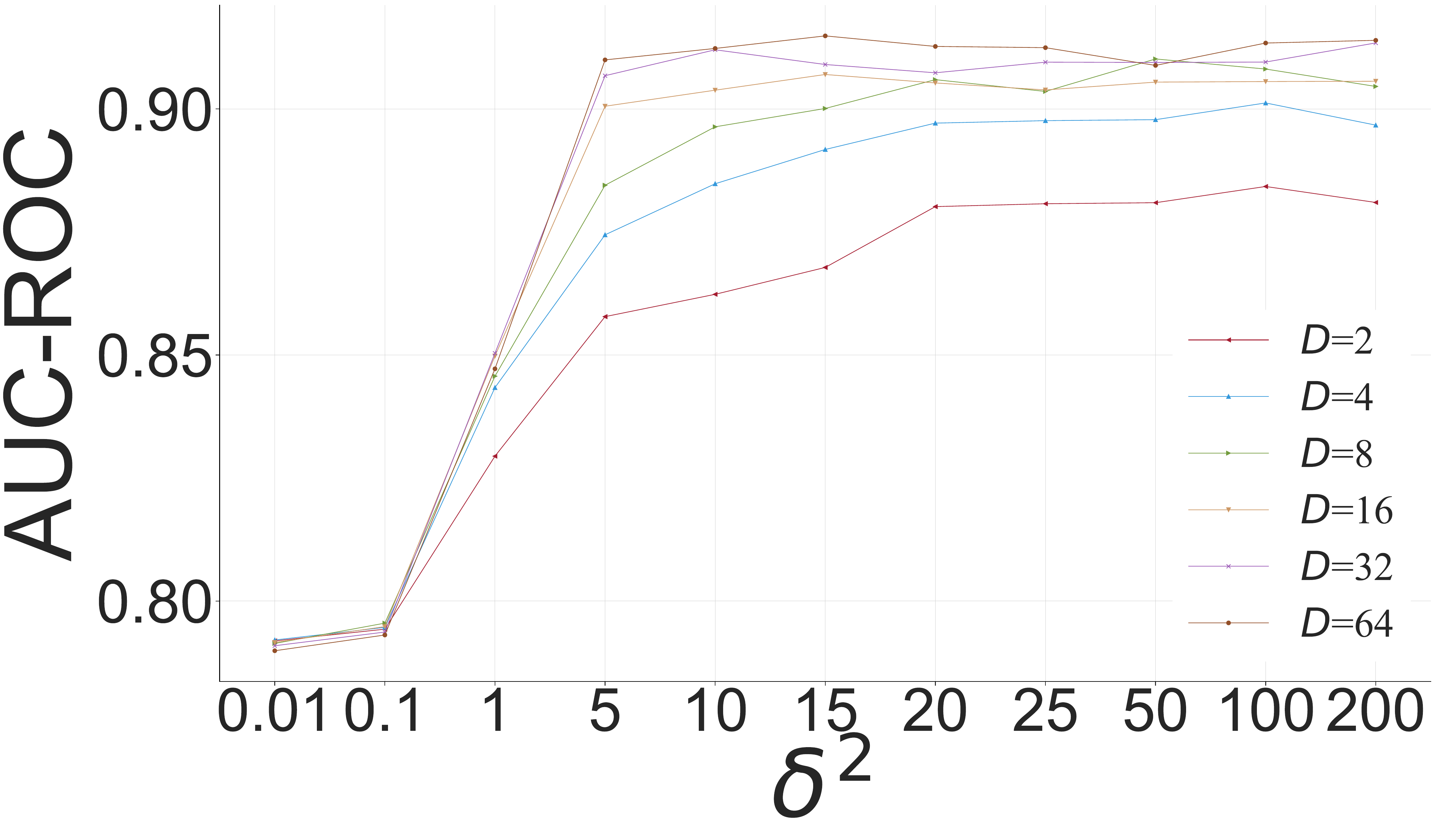} }}%
  \hfill
    \subfloat[\textsl{AstroPh}]{{ \includegraphics[width=0.23\columnwidth]{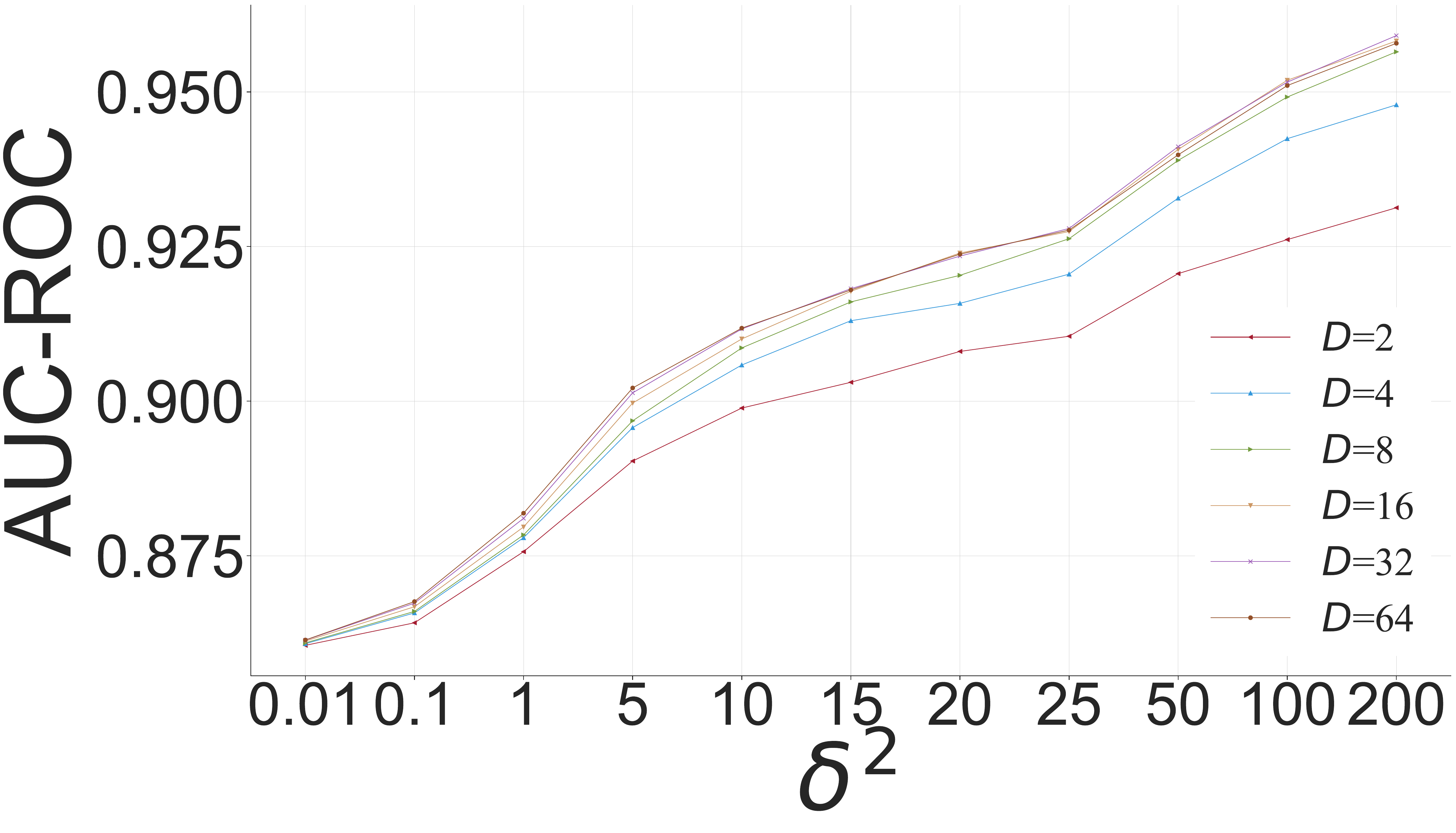} }}%
\hfill 
  \subfloat[\textsl{Facebook}]{{ \includegraphics[width=0.23\columnwidth]{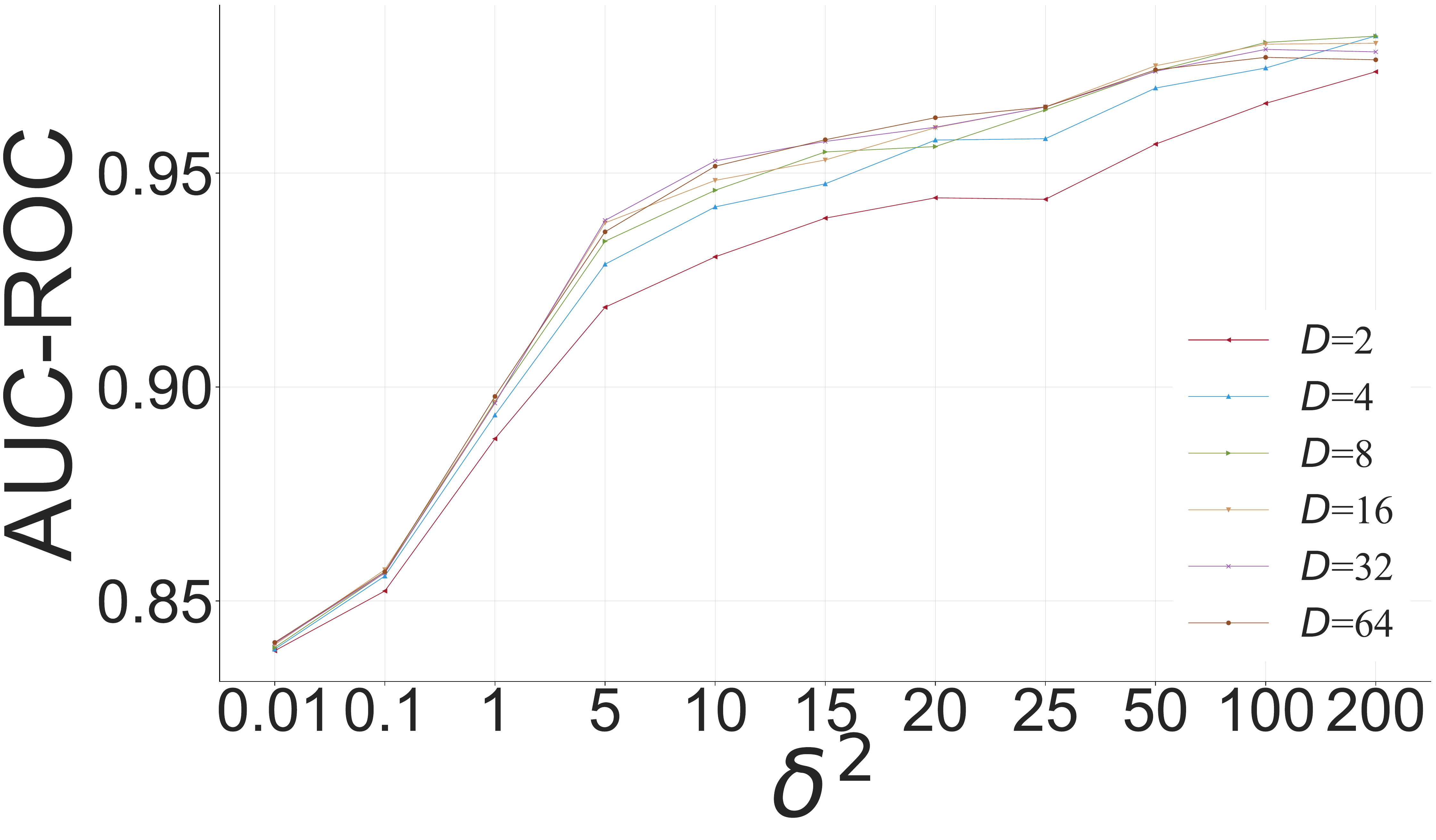} }}%
\hfill 
  \subfloat[\textsl{GrQc}]{{ \includegraphics[width=0.23\columnwidth]{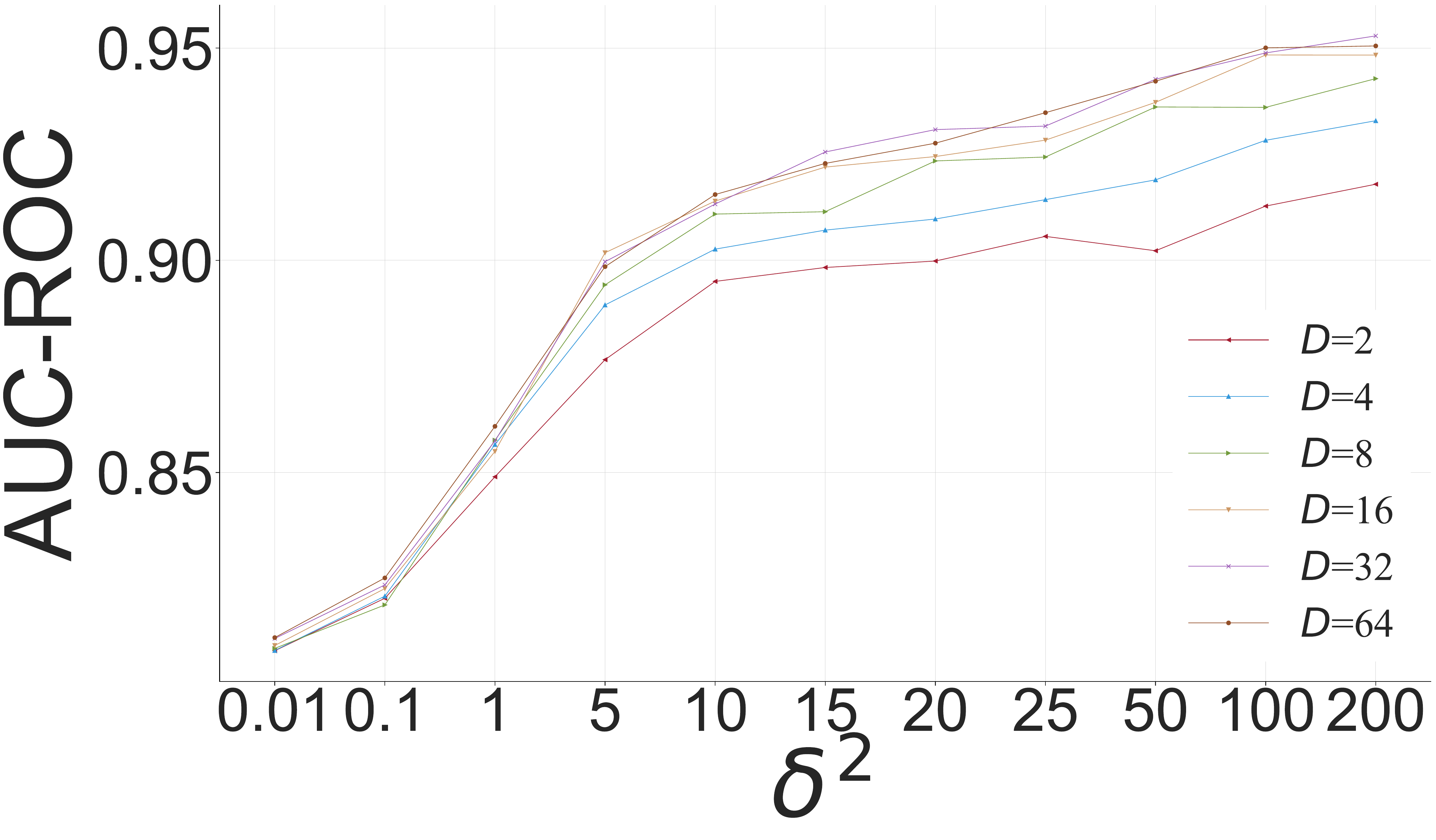} }}%
\hfill
  \subfloat[\textsl{HepTh}]{{ \includegraphics[width=0.23\columnwidth]{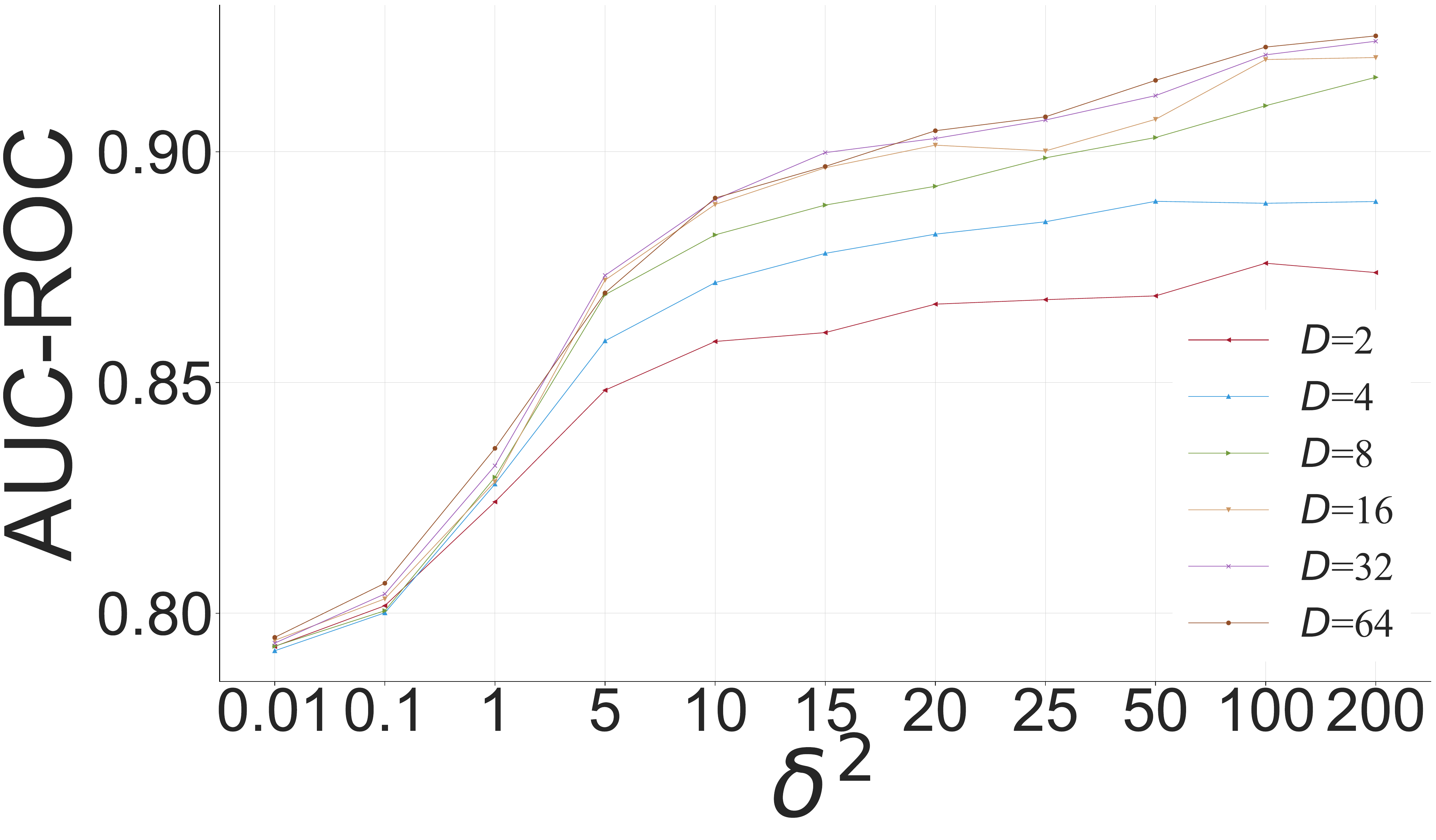} }}%
  \caption{AUC-ROC scores as a function of $\delta^2$ (simplex size) across dimensions for \textsc{\modelabbrv}. Increasing $\delta^2$ (simplex volume) leads to higher performance as the model becomes more flexible until saturation (unconstrained $\textsc{LDM}$ regime). Top row: $p = 2$ model specification. Bottom row $p = 1$ model specification.}\label{fig:roc_d}
\end{figure}

 \textbf{Performance and simplex sizes:} Fig \ref{fig:roc_d} displays the AUC-ROC scores in terms of link prediction performance as a function of $\delta^2$ for various latent dimensions, and networks, and both $p=1$ and $p=2$. As expected, we here understand that small $\delta$ values provide the minimum scores. This is a direct consequence of the fact that homophily properties are not adequately met (except within clusters) due to the very small simplex volumes that these low $\delta$ values constrain the latent space to. If we think of \textsc{\modelabbrv} with $p=2$ as a positive Eigenmodel, we can also see how the positivity constraint on the $\Lambda$ diagonal matrix hinders the expression of stochastic equivalence, which would boost performance even on low simplex volumes. As we increase $\delta$ values, we naturally approach the performance of an unconstrained \textsc{LDM}. Comparing the case of p=2 (squared), and p=1 (simple) for the $\ell^2$-norm, we observe that the former reaches performance saturation more rapidly.
 
\begin{figure} [!t]
  \centering
  \subfloat[\textsl{AstroPh}]{{ \includegraphics[width=0.23\columnwidth]{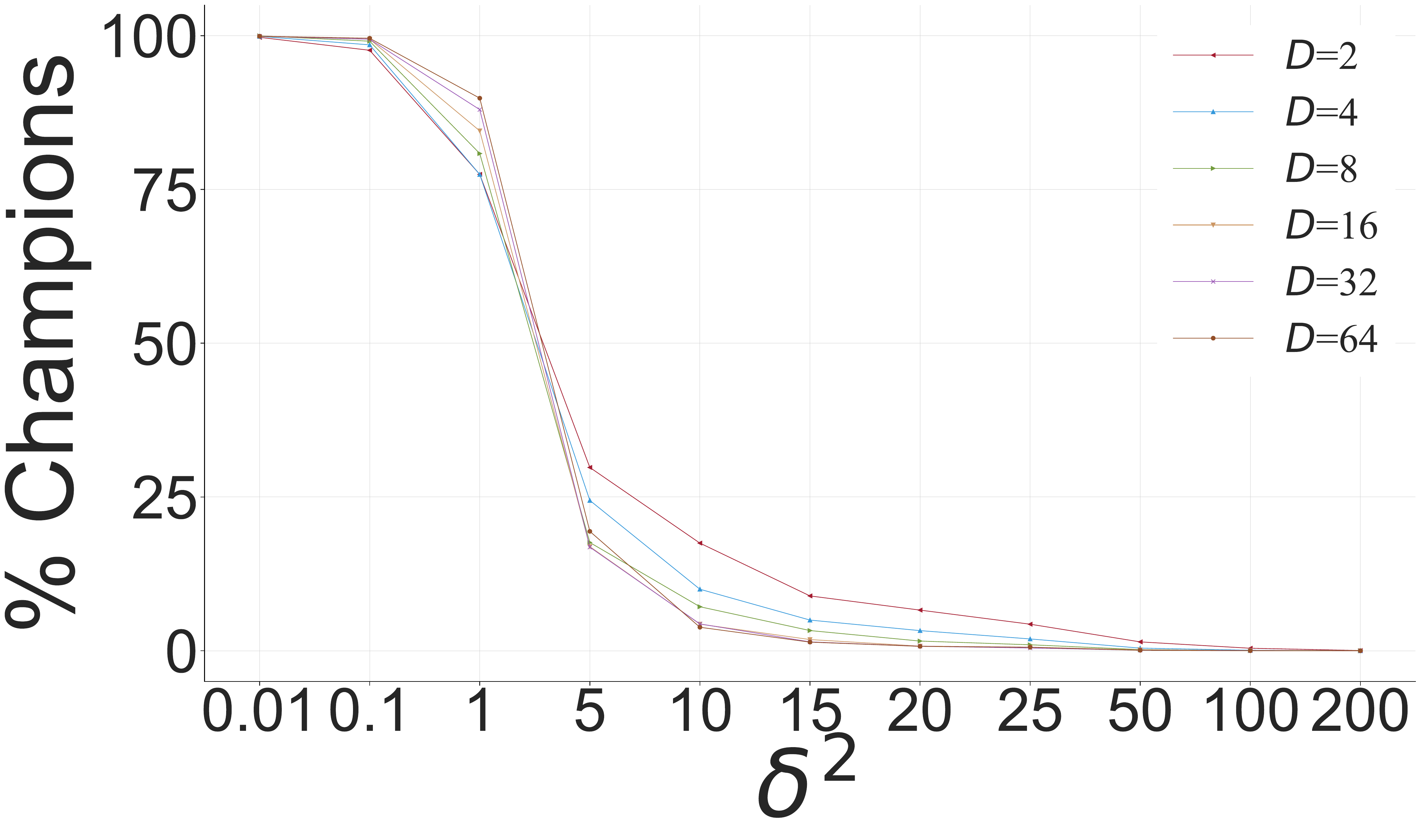} }}%
  \hfill
  \subfloat[\textsl{Facebook}]{{ \includegraphics[width=0.23\columnwidth]{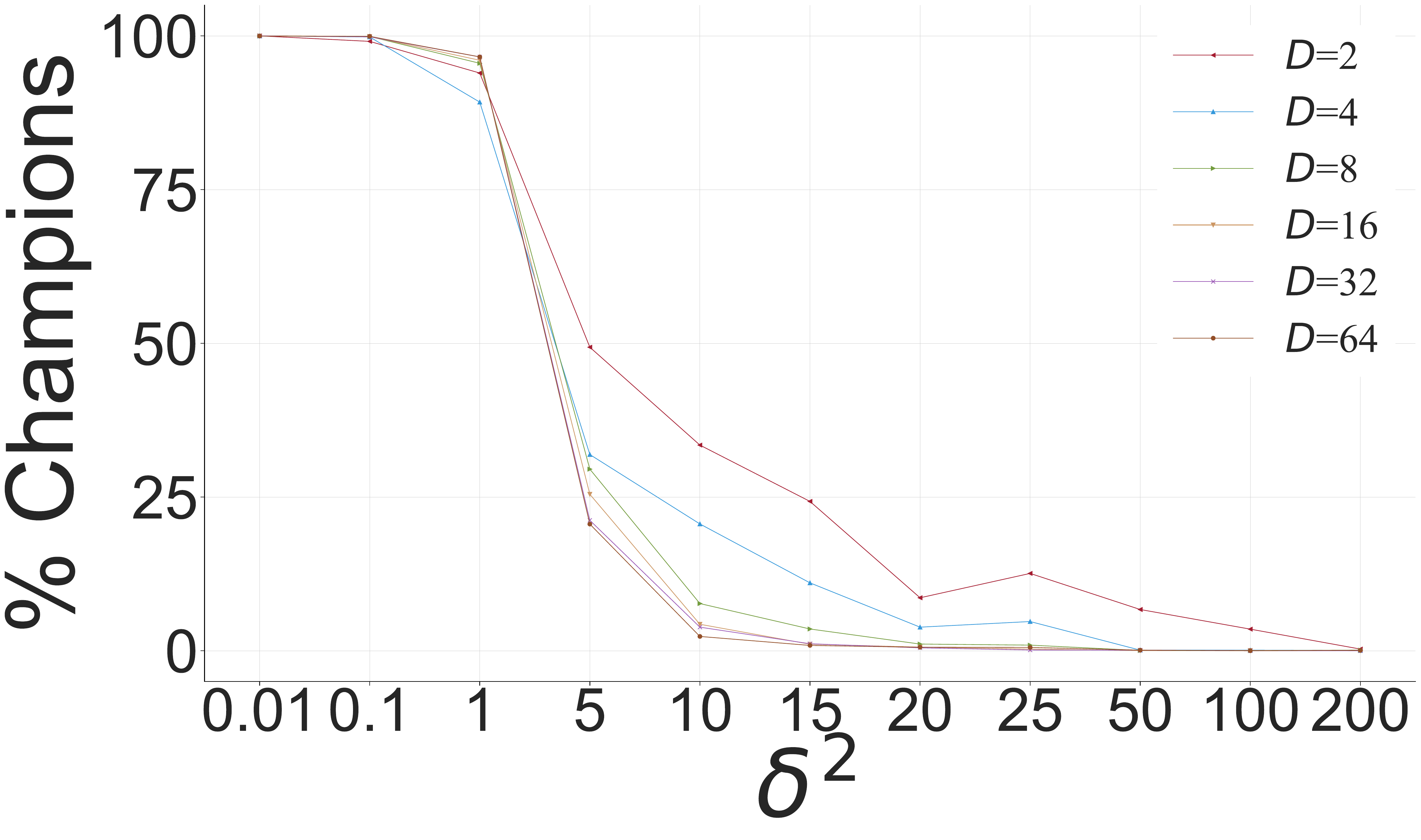} }}%
  \hfill
  \subfloat[\textsl{GrQc}]{{ \includegraphics[width=0.23\columnwidth]{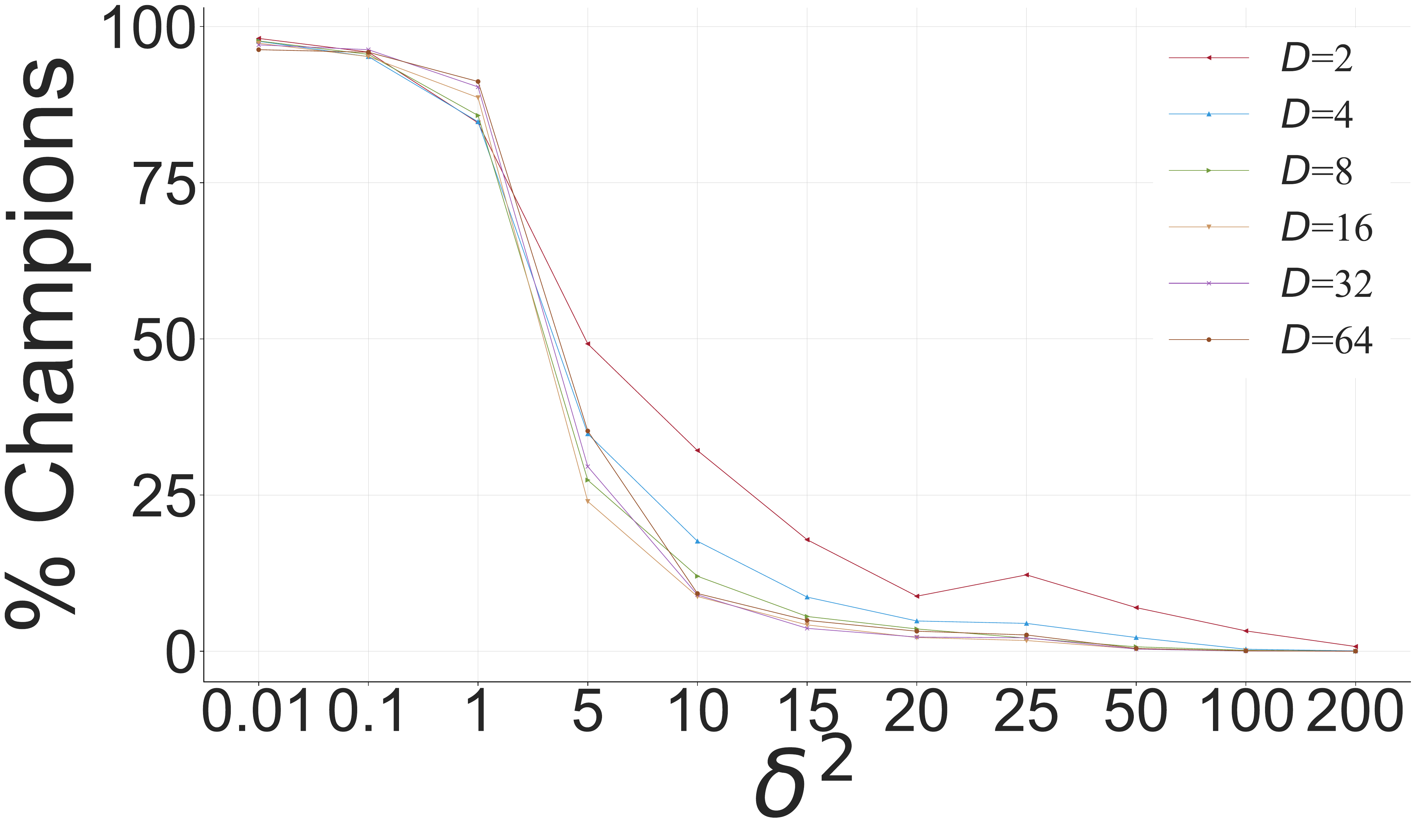} }}%
  \hfill
  \subfloat[\textsl{HepTh}]{{ \includegraphics[width=0.23\columnwidth]{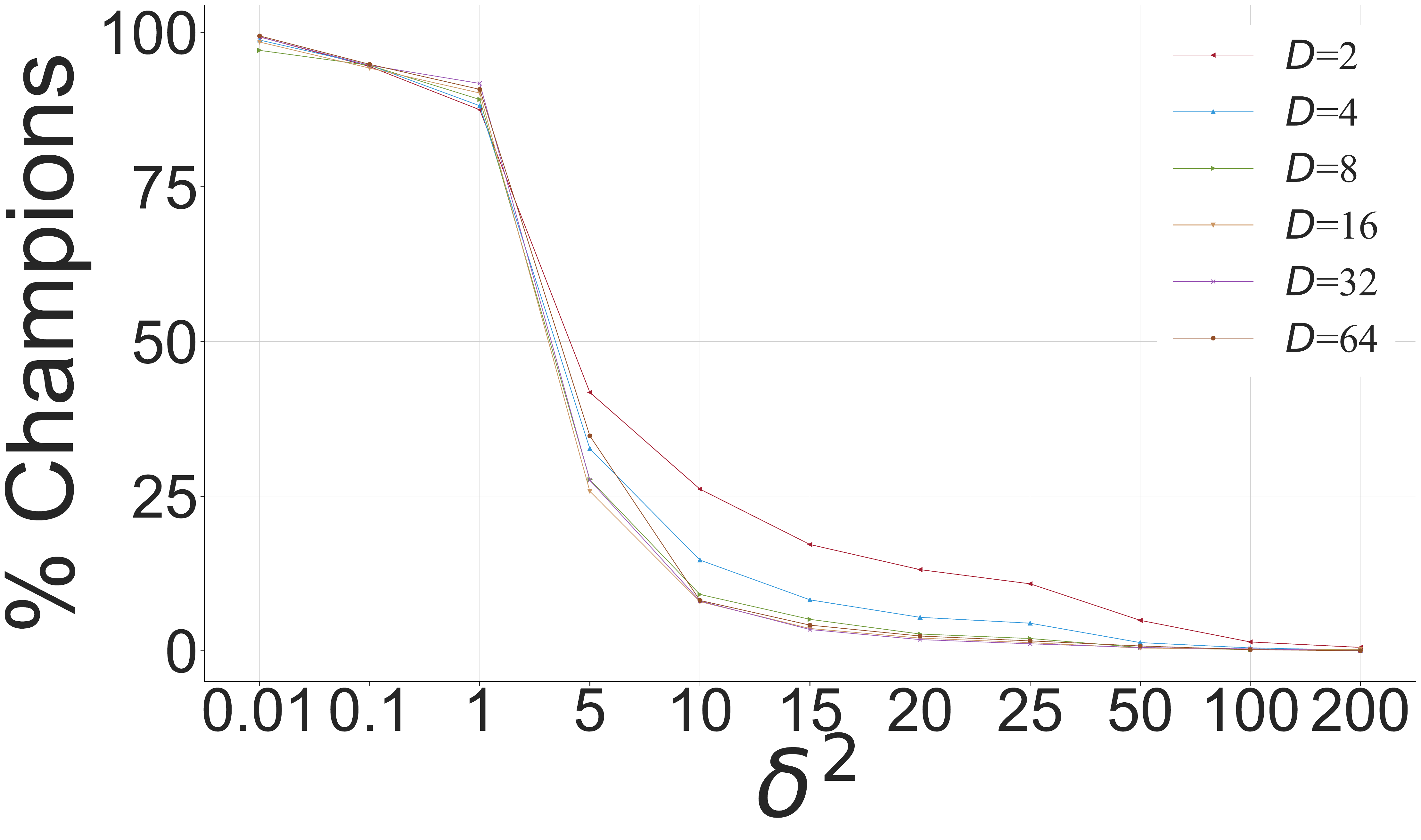} }}%
  \hfill
    \subfloat[\textsl{AstroPh}]{{ \includegraphics[width=0.23\columnwidth]{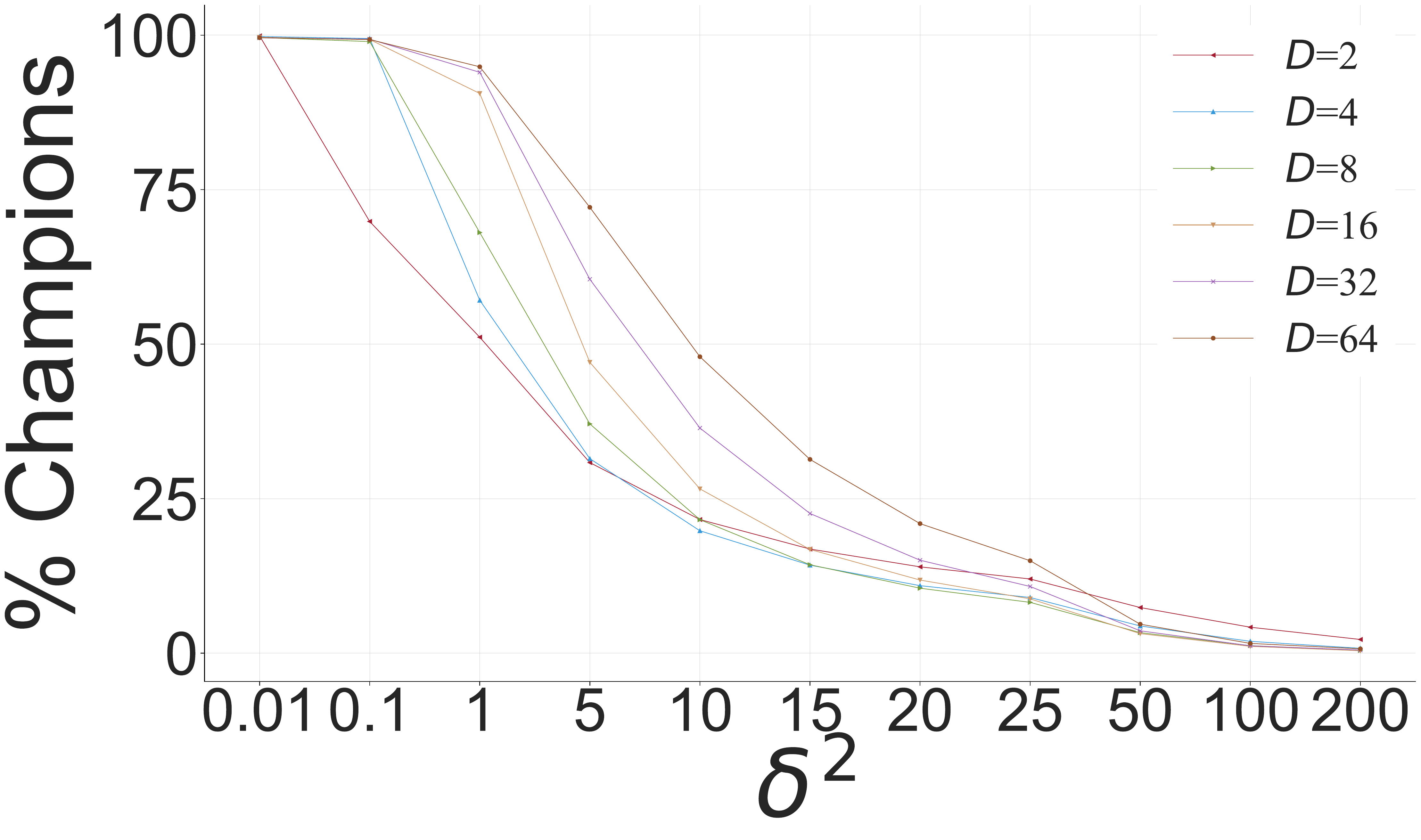} }}%
  \hfill
  \subfloat[\textsl{Facebook}]{{ \includegraphics[width=0.23\columnwidth]{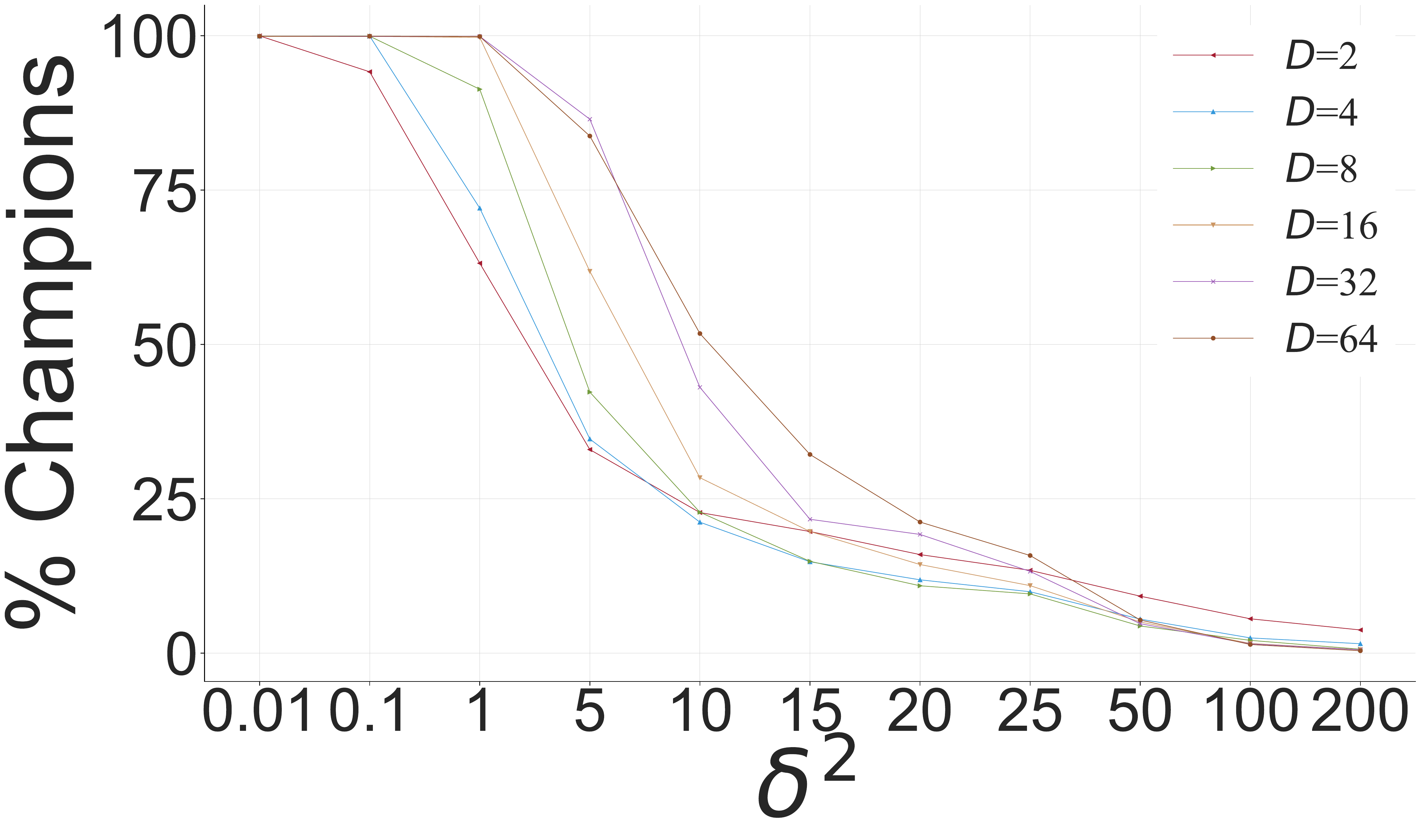} }}%
  \hfill
  \subfloat[\textsl{GrQc}]{{ \includegraphics[width=0.23\columnwidth]{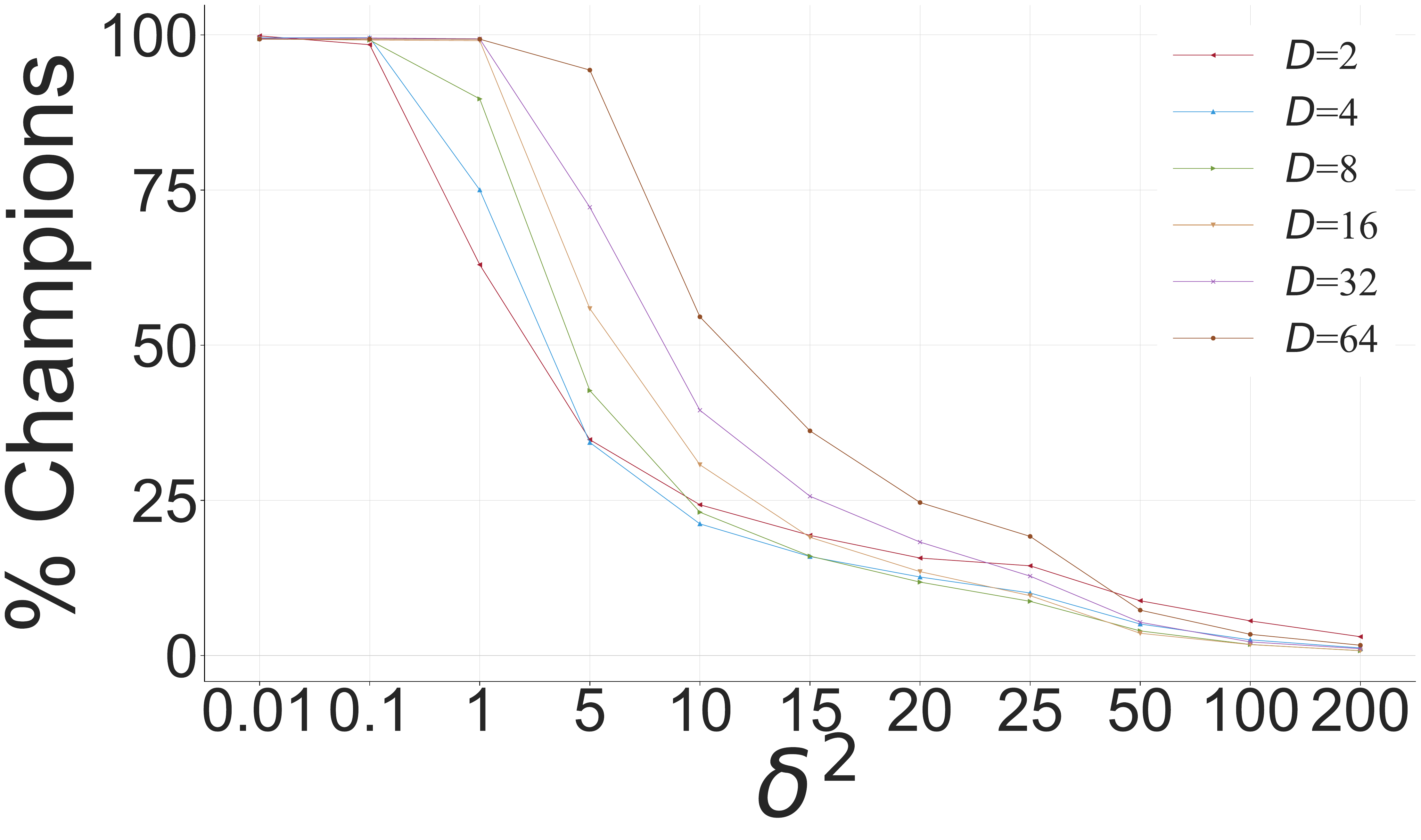} }}%
  \hfill
  \subfloat[\textsl{HepTh}]{{ \includegraphics[width=0.23\columnwidth]{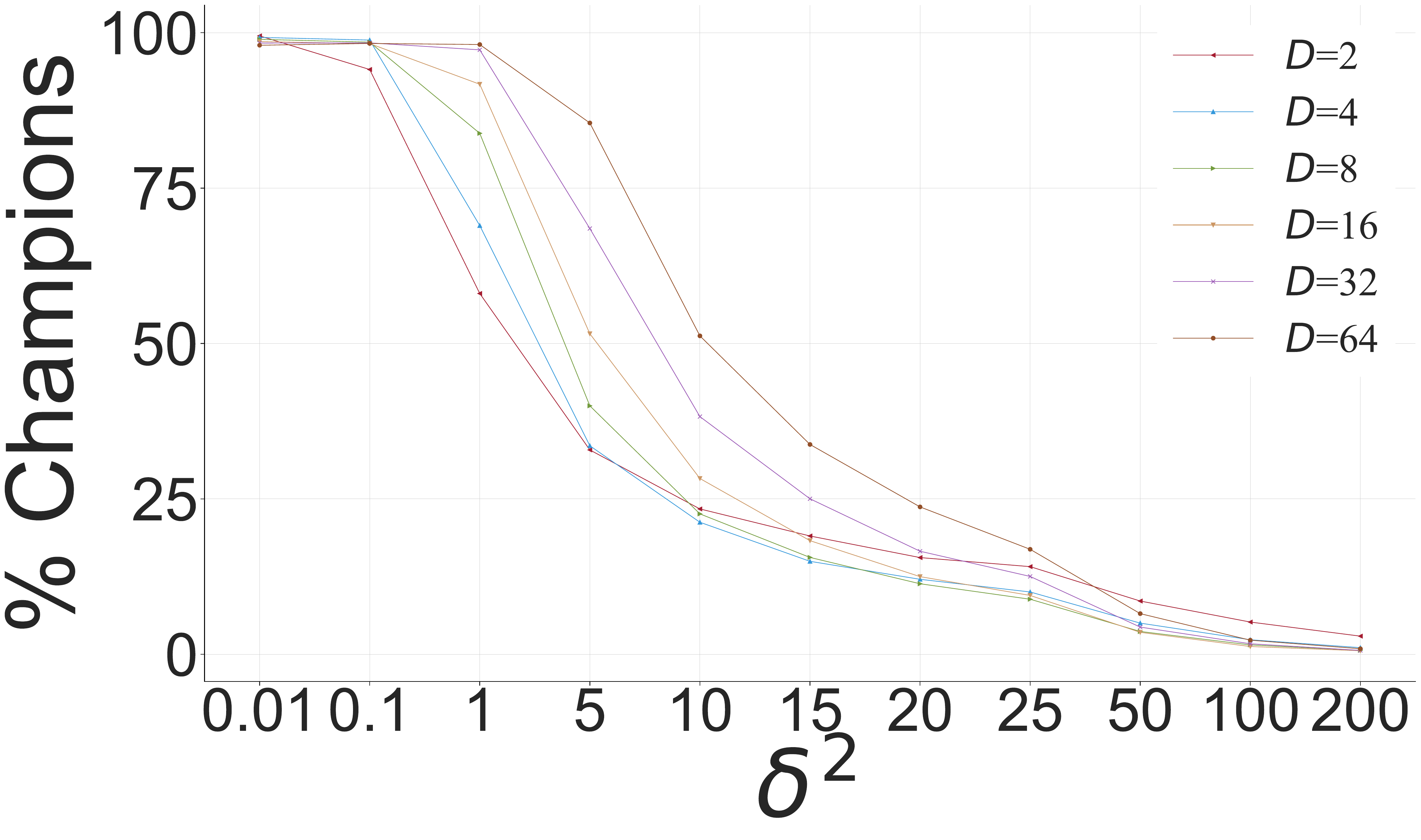} }}%
  \caption{Total community champions (\%) in terms of $\delta^2$ (simplex size) across dimensions for \textsc{\modelabbrv}. Decreasing $\delta^2$ (simplex volumes) leads to a higher percentage of nodes positioned on the simplex corners (equivalent to hard clustering) until all nodes are pushed on the corners for very small volumes. Top row: $p = 2$ model specification. Bottom row $p = 1$ model specification.}  \label{fig:phase_transitions}
\end{figure}

\begin{figure}[!t]
  \centering
  \subfloat[\textsl{GrQc} $(p=2)$]{{ \includegraphics[width=0.23\columnwidth]{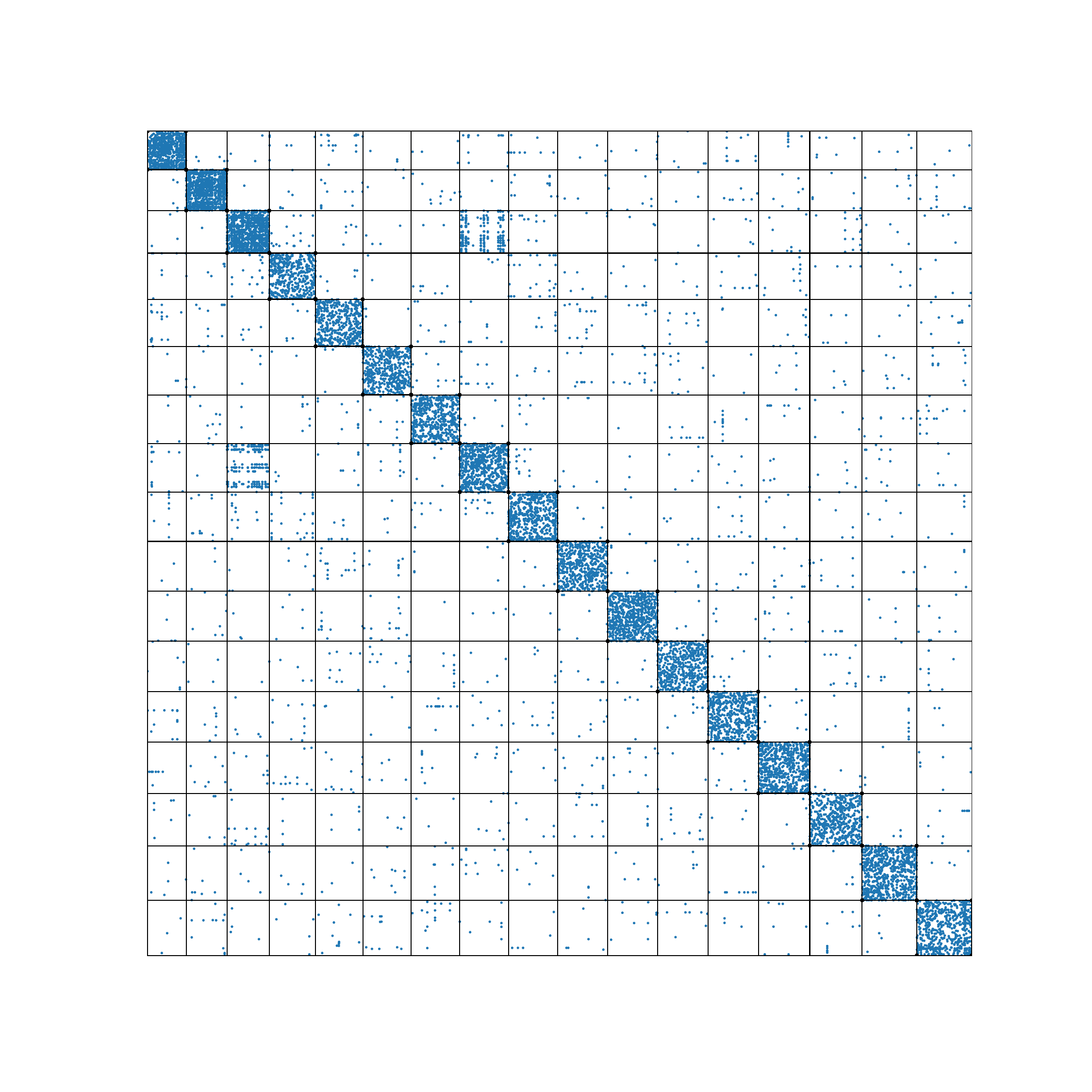} }}%
\hfill
  \subfloat[\textsl{HepTh} $(p=2)$]{{ \includegraphics[width=0.23\columnwidth]{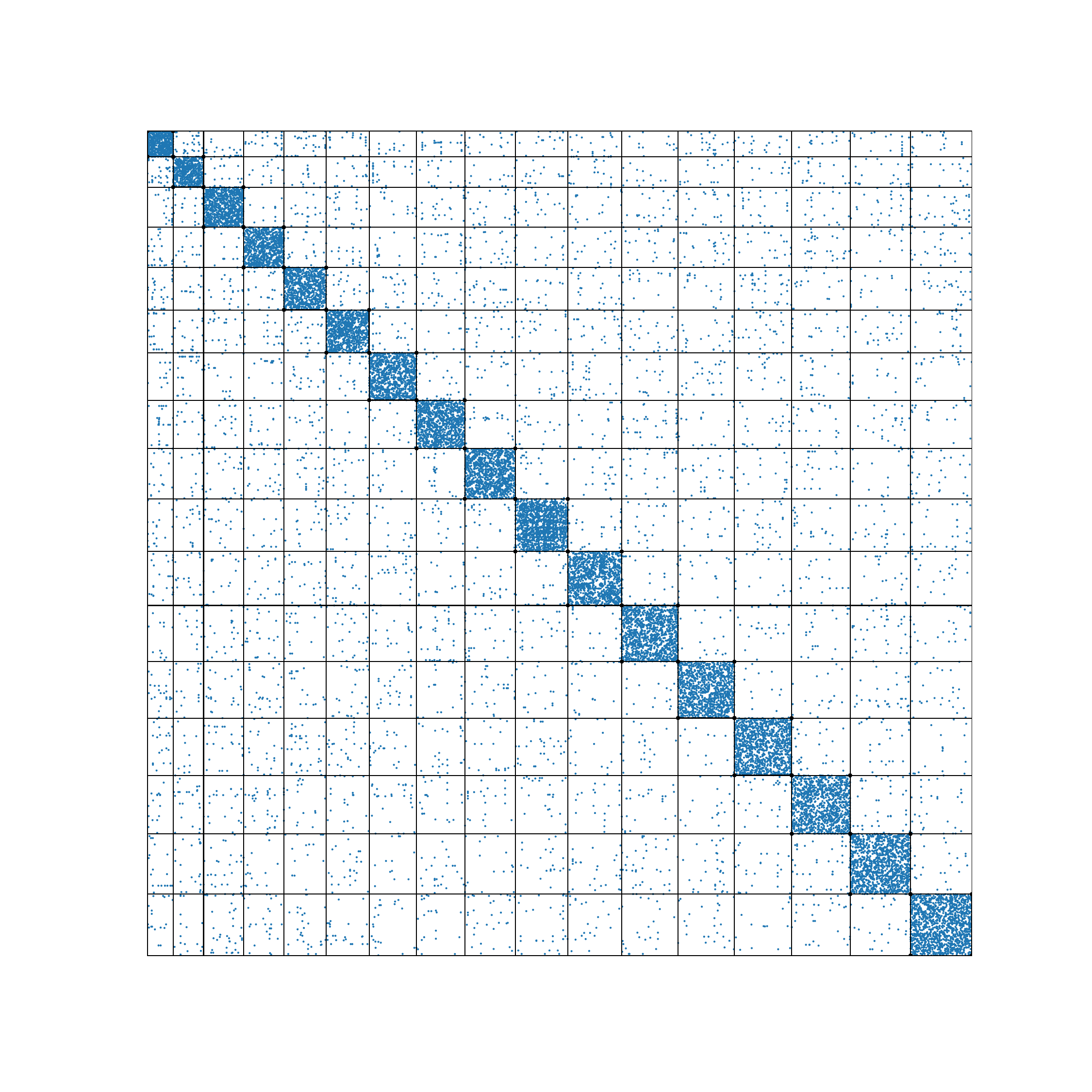} }}%
\hfill
  \subfloat[\textsl{GrQc} $(p=1)$]{{ \includegraphics[width=0.23\columnwidth]{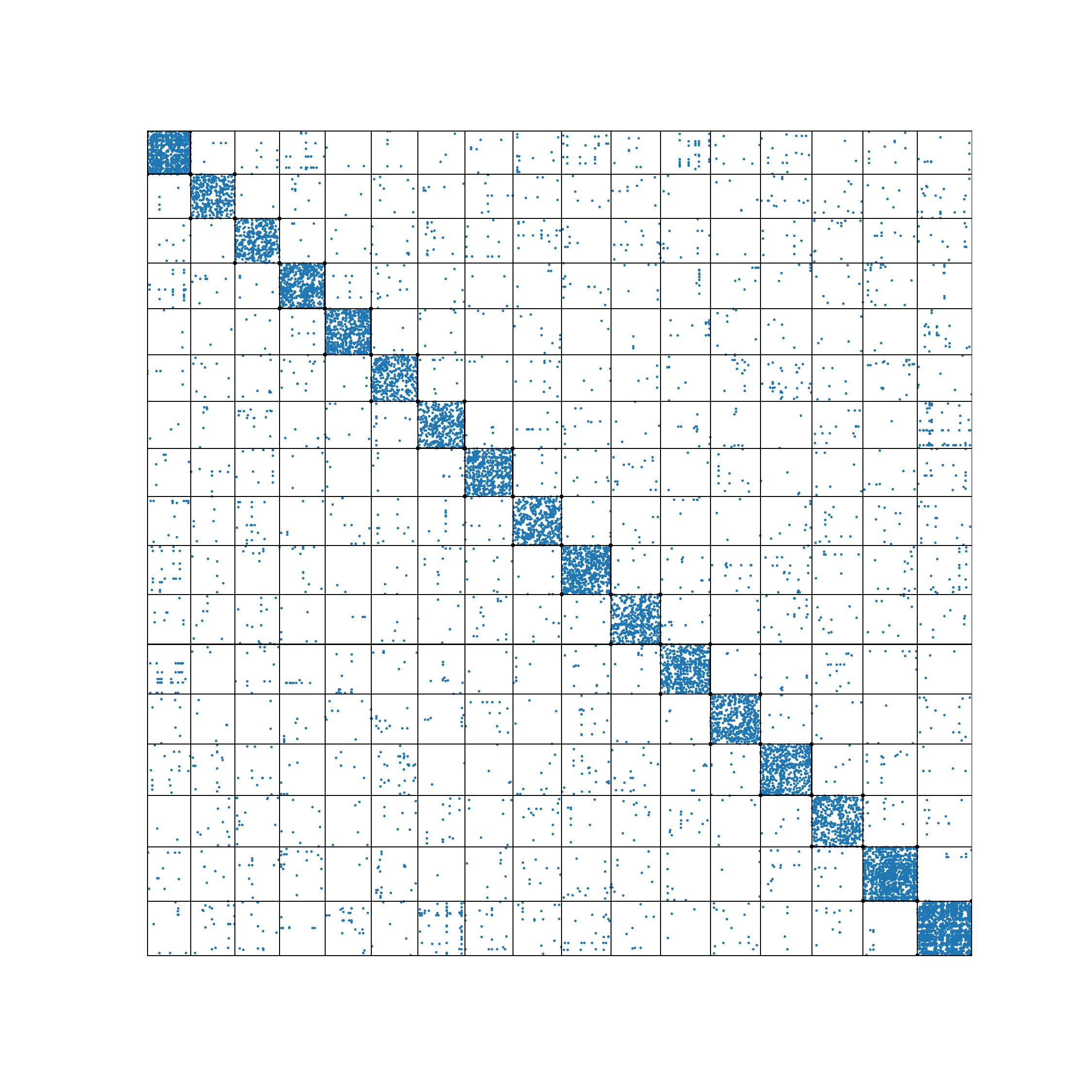} }}%
\hfill
  \subfloat[\textsl{HepTh} $(p=1)$]{{ \includegraphics[width=0.23\columnwidth]{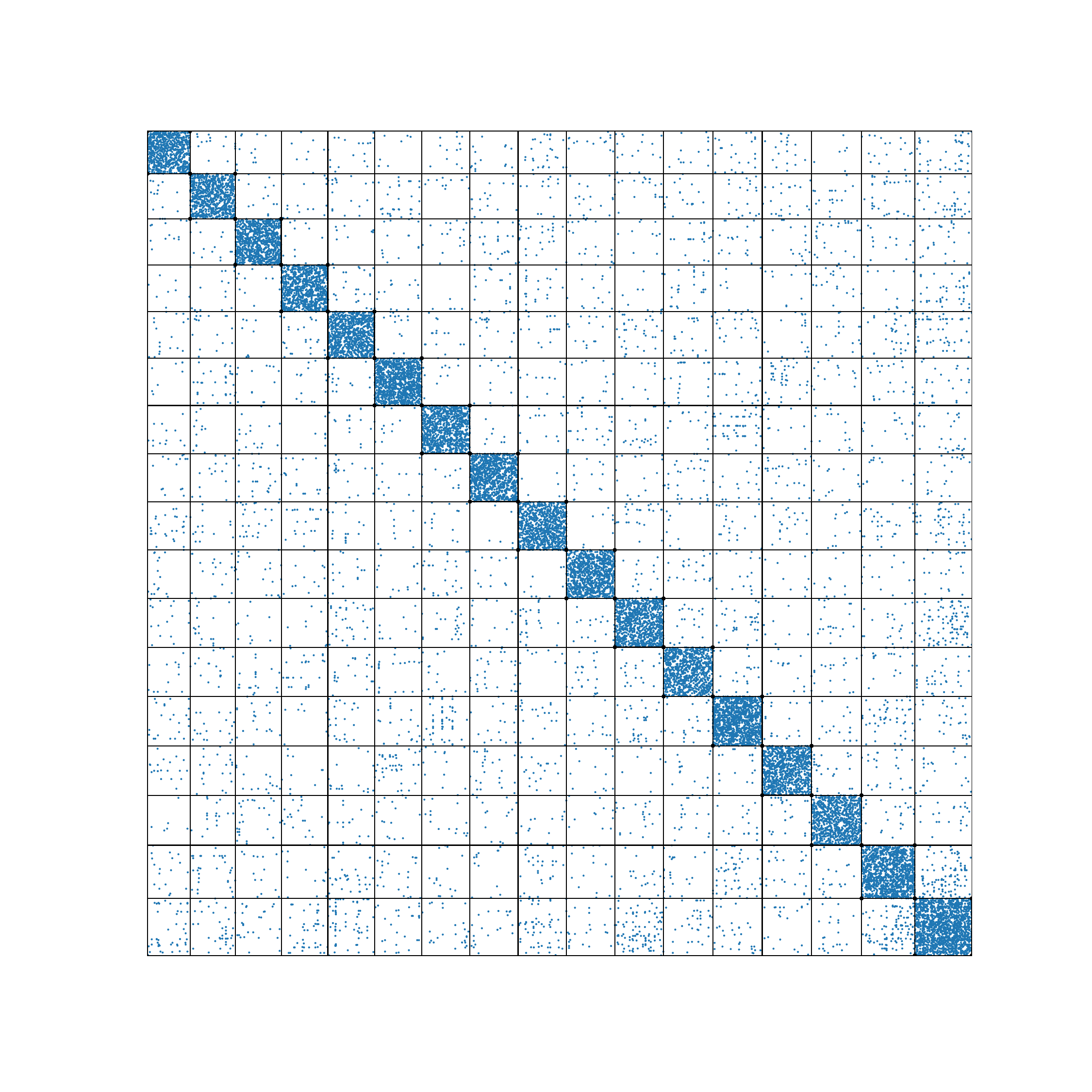} }}%
  \caption{Ordered adjacency matrices based on the memberships of a $D=16$ dimensional \textsc{\modelabbrv} with $\delta$ values ensuring identifiability, empirically showcasing community extraction and identification.}\label{fig:adj}
\end{figure}

 \textbf{Type and quality of latent memberships:} We here study how the size of the simplex affects the membership types of \textsc{\modelabbrv}. Fig \ref{fig:phase_transitions} illustrates how the percentage of community champions (nodes assigned to simplex corners) for \textsc{\modelabbrv} as a function of $\delta^2$ and for different latent dimensions. When $\delta$ is small, almost all nodes are exclusively assigned to a simplex corner, resulting in hard assignments to clusters. As $\delta$ increases, more nodes are assigned with mixed memberships, while the number of champions decreases to zero for large $\delta$ values in all dimension cases. The decrease in community champions is steeper for $p=2$ compared to $p=1$. This also explains why the squared $\ell^2$ choice leads to faster convergence in AUC-ROC, as the model converges faster to the classic \textsc{LDM}. It is evident that the $p=2$ \textsc{\modelabbrv} requires smaller simplex volumes to be identifiable. In Fig \ref{fig:adj}, we provide the reorganized adjacency matrices with community allocations given by \textsc{\modelabbrv}, showing how the model successfully uncovers latent communities and produces part-based network representations while identifiability is ensured by choosing appropriate $\delta$ values, or equivalently appropriate simplex volumes. (for mixed-memberships nodes are assigned to the cluster in which they express the maximum membership)

 \textbf{Experiments using real ground-truth communities:} To evaluate the effectiveness of \textsc{HM-LDM} in identifying meaningful communities, we conduct experiments using four networks with known ground-truth community labels. For NMF-based methods, including our own, we assess the algorithms' ability to identify correct structures by comparing the inferred memberships with the ground-truth labels. We set the number of latent dimensions equal to the total number of communities. For GRL approaches that do not provide memberships, we extract latent embeddings and use {\it k-means} to assign communities. We report the Normalized Mutual Information (NMI) score and Adjusted Rand Index (ARI), which are well-established measures for community quality assessment \cite{com_metrics}. We tune all baseline methods separately for each network in terms of their hyperparameters. In contrast, for \textsc{HM-LDM}, we do not perform any tuning and just set $\delta=1$ for all networks, resulting in informative and mostly hard cluster assignments. We report scores averaged over five independent runs of the Adam optimizer, each of which includes five additional runs, selecting the model with the lowest training loss to avoid the effect of local minimas. We summarize our findings in Table \ref{tab:nmi_ari}, where we witness a mostly favorable or on-par performance of \textsc{\modelabbrv} with all of the competitive baselines for the NMI metric. For the ARI metric, we observe that our framework significantly outperforms the baselines for all of the considered networks.

\begin{table*}[!t]
\centering
\caption{Normalized Mutual Information (NMI) and Adjusted Rand Index (ARI) scores for networks with ground-truth communities.}
\label{tab:nmi_ari}
\resizebox{0.6\textwidth}{!}{%
\begin{tabular}{rcccccccc}\toprule
\multicolumn{1}{l}{} & \multicolumn{2}{c}{\textsl{Amherst}} & \multicolumn{2}{c}{\textsl{Rochester}} & \multicolumn{2}{c}{\textsl{Mich}}& \multicolumn{2}{c}{\textsl{Hamilton}}\\\cmidrule(rl){2-3}\cmidrule(rl){4-5}\cmidrule(rl){6-7}\cmidrule(rl){8-9}
\multicolumn{1}{c}{Metric} & NMI & ARI & NMI & ARI & NMI & ARI & NMI & ARI
\\\cmidrule(rl){1-1}\cmidrule(rl){2-2}\cmidrule(rl){3-3}\cmidrule(rl){4-4}\cmidrule(rl){5-5}\cmidrule(rl){6-6}\cmidrule(rl){7-7}\cmidrule(rl){8-8}\cmidrule(rl){9-9}
\textsc{DeepWalk}\cite{deepwalk-perozzi14}       &.498 	&.347  &.348  & .205	&.207  &.157	 & .447   &.303	  \\
\textsc{Node2Vec}\cite{node2vec-kdd16}       &.535 	&.375  & .364 & .223	& .217  &.161	 &.481   &.348	  \\
\textsc{LINE}  \cite{line}         & .549	&.452  & .365 & .217	&\textbf{.249}   &.192	 &.499  &.411		  \\
\textsc{NetMF} \cite{netmf-wsdm18}       &.491 	&.330  &.377  & .243	& .237  &.136	 &.456    &	.297	  \\
\textsc{NetSMF} \cite{netsmf-www2019}        &\underline{.562} 	&.408  &\underline{.381}  & .228	& \underline{.242}  &.169	 & .494   &	.391	  \\
\textsc{LouvainNE}\cite{louvainNE-wsdm20}      & \underline{.562}	&.395  &  .347 &.204	&.175  &.114 	 &.475	  &	.334  \\
\textsc{ProNE}\cite{prone-ijai19}     & .536	& .443 &.356  & .312	&.229   &\underline{.200}	 &.478    &	.396	  \\\midrule
\textsc{NNSED}\cite{NNSED}     &.295	& .243 &.168   &.116 	& .064  &.035	 & .335   &	.285 \\
\textsc{MNMF}\cite{MNMF}      & .542	&.362  & .324 & .171	&.188   &.102	 & .466   &	.287	  \\
\textsc{BigClam}\cite{nmf3}   & .091	&.066	& .028  & .022	&.024	&.015   & .053  &.041 \\
\textsc{SymmNMF}\cite{SymmNMF}  &\textbf{.596} 	&.397	& .308  &.175 	&.207	& .088  & .437  &.341   \\\midrule
\textsc{HM-LDM($p=1$)} & \underline{.562}	&\underline{.502}	&\textbf{.400}   &\textbf{.392} 	&.228	&\textbf{.205}   & \textbf{.527}  &  \underline{.485}
\\
\textsc{HM-LDM($p=2$)} &.539 	&\textbf{.506}  &\underline{.384}  & \underline{.373}	&.217 &.183	 &\underline{.507}    &	\textbf{.504}	
\\\bottomrule    
\end{tabular}%
}
\end{table*}

\textbf{Comparison with the \textsc{LDM}:} We explore the performance of \textsc{HM-LDM} in comparison to the classical LDM with random effects, considering normal and squared $\ell^2$-norms denoted as \textsc{LDM-Re} and \textsc{LDM-Re}-$(\ell^2)^2$, accordingly. We evaluate the models, based on link prediction and clustering tasks and report the scores in Table \ref{tab:auc_roc_comp} and Table \ref{tab:nmi_ari_comp}. The results show that despite constraining the latent space into the $D-$simplex with volumes ensuring identifiable solutions, we only observe a slight decrease in AUC-ROC scores. In contrast, the \textsc{\modelabbrv} yields favorable NMI scores for community detection and considerably higher ARI scores when compared to classic LDM. For sufficiently large $\delta$ values (i.e. $\delta^2=10^3$), link-prediction performance for \textsc{\modelabbrv} reached the one of the unconstrained LDM, but the clustering scores of the latter decrease significantly. This is since for large simplex volumes, the \textsc{\modelabbrv} closely approximates the LDM at the expense of model and structure identifiability.

\begin{figure}[!b]
\centering
\subfloat[$p=1$, $\delta=1$]{{ \includegraphics[scale=0.2]{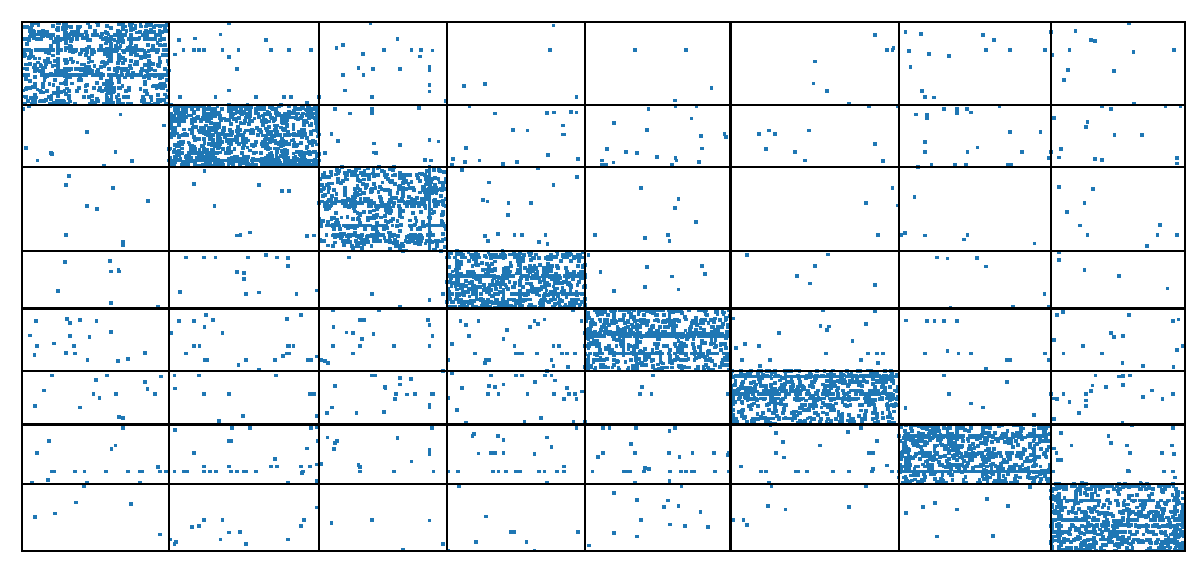} }}%
\hspace{0.5cm}
\subfloat[$p=2$, $\delta=1$]{{ \includegraphics[scale=0.2]{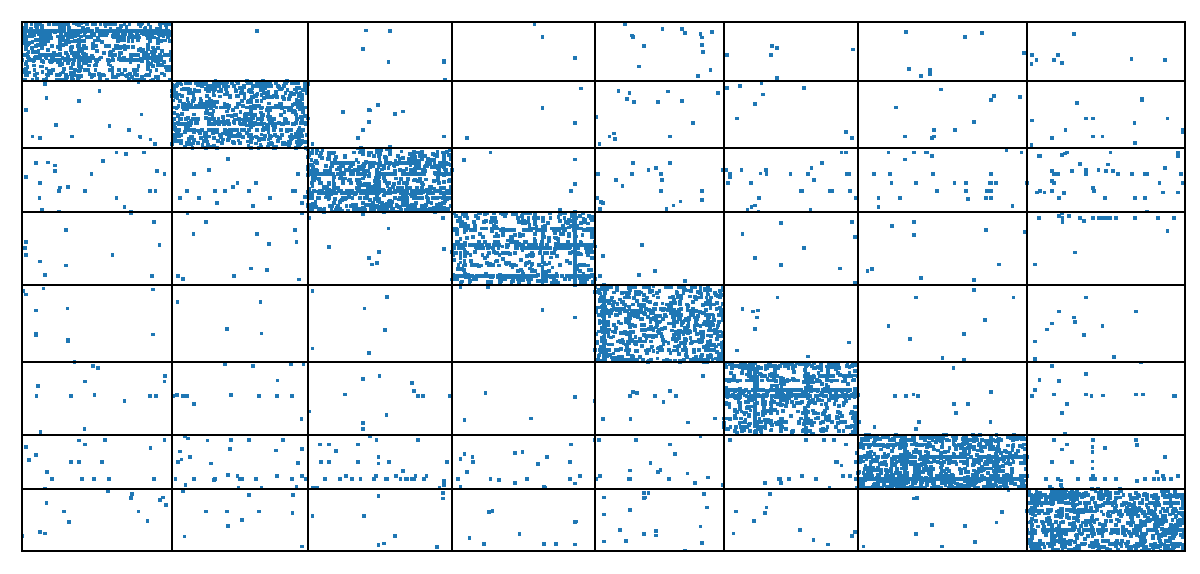} }}
\caption{\textsl{Drug-Gene} ordered adjacency matrices based on \textsc{\modelabbrv} with $D=8$, empirically showcasing community extraction and identification extended to bipartite networks.}\label{fig:bip}
\end{figure}

\textbf{Extension to bipartite networks:} 
We can trivially extend the \textsc{\modelabbrv} model to account for unsigned bipartite networks \cite{hbdm}. Such an extension is achieved by defining and introducing a different set of latent variables for the two disjoint sets of nodes, as present in a bipartite structure. In addition, the \textsc{\modelabbrv}(p=2) model simply extends the symmetric NMF operation, obtained for the undirected networks, to the non-symmetric NMF specification. In Fig \ref{fig:bip}, we provide the re-ordered adjacency matrix with respect to the community allocations defined by the learned embeddings of \textsc{\modelabbrv} for a \textsl{Drug-Gene} \cite{snapnets} network ($|\mathcal{V}|=7,341|$, $|\mathcal{E}|=15,138$) where we observe a clear block structure. Importantly, the \textsc{\modelabbrv} offers identifiable joint embedding representations, mixed memberships, and community discovery for bipartite networks, tasks in general considered to be non-trivial and arduous.

\begin{table*}[!t]
\centering
\caption{\textsc{\modelabbrv} and \textsc{LDM-Re} comparison for the link prediction task.}
\label{tab:auc_roc_comp}
\resizebox{0.9\textwidth}{!}{%
\begin{tabular}{lcccccccccccc}\toprule
\multicolumn{1}{l}{} & \multicolumn{3}{c}{\textsl{AstroPh}} & \multicolumn{3}{c}{\textsl{GrQc}} & \multicolumn{3}{c}{\textsl{Facebook}}& \multicolumn{3}{c}{\textsl{HepTh}}\\\cmidrule(rl){2-4}\cmidrule(rl){5-7}\cmidrule(rl){8-10}\cmidrule(rl){11-13}
\multicolumn{1}{c}{Dimension ($D$)} & $8$ & $16$ & $32$ & $8$ & $16$ & $32$ & $8$ & $16$ & $32$& $8$ & $16$ & $32$ \\\cmidrule(rl){1-1}\cmidrule(rl){2-2}\cmidrule(rl){3-3}\cmidrule(rl){4-4}\cmidrule(rl){5-5}\cmidrule(rl){6-6}\cmidrule(rl){7-7}\cmidrule(rl){8-8}\cmidrule(rl){9-9}\cmidrule(rl){10-10}\cmidrule(rl){11-11}\cmidrule(rl){12-12}\cmidrule(rl){13-13}
\textsc{LDM-Re}  &.973  	&.974	&.979  & .949	&.952	&.954 & .993 & .994& .992   &.920 &.923 &.923 \\ 
\textsc{HM-LDM}($p=1,\delta^2=\text{identifiable}$)  & .956	&.952	&.952   &.944	&.948	&.951   & .982  & .979 & .974   &.916  & .921 &.924
\\
\textsc{HM-LDM}($p=1,\delta^2=10^3$) &.967 &  .967 & .965 & .956 & .955 & .951 & .985 & .986 & .987 & .932 & .931 & .926
\\\midrule
\textsc{LDM-Re}-$(\ell^2)^2$    &.979  &.978  &.976    & .944	&.944	& .945 & .990 & .990 & .991 & .913 &.912 &.909\\ 
\textsc{\modelabbrv}($p=2,\delta^2=\text{identifiable}$)      &.972   &.973  & .963 &.940 & .942 & .946 & .992   & .993     & .993     &.908 &.910 &.911
\\
\textsc{\modelabbrv}($p=2,\delta^2=10^3$) & .984& .983 &  .980 & .948 & .946 & .946 & .991 & .991 & .992 & .920 & .918 & .913
\\\bottomrule    
\end{tabular}%
 }
\end{table*}

\begin{table*}[!t]
\centering
\caption{\textsc{\modelabbrv} and \textsc{LDM-Re} comparison for the clustering task.}
\label{tab:nmi_ari_comp}
\resizebox{0.8\textwidth}{!}{%
\begin{tabular}{lcccccccc}\toprule
\multicolumn{1}{l}{} & \multicolumn{2}{c}{\textsl{Amherst}} & \multicolumn{2}{c}{\textsl{Rochester}} & \multicolumn{2}{c}{\textsl{Mich}}& \multicolumn{2}{c}{\textsl{Hamilton}}\\\cmidrule(rl){2-3}\cmidrule(rl){4-5}\cmidrule(rl){6-7}\cmidrule(rl){8-9}
\multicolumn{1}{c}{Metric} & NMI & ARI & NMI & ARI & NMI & ARI & NMI & ARI  \\\cmidrule(rl){1-1}\cmidrule(rl){2-2}\cmidrule(rl){3-3}\cmidrule(rl){4-4}\cmidrule(rl){5-5}\cmidrule(rl){6-6}\cmidrule(rl){7-7}\cmidrule(rl){8-8}\cmidrule(rl){9-9}
\textsc{LDM-Re}    & .548	&.366   &.391   &.212  	&.230    &.132	 & .491  & .320	  \\ \textsc{HM-LDM}($p=1,\delta^2=\text{identifiable}$) & .562	&.502	&.400   &.392 	&.228	&.205   & .527  &  .485\\
\textsc{HM-LDM}($p=1,\delta^2=10^3$) & .439	&.386	& .308 &.303	&.176	& .133 & .405  &  .377
\\
\midrule
\textsc{LDM-Re}-$(\ell^2)^2$ &.546	&.370   &.393   &.211  	&.231    &.137 	 &.497     &.327 	  \\
\textsc{\modelabbrv}($p=2,\delta^2=\text{identifiable}$) &.539 	&.506  &.384 & .373	&.217 &.183	 &.507   &	.504	 
\\
\textsc{HM-LDM}($p=2,\delta^2=10^3$) &.240 	&.133  &.206  & .119	& .116 &.056	 & .232  &.209		 
\\
\bottomrule    
\end{tabular}%
}
\end{table*}

\subsection{Signed Networks Experiments}

  For the signed network experiments, we introduce four networks of varying sizes and structures. (\textbf{i}) \textsl{Reddit} which uses hyperlinks to create directed edges between communities belonging to the social network platform \cite{dataset_reddit}. (\textbf{ii}) \textsl{wikiElec} and (\textbf{iii}) its more recent version \textsl{wikiRfa} which follow election procedures carried out through multiple timelines and convey voting information as links about users to describe positive, neutral, and opposing views for potential users to be elected administrators on Wikipedia. \cite{dataset_wikielec,dataset_wikirfa}. (\textbf{iv}) \textsl{Twitter} is an undirected network with positive and negative links obtained from user tweets about the referendum concerning the reform of the Italian Constitution back in 2016 \cite{dataset_twitter}. 

The performance of \textsc{sHM-LDM} is compared against seven graph representation learning baselines, eligible for analyzing signed networks: (\textbf{i}) \textsc{POLE} \cite{pole} where embeddings are based on the decomposition of an auto-covariance matrix created through signed random walks, (\textbf{ii}) \textsc{SLF} \cite{slf} that creates representations based on latent factors capable of describing both positive and negative connections, (\textbf{iii}) \textsc{SiGAT} \cite{sigat} a graph neural network model learning node embeddings through a graph attention mechanism, (\textbf{iv}) \textsc{SIDE} \cite{side} utilizing truncated random walks under a general likelihood expression for signed relationships modeling both positive and negative ties, (\textbf{v}) \textsc{SigNet} \cite{signet} a deep neural network using a similarity measure through the Hadamard product able to describe signed proximity between a pair of nodes,  (\textbf{vi}) \textsl{SLDM} and  (\textbf{vii}) \textsl{SLIM} models \cite{slim} which define an unconstrained and a constrained to polytopes latent distance model, respectively. Both of these two models utilize the Skellam distribution as the \textsc{sHM-LDM} which constrains the model to the $D-$simplex while \textsl{SLIM} operates on the inferred convex-hull of the latent space.

\begin{table}[!t]
\centering
\caption{Binary operators considered for designing feature vectors (edge features). The notation, $f(v)_d$ denotes the $d$'th coordinate of the embedding vector of node $v$.}
\label{tab:bin_oper}
\begin{tabular}{c|c|c}
\textbf{Operator} & \textbf{Symbol} & \textbf{Definition} \\
\hline
Average & $\boxplus$ & $[f(u) \boxplus f(v)]_d = ({f(u)_d+f(v)_d}) / {2}$ \\
Hadamard & $\boxdot$ & $[f(u) \boxdot f(v)]_{d} = f(u)_d \cdot f(v)_d$ \\
Weighted-L1 & $\|\cdot\|_{{1}}$ & $\|f(u) - f(v)\|_{{1}_{d}} = |f(u)_d - f(v)_d|$ \\
Weighted-L2 & $\|\cdot\|_{{2}}$ & $\|f(u) - f(v)\|_{{2}_{d}} = |f(u)_d - f(v)_d|^2$ \\
Concatenate & $\oplus$ & $[f(u) \oplus f(v)]_{d} = \big(f(u)_d, \ f(v)_d\big)$ \\
\hline
\end{tabular}
\end{table}

\begin{table*}[!t]
\centering
\caption{Area Under Curve (AUC-ROC) scores for representation size of $D=8$ and $\delta$ values ensuring identifiability. ("x" denotes a baseline that was unable to run due to high memory/runtime complexity)}
\label{tab:auc_roc_signed}
\resizebox{0.9\textwidth}{!}{%
\begin{tabular}{rccccccccccccccccccccccccc}\toprule
\multicolumn{1}{l}{} & \multicolumn{3}{c}{\textsl{WikiElec}} & \multicolumn{3}{c}{\textsl{WikiRfa}} & \multicolumn{3}{c}{\textsl{Twitter}}& \multicolumn{3}{c}{\textsl{Reddit}} \\\cmidrule(rl){2-4}\cmidrule(rl){5-7}\cmidrule(rl){8-10}\cmidrule(rl){11-13}
\multicolumn{1}{r}{Task} & $p@n$ & $p@z$ & $n@z$ & $p@n$ & $p@z$ & $n@z$ & $p@n$ & $p@z$ & $n@z$ & $p@n$ & $p@z$ & $n@z$  \\\cmidrule(rl){1-1}\cmidrule(rl){2-2}\cmidrule(rl){3-3}\cmidrule(rl){4-4}\cmidrule(rl){5-5}\cmidrule(rl){6-6}\cmidrule(rl){7-7}\cmidrule(rl){8-8}\cmidrule(rl){9-9}\cmidrule(rl){10-10}\cmidrule(rl){11-11}\cmidrule(rl){12-12}\cmidrule(rl){13-13}
\textsc{POLE}\cite{pole}     &.809 &.896 &.853 &.904 &.921 &.767 &.965 &.902 &.922 &x &x &x\\
\textsc{SLF}\cite{slf}    &\textbf{.888} &.954 &\textbf{.952} &\textbf{.971} &.963 &.961 &.914 &.877 &.968 &\textbf{.729} &\textbf{.955} &.968\\
\textsc{SiGAT}\cite{sigat}     &.874 &.775 &.754 &.944 &.766 &.792 &\textbf{.998} &.875 &.963 &\underline{.707} &.682 &.712\\
\textsc{SIDE}\cite{side}      &.728 &.866 &.895 &.869 &.861 &.908 &.799 &.843 &.910 &.653 &.830 &.892\\
\textsc{SigNet}\cite{signet}    &.841 &.774 &.635 &.920 &.736 &.717 &.968 &.719 &.891 &.646 &.547 &.623\\
\textsc{SLDM}\cite{slim}     & \underline{.876} &\textbf{.969} &.936 &\underline{.963} &\textbf{.982} &\underline{.963} &.986 &\underline{.962} &\underline{.973} &.648 &.951 &.975
\\
\textsc{SLIM}\cite{slim}  &.862 &.965 &.935 &.956 &\underline{.980} &.960 &\underline{.988} &\textbf{.963} &.972 &.667 &\textbf{.955} &\textbf{.978}\\
\midrule
\textsc{sHM-LDM}(p=1) &.872  &.963 &\underline{.938} &.959 &.977 &\underline{.963} &.978 &.958 &\textbf{.976} &.642 &.951 &\underline{.977}\\
\textsc{sHM-LDM}(p=2)  &.872  & \underline{.966}&.937 &.960 &.975 &\textbf{.964} &.977 &.958 &\underline{.973} &.610 &\underline{.953} &.976 
\\\bottomrule    
\end{tabular}%
}
\end{table*}

\subsection{Signed Link prediction}
We follow the same evaluation procedure as in \cite{slim} and define two settings considering link prediction, in order to benchmark \textsc{sHM-LDM}'s predictive capability against the considered baselines. For that, we randomly choose $20\%$ of the total network links/cases (both positive and negative) which are then zeroed out with the constraint that the residual signed network stays connected. Furthermore, an equal size of disconnected pairs in the original networks is also drawn to act as the controls in the prediction tasks. The combined samples of removed links and drawn controls define the test set for each network. All models are trained on the residual networks for each dataset.

\textbf{Performance evaluation:} For our methods, as well as, for \textsl{SLDM} and \textsl{SLIM} we learn a logistic regression model with inputs given by both the rates and log-rates, as defined by the Skellam distribution, i.e. $\chi_{ij}=\big [\lambda_{ij}^{+},\lambda_{ij}^{-},\log \lambda_{ij}^{+},\log \lambda_{ij}^{-}\big ]$. It is argued in \cite{slim} that the Skellam distribution operates on both the ratios and products of the rates during inference. Consequently, training a logistic regression based on rates and log rates allow for learning linear and non-linear mappings based on both the rates as well as their products and ratios due to the log transformation. For the performance evaluation of the baselines, we consider five binary operators, as established in the GRL literature. These include the \{average, Hadamard product, weighted L1, weighted L2, concatenate\}, as shown in Table \ref{tab:bin_oper}. These are utilized to construct five different feature vectors, used to train multiple logistic regression models for each task. For every baseline defining multiple feature vectors, we choose the logistic regression model that returns the maximum performance for each individual task. Lastly, for each link prediction task, we consider the robust against class imbalance metric, area-under-curve of the receiver operating characteristic (AUC-ROC).

\textbf{Task 1: Link sign prediction.} For the first task we consider only the links/cases of the test set for each network. After training, each model is provided with the test set link pairs and evaluated in its ability to predict the sign of the removed links. The AUC-ROC results are summarized in Table \ref{tab:auc_roc_signed} where the link sign prediction is represented as $p@n$. We mostly observe favorable or on-par results and performance against the baselines. More specifically, comparing to the \textsl{SLDM} and \textsl{SLIM}, our models despite defining a more constrained latent space (recall that  $\mathbf{A}=\delta\cdot\mathbf{I}$ for \textsc{sHM-LDM}) the obtained results shows identical or on-par performance.

\textbf{Task 2: Signed link prediction.} The second task is more difficult and evaluates the performance of a model in its ability to both predict the sign, as well as, the presence of a link. For that, the whole test set is used to create two test subsets. The first contains the controls and positive links while the second the controls and the negative links. The models then are asked to distinguish controls from positive cases and controls from negative cases, respectively. We denote these tasks accordingly as $p@z$ and $n@z$ and AUC-ROC scores are provided in Table \ref{tab:auc_roc_signed}. Once more, the \textsl{sHM-LDM} frameworks provide favorable or on-par performance against the baselines and especially to the \textsl{SLDM} and \textsl{SLIM} models.

\begin{figure*}
\centering
\begin{subfigure}{0.24\textwidth}
    \includegraphics[width=0.8\textwidth]{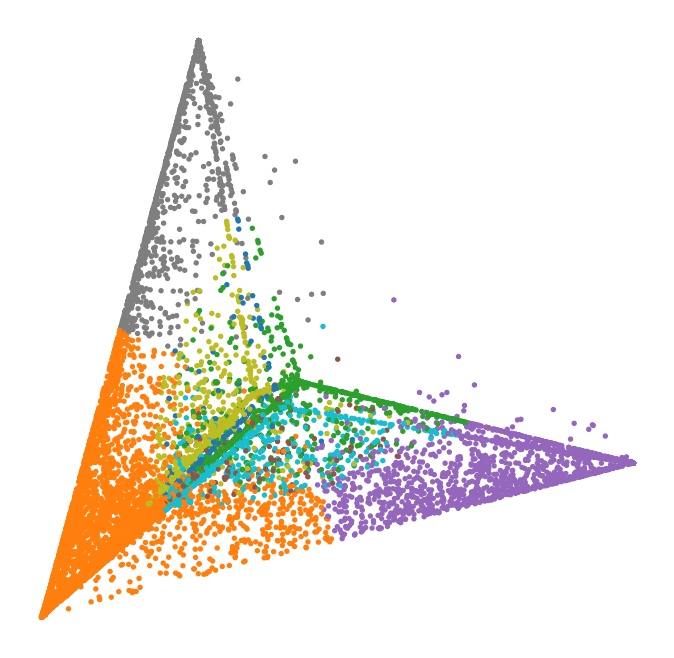}
    \caption{PCA $(D=8)$}
    
\end{subfigure}
\hfill
\begin{subfigure}{0.24\textwidth}
    \includegraphics[width=0.8\textwidth]{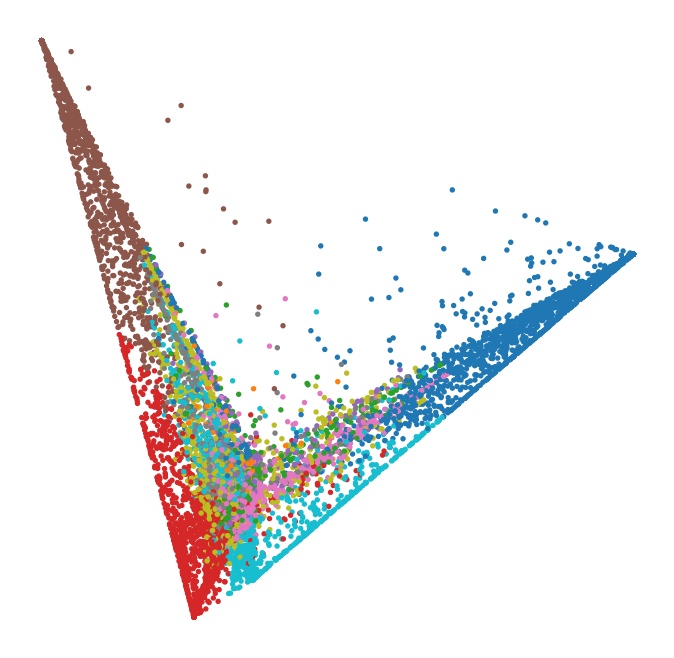}
    \caption{PCA $(D=16)$}
    
\end{subfigure}
\hfill
\begin{subfigure}{0.24\textwidth}
    \includegraphics[width=0.8\textwidth]{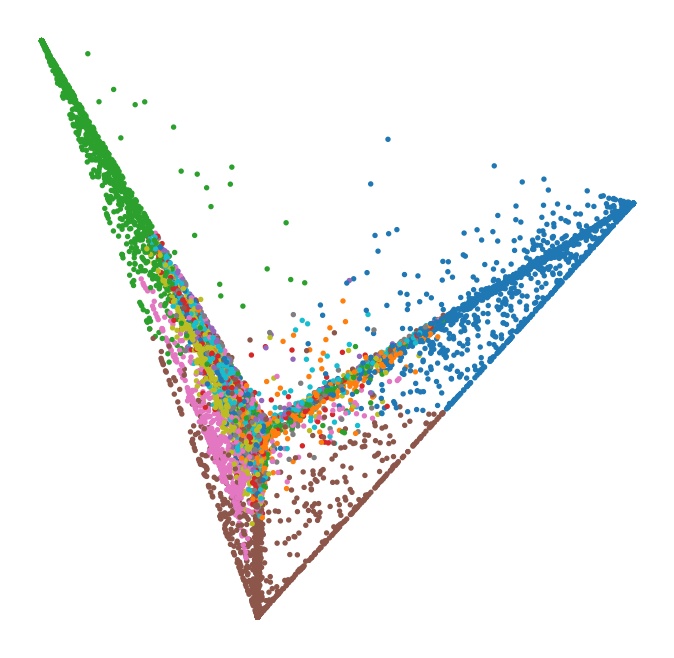}
    \caption{PCA $(D=32)$}
    
\end{subfigure}
\begin{subfigure}{0.24\textwidth}
    \includegraphics[width=0.8\textwidth]{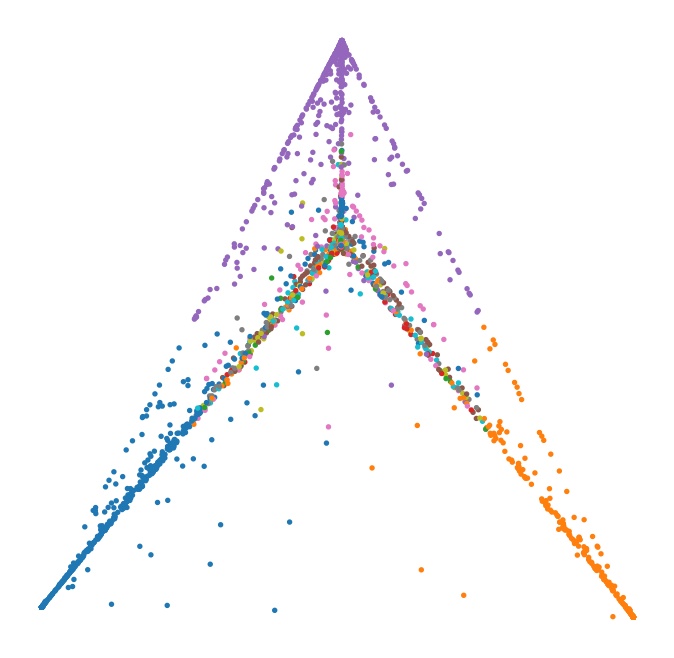}
    \caption{PCA $(D=64)$}
    
\end{subfigure}
\hfill
\begin{subfigure}{0.24\textwidth}
    \includegraphics[width=\textwidth]{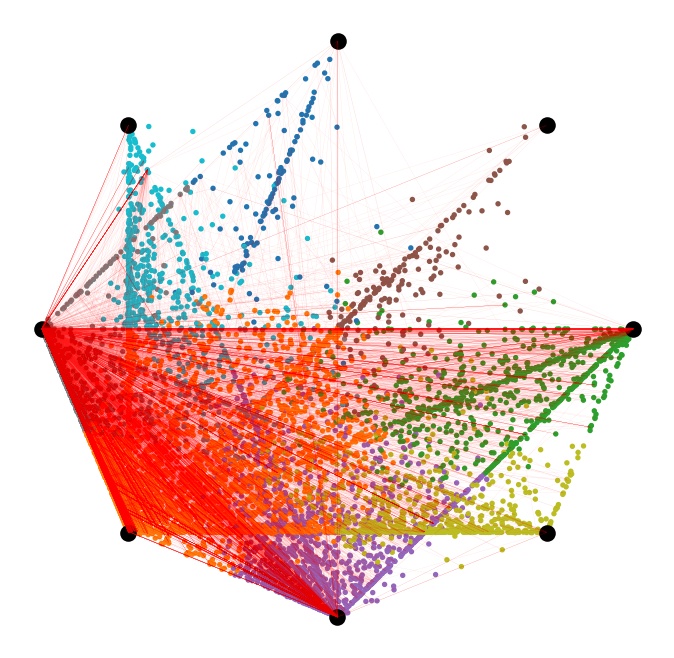}
    \caption{\textsc{NCP} $(D=8)$}
    
\end{subfigure}
\hfill
\begin{subfigure}{0.24\textwidth}
    \includegraphics[width=\textwidth]{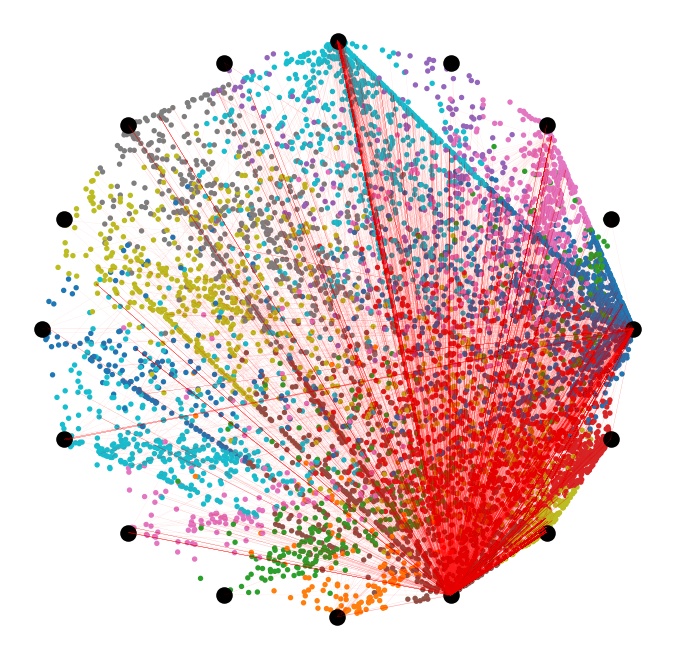}
    \caption{\textsc{NCP} $(D=16)$}
    
\end{subfigure}
\begin{subfigure}{0.24\textwidth}
    \includegraphics[width=\textwidth]{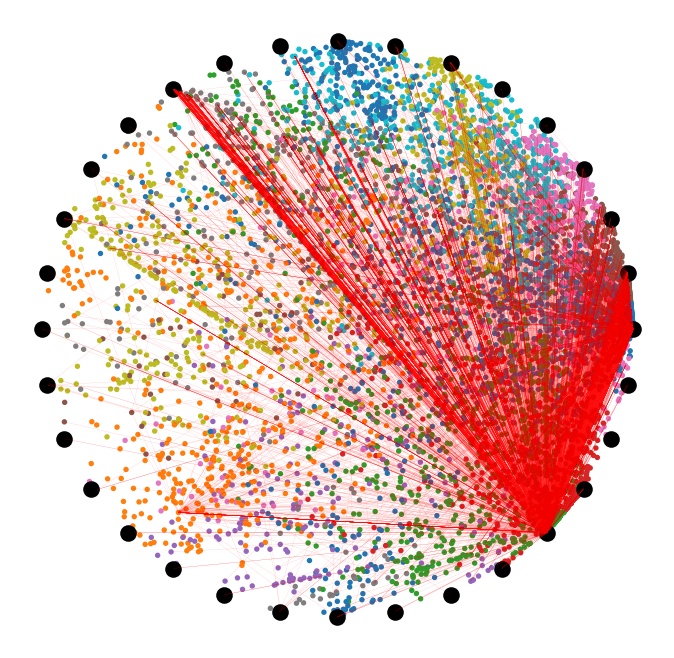}
    \caption{\textsc{NCP} $(D=32)$}
    
\end{subfigure}
\hfill
\begin{subfigure}{0.24\textwidth}
    \includegraphics[width=\textwidth]{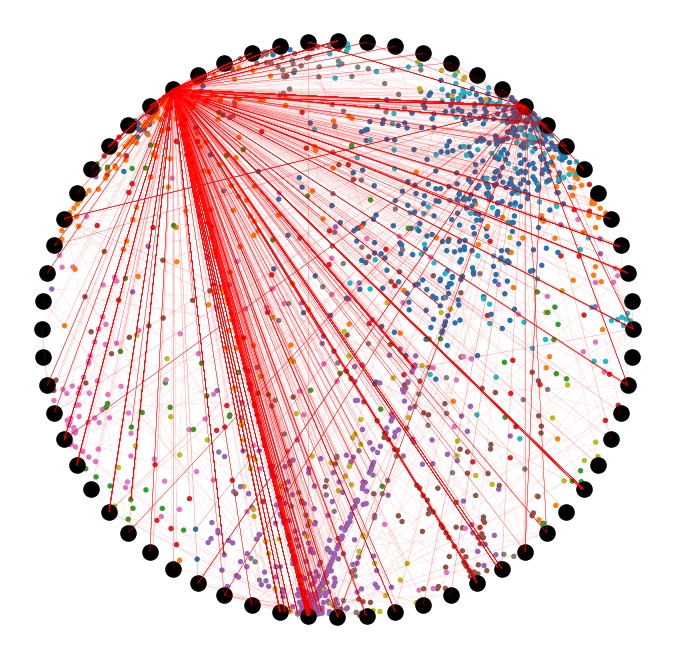}
    \caption{\textsc{NCP} $(D=64)$}
    
\end{subfigure}
\hfill
\begin{subfigure}{0.24\textwidth}
    \includegraphics[width=\textwidth]{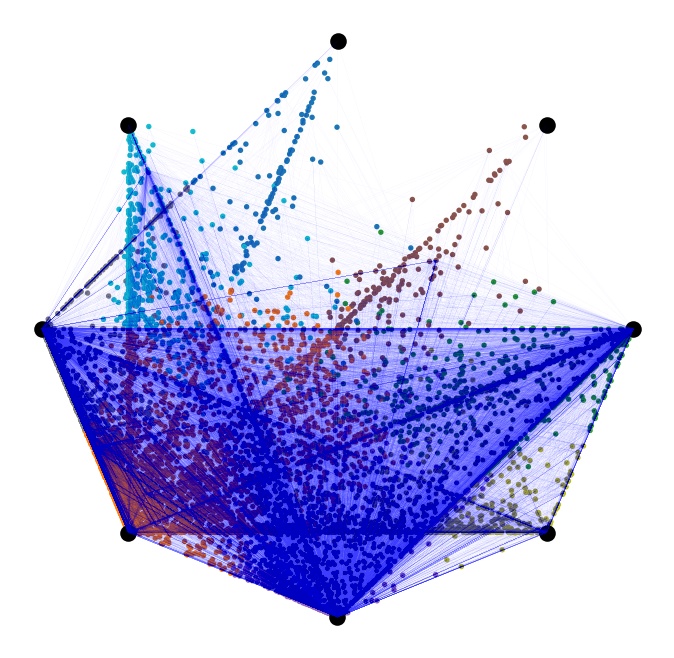}
    \caption{\textsc{PCP} $(D=8)$}
    
\end{subfigure}
\begin{subfigure}{0.24\textwidth}
    \includegraphics[width=\textwidth]{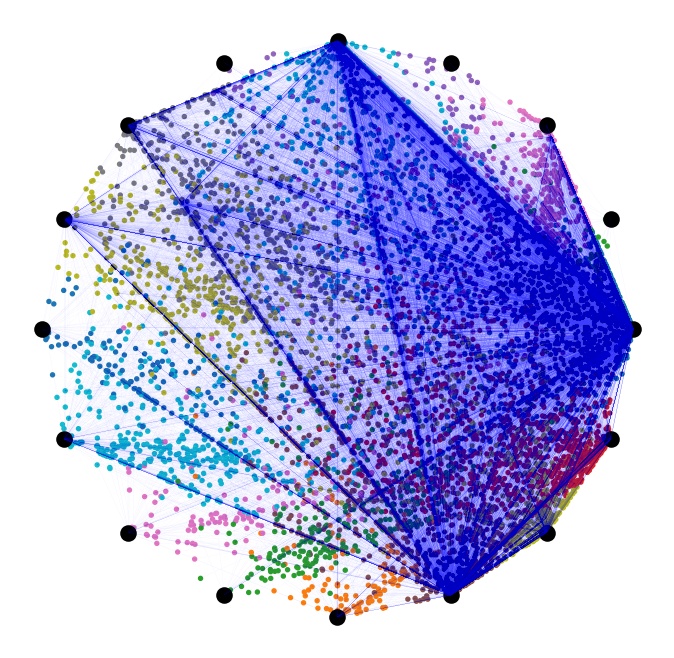}
    \caption{\textsc{PCP} $(D=16)$}
    
\end{subfigure}
\hfill
\begin{subfigure}{0.24\textwidth}
    \includegraphics[width=\textwidth]{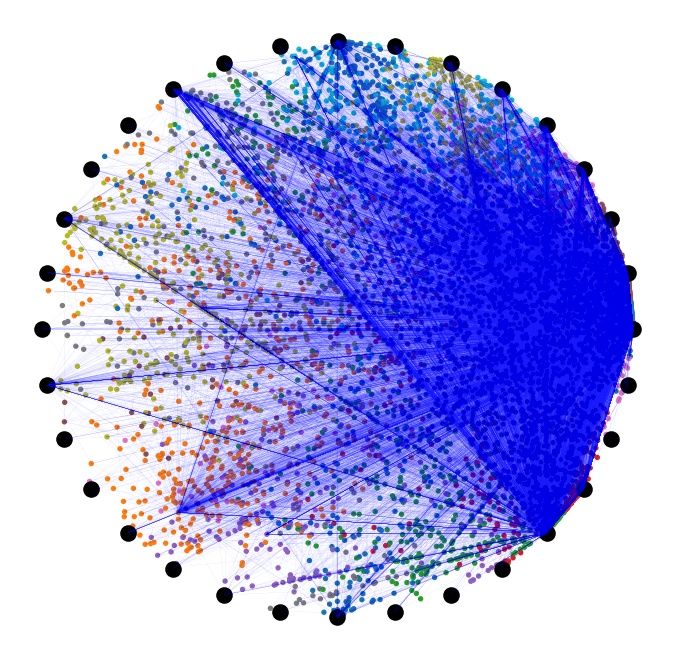}
    \caption{\textsc{PCP} $(D=32)$}
    
\end{subfigure}
\hfill
\begin{subfigure}{0.24\textwidth}
    \includegraphics[width=\textwidth]{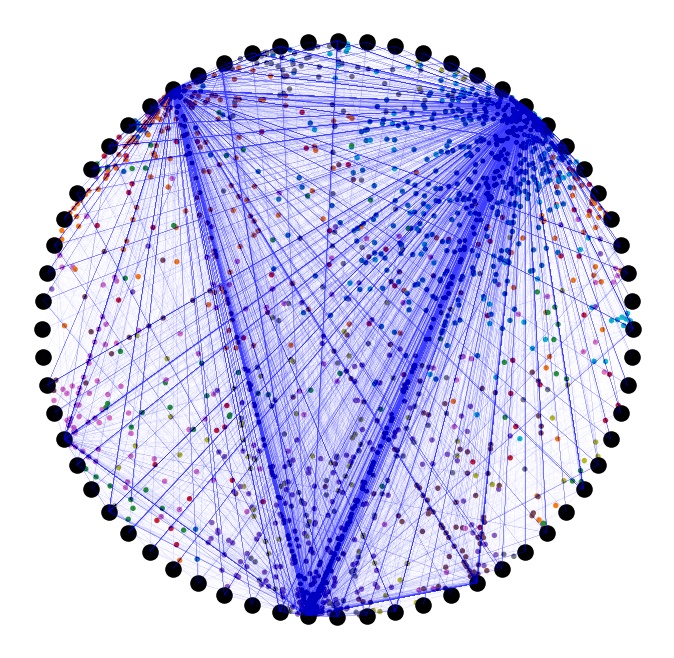}
    \caption{\textsc{PCP} $(D=64)$}
    
\end{subfigure}
\hfill
\begin{subfigure}{0.24\textwidth}
    \includegraphics[width=0.9\textwidth]{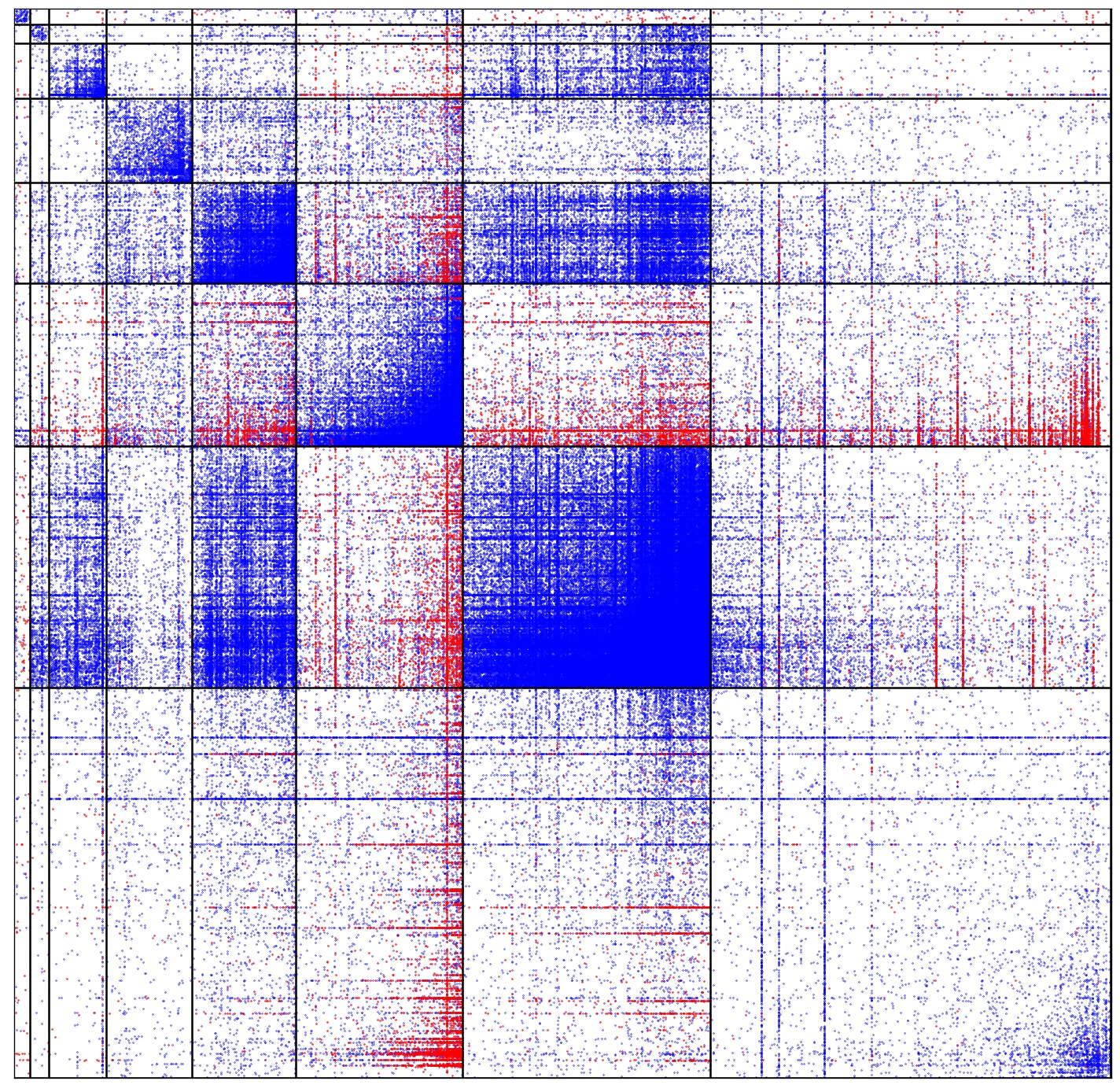}
    \caption{\textsc{OrA} $(D=8)$}
    
\end{subfigure}
\hfill
\begin{subfigure}{0.24\textwidth}
    \includegraphics[width=0.9\textwidth]{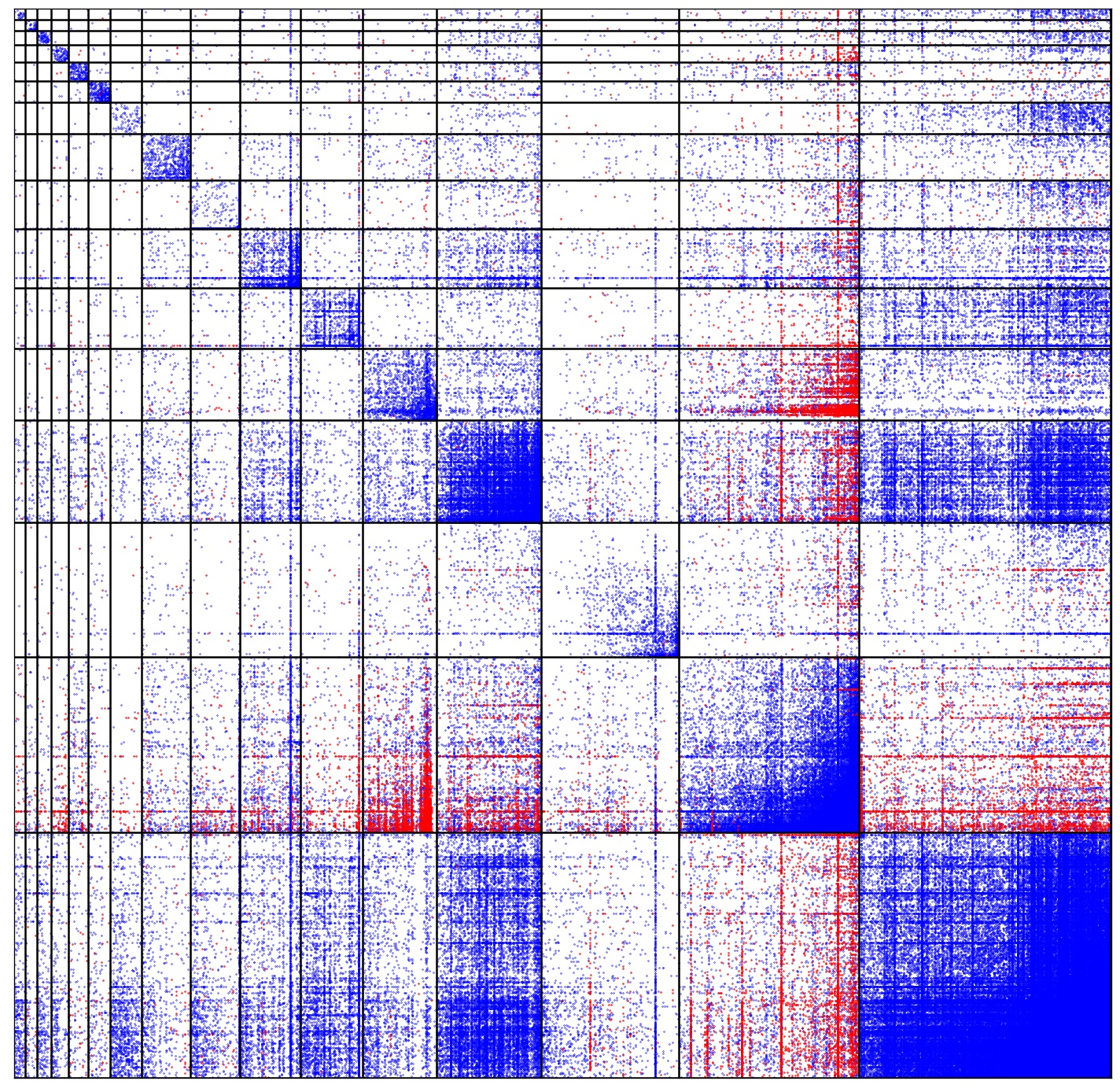}
    \caption{\textsc{OrA} $(D=16)$}
    
\end{subfigure}
\begin{subfigure}{0.24\textwidth}
    \includegraphics[width=0.9\textwidth]{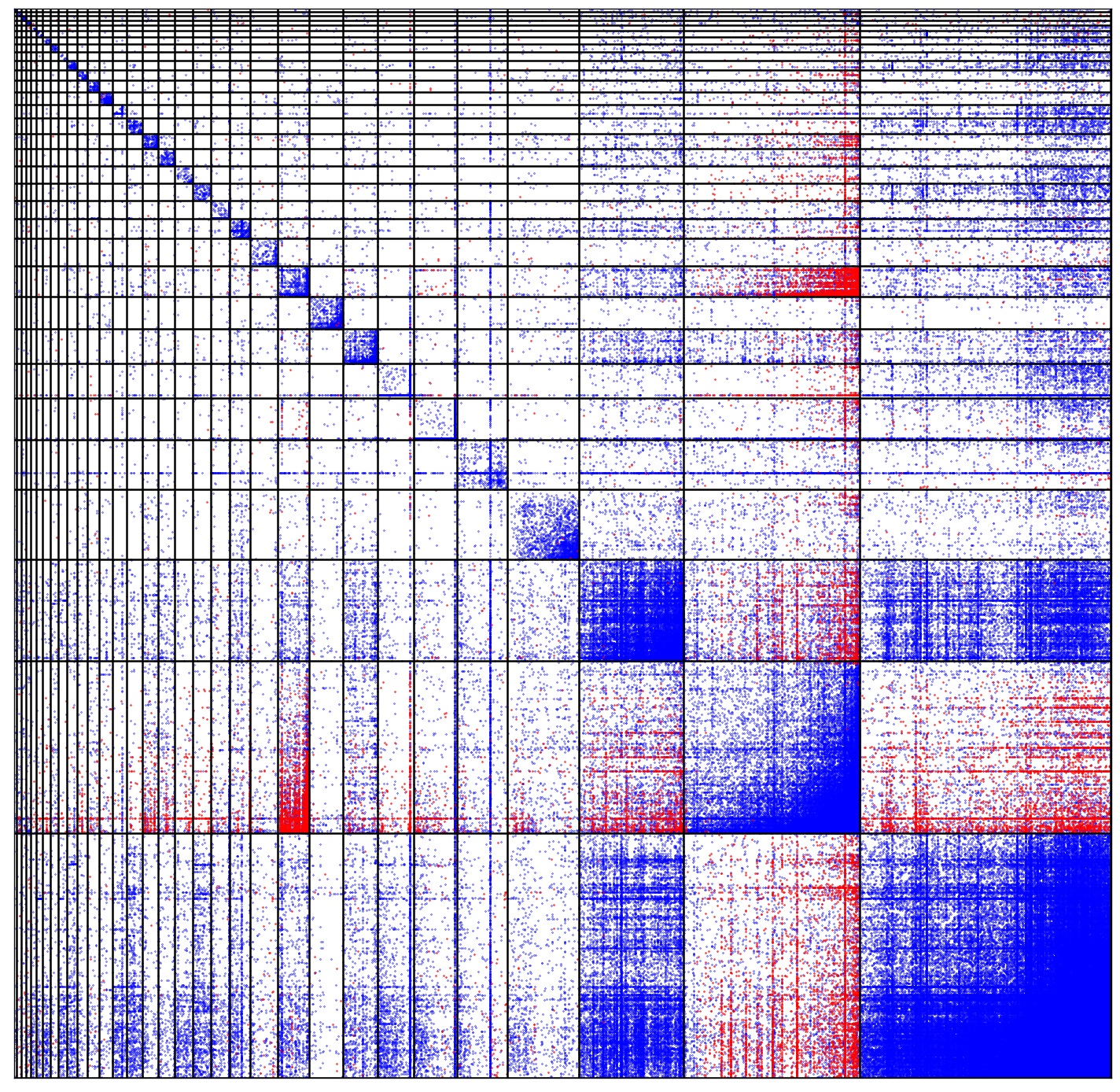}
    \caption{\textsc{OrA} $(D=32)$}
    
\end{subfigure}
\hfill
\begin{subfigure}{0.24\textwidth}
    \includegraphics[width=0.9\textwidth]{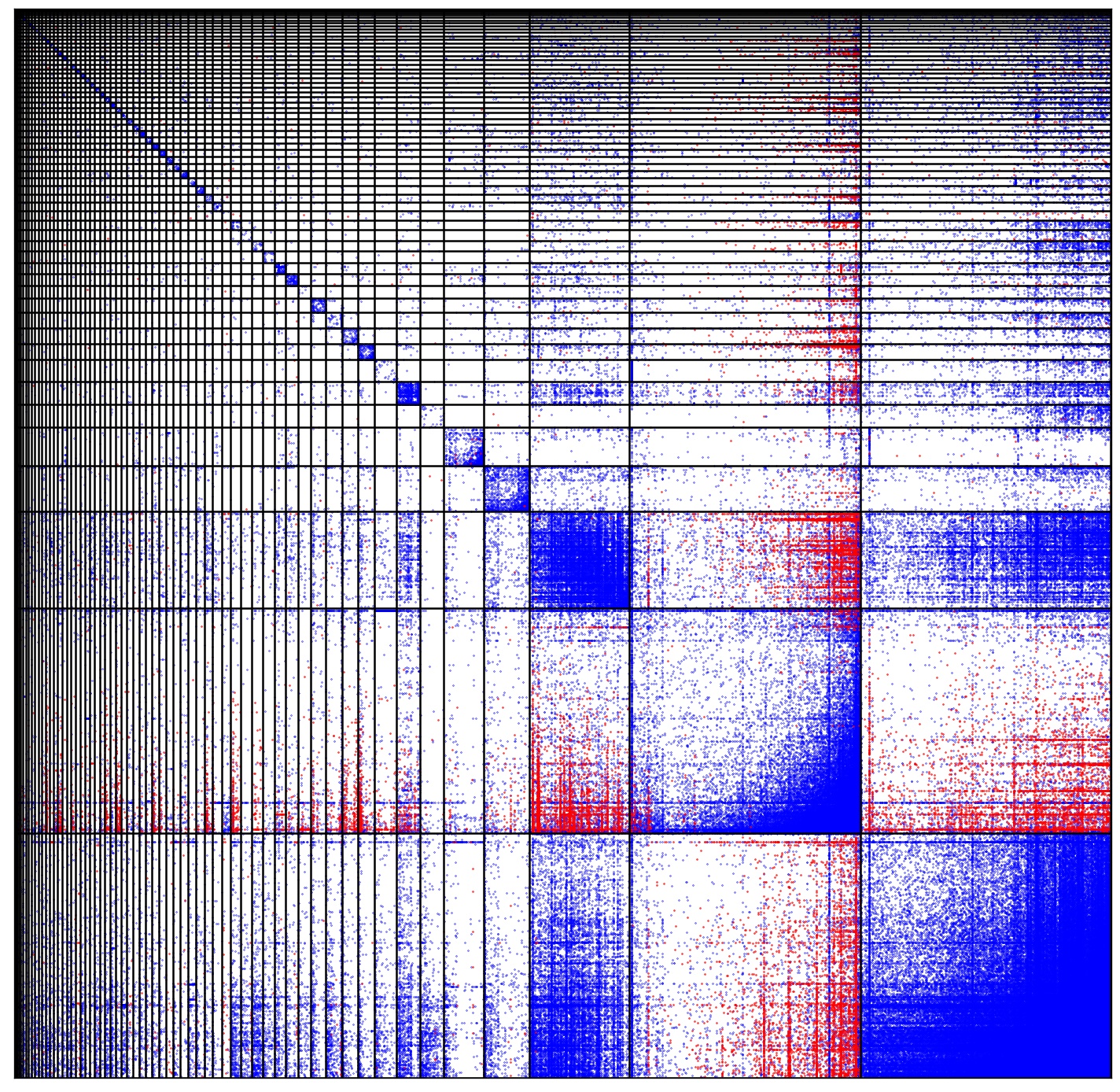}
    \caption{\textsc{OrA} $(D=64)$}
    
\end{subfigure}
\caption{\textbf{\textsc{sHM-LDM}(p=2)}: \textsl{Twitter} Network---Inferred simplex visualizations and ordered adjacency matrices for various dimensions $D$ and with simplex side lengths $\delta$ ensuring identifiability. The first row shows the latent space projection to the first two Principal Components---The second row provides a Negative Circular Plot (\textsc{NCP}) with red lines showcasing negative links between nodes---The third row shows a Positive Circular Plot (\textsc{PCP}) with the blue lines denoting positive links between node pairs---The fourth and final row shows the Ordered Adjacency (\textsc{OrA}) matrices sorted based on the memberships $\mathbf{w}_i$, in terms of maximum simplex corner responsibility, and internally according to the magnitude of the corresponding corner assignment for their reconstruction.}
\label{fig:soc_viz_p2}
\end{figure*}

\begin{figure*}
\centering
\begin{subfigure}{0.24\textwidth}
    \includegraphics[width=0.8\textwidth]{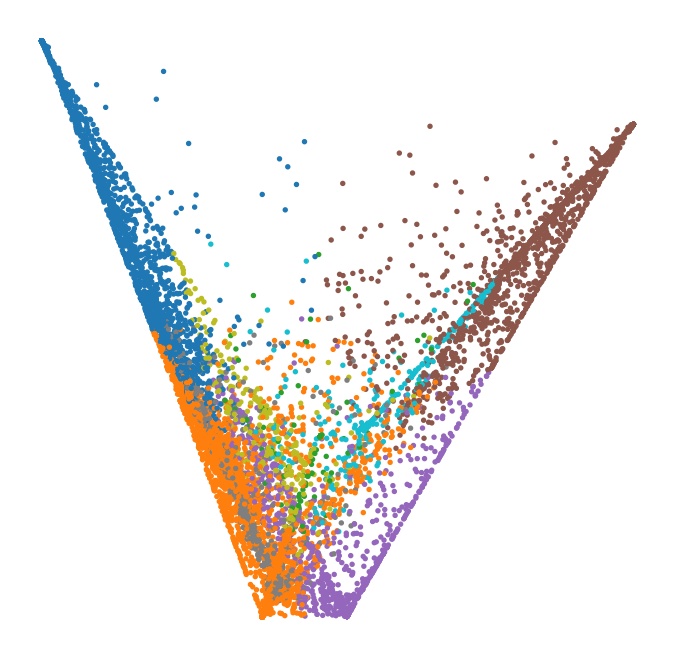}
    \caption{PCA $(D=8)$}
    
\end{subfigure}
\hfill
\begin{subfigure}{0.24\textwidth}
    \includegraphics[width=0.8\textwidth]{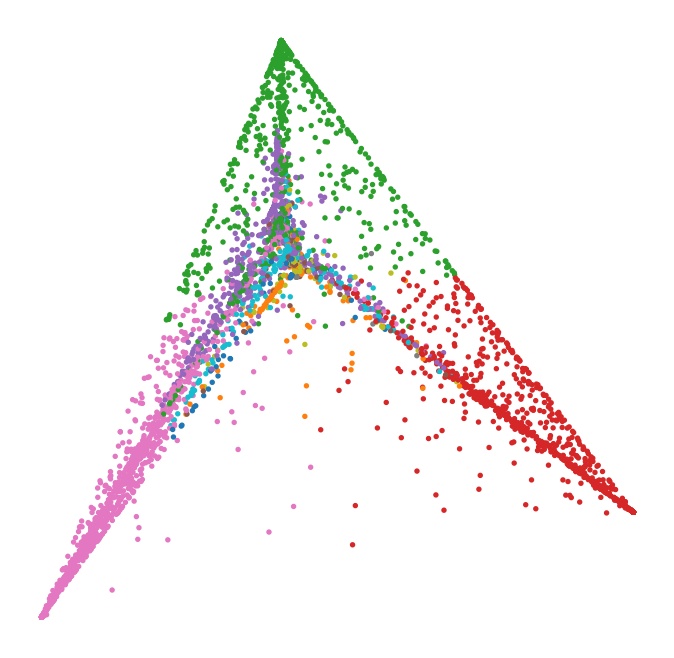}
    \caption{PCA $(D=16)$}
    
\end{subfigure}
\hfill
\begin{subfigure}{0.24\textwidth}
    \includegraphics[width=0.8\textwidth]{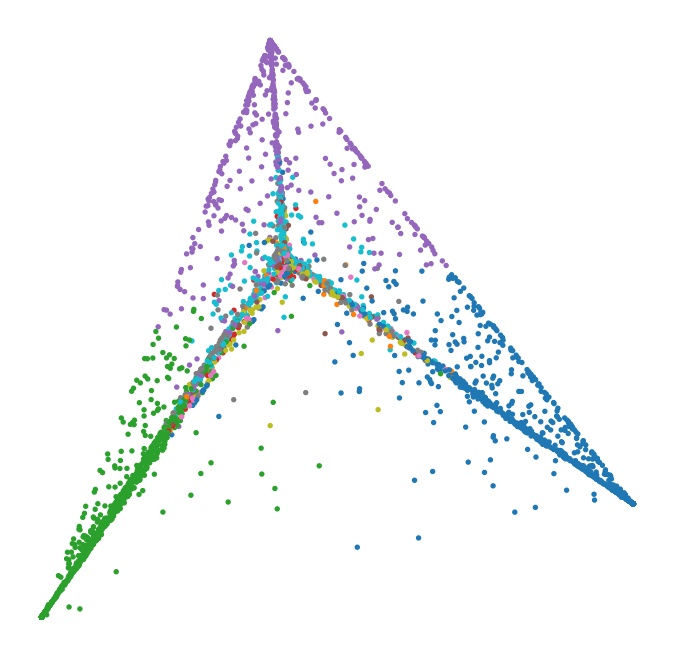}
    \caption{PCA $(D=32)$}
    
\end{subfigure}
\begin{subfigure}{0.24\textwidth}
    \includegraphics[width=0.8\textwidth]{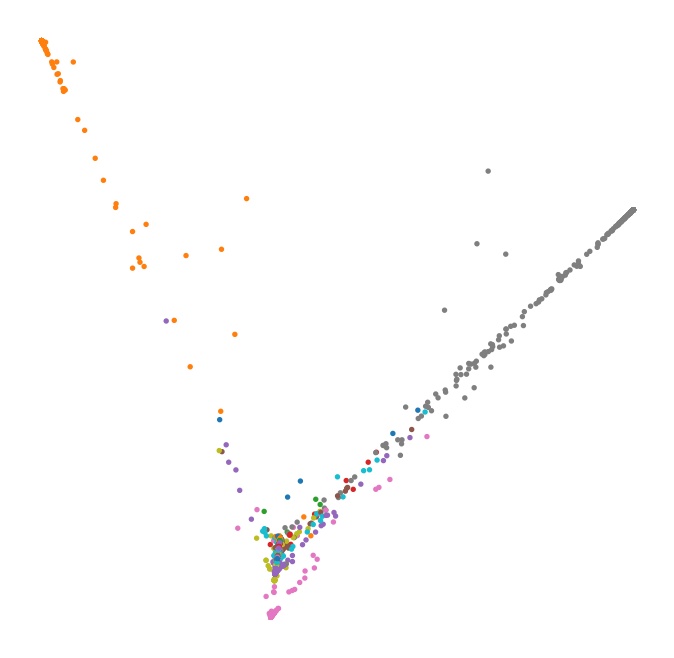}
    \caption{PCA $(D=64)$}
    
\end{subfigure}
\hfill
\begin{subfigure}{0.24\textwidth}
    \includegraphics[width=\textwidth]{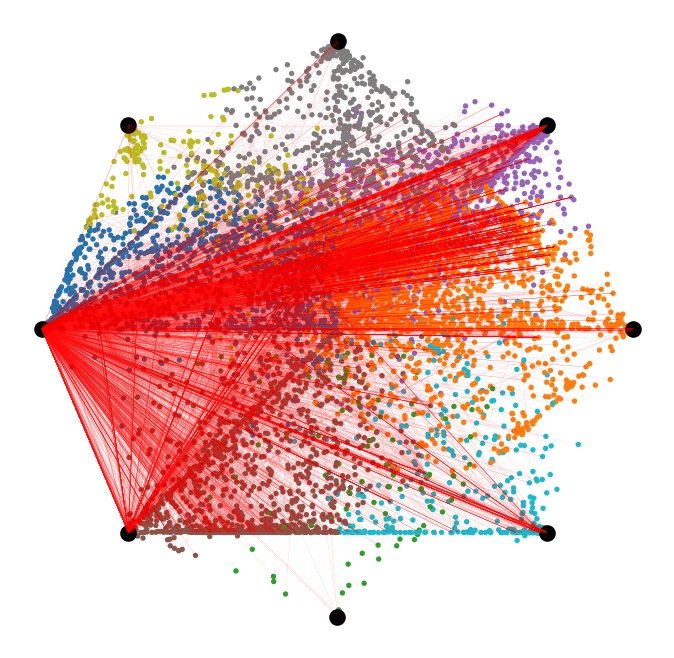}
    \caption{\textsc{NCP} $(D=8)$}
    
\end{subfigure}
\hfill
\begin{subfigure}{0.24\textwidth}
    \includegraphics[width=\textwidth]{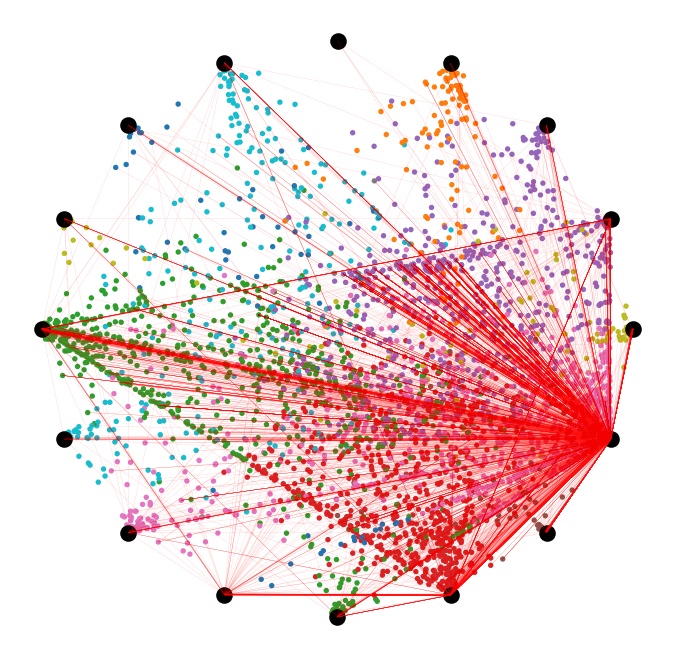}
    \caption{\textsc{NCP} $(D=16)$}
    
\end{subfigure}
\begin{subfigure}{0.24\textwidth}
    \includegraphics[width=\textwidth]{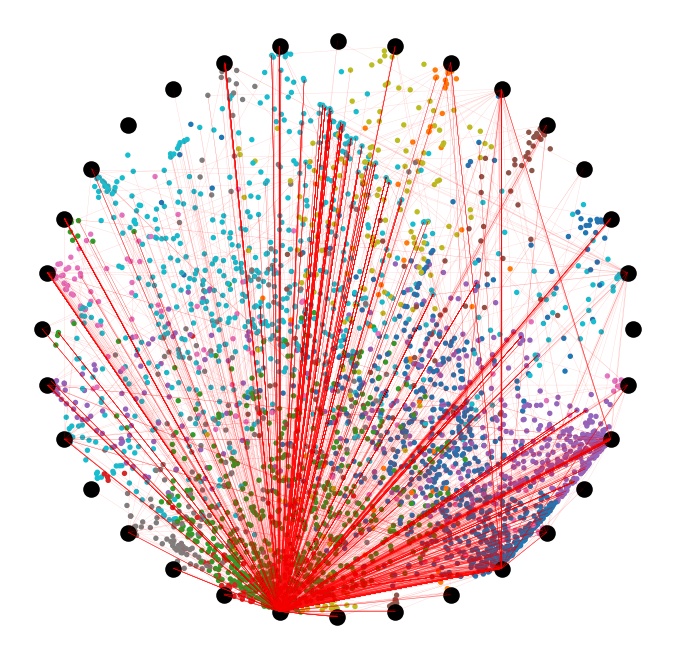}
    \caption{\textsc{NCP} $(D=32)$}
    
\end{subfigure}
\hfill
\begin{subfigure}{0.24\textwidth}
    \includegraphics[width=\textwidth]{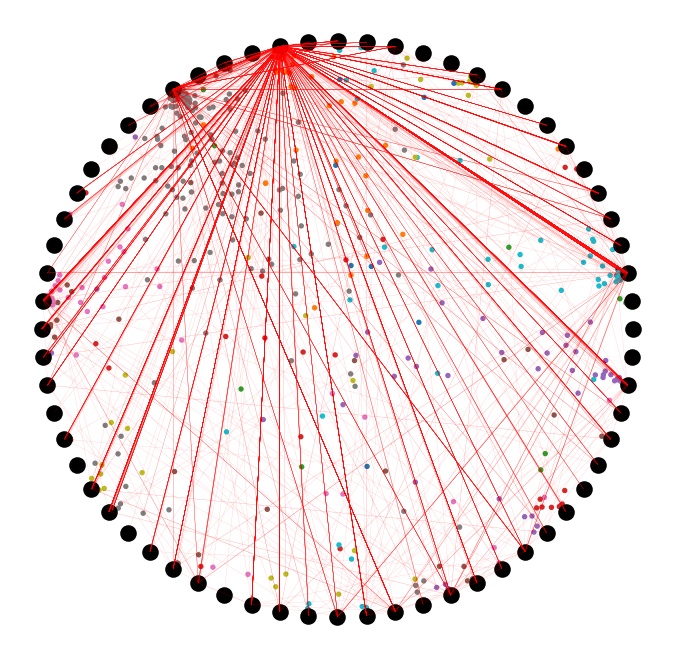}
    \caption{\textsc{NCP} $(D=64)$}
    
\end{subfigure}
\hfill
\begin{subfigure}{0.24\textwidth}
    \includegraphics[width=\textwidth]{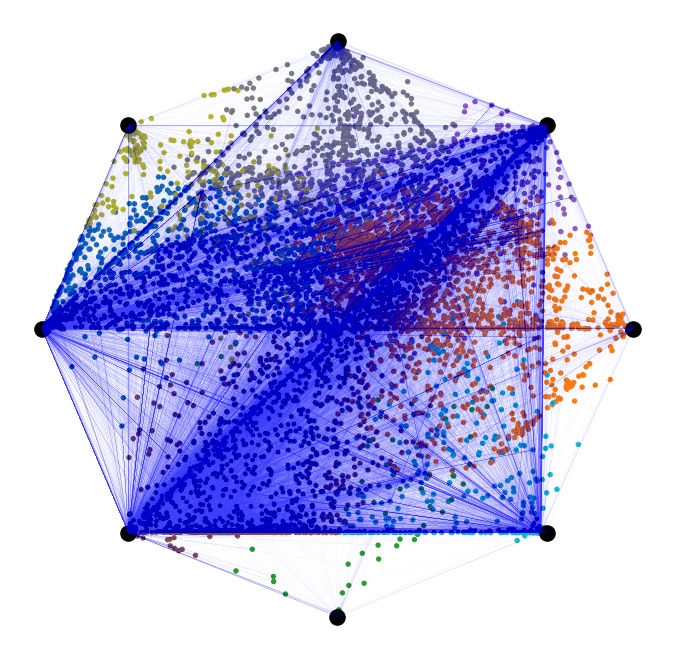}
    \caption{\textsc{PCP} $(D=8)$}
    
\end{subfigure}
\begin{subfigure}{0.24\textwidth}
    \includegraphics[width=\textwidth]{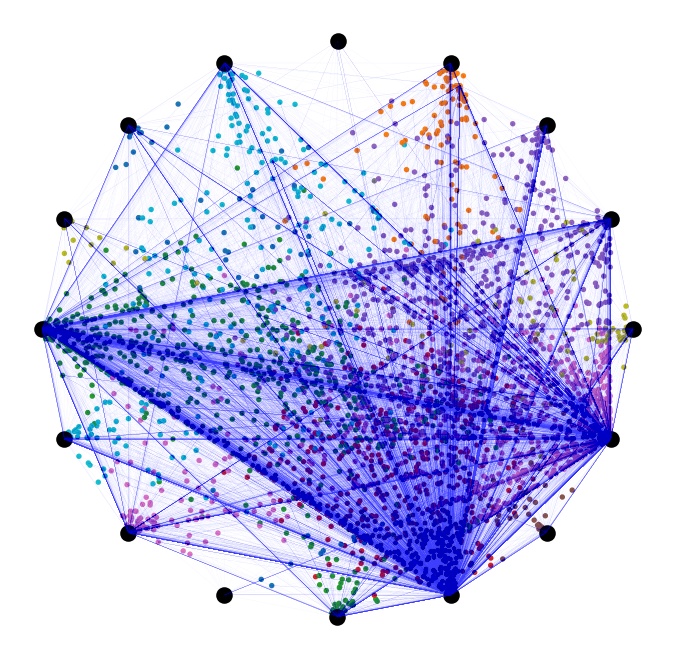}
    \caption{\textsc{PCP} $(D=16)$}
    
\end{subfigure}
\hfill
\begin{subfigure}{0.24\textwidth}
    \includegraphics[width=\textwidth]{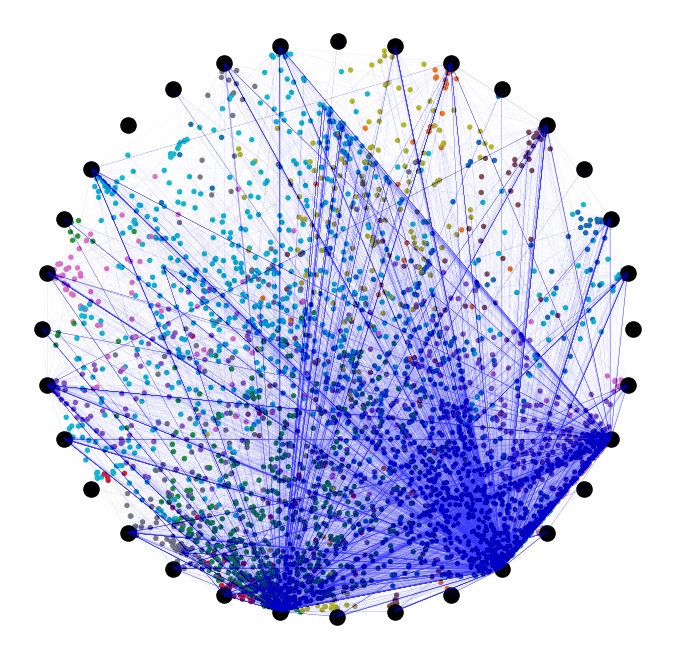}
    \caption{\textsc{PCP} $(D=32)$}
    
\end{subfigure}
\hfill
\begin{subfigure}{0.24\textwidth}
    \includegraphics[width=\textwidth]{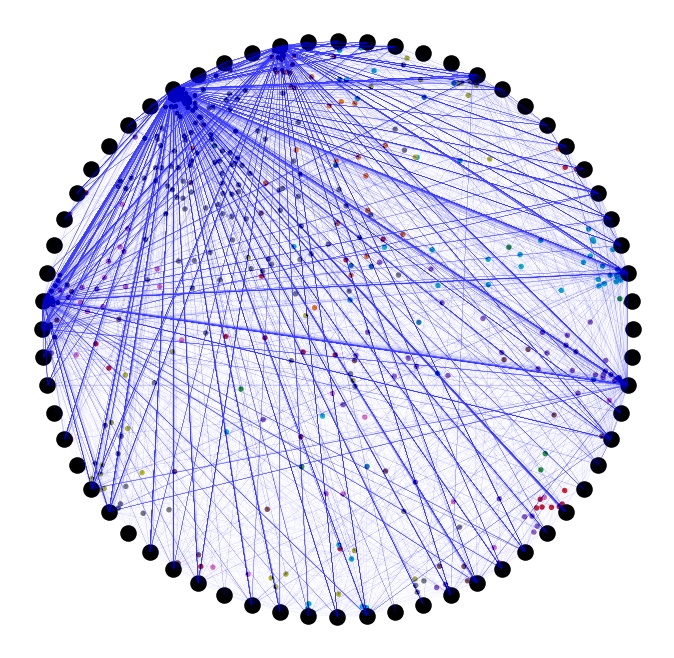}
    \caption{\textsc{PCP} $(D=64)$}
    
\end{subfigure}
\hfill
\begin{subfigure}{0.24\textwidth}
    \includegraphics[width=0.9\textwidth]{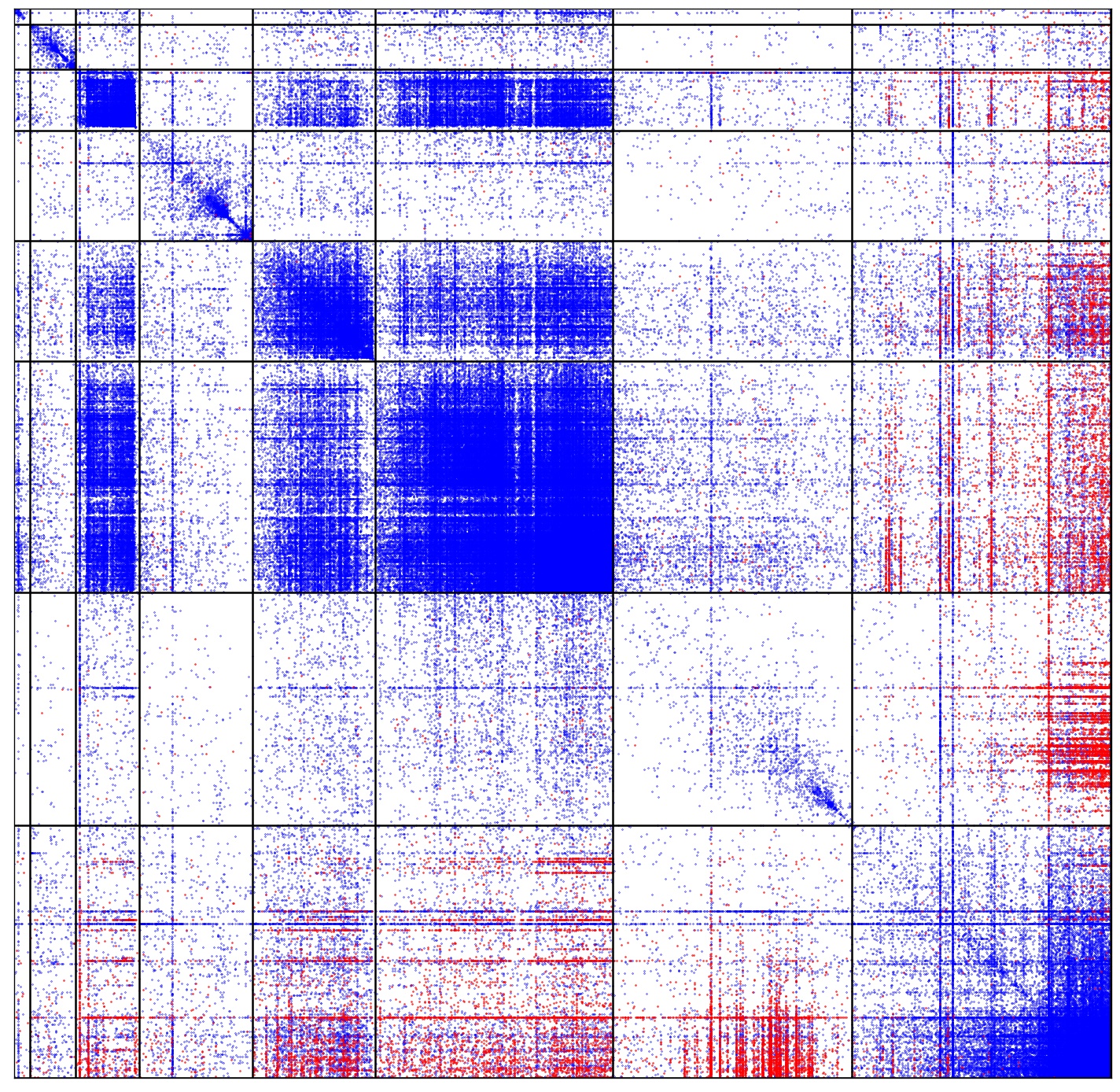}
    \caption{\textsc{OrA} $(D=8)$}
    
\end{subfigure}
\hfill
\begin{subfigure}{0.24\textwidth}
    \includegraphics[width=0.9\textwidth]{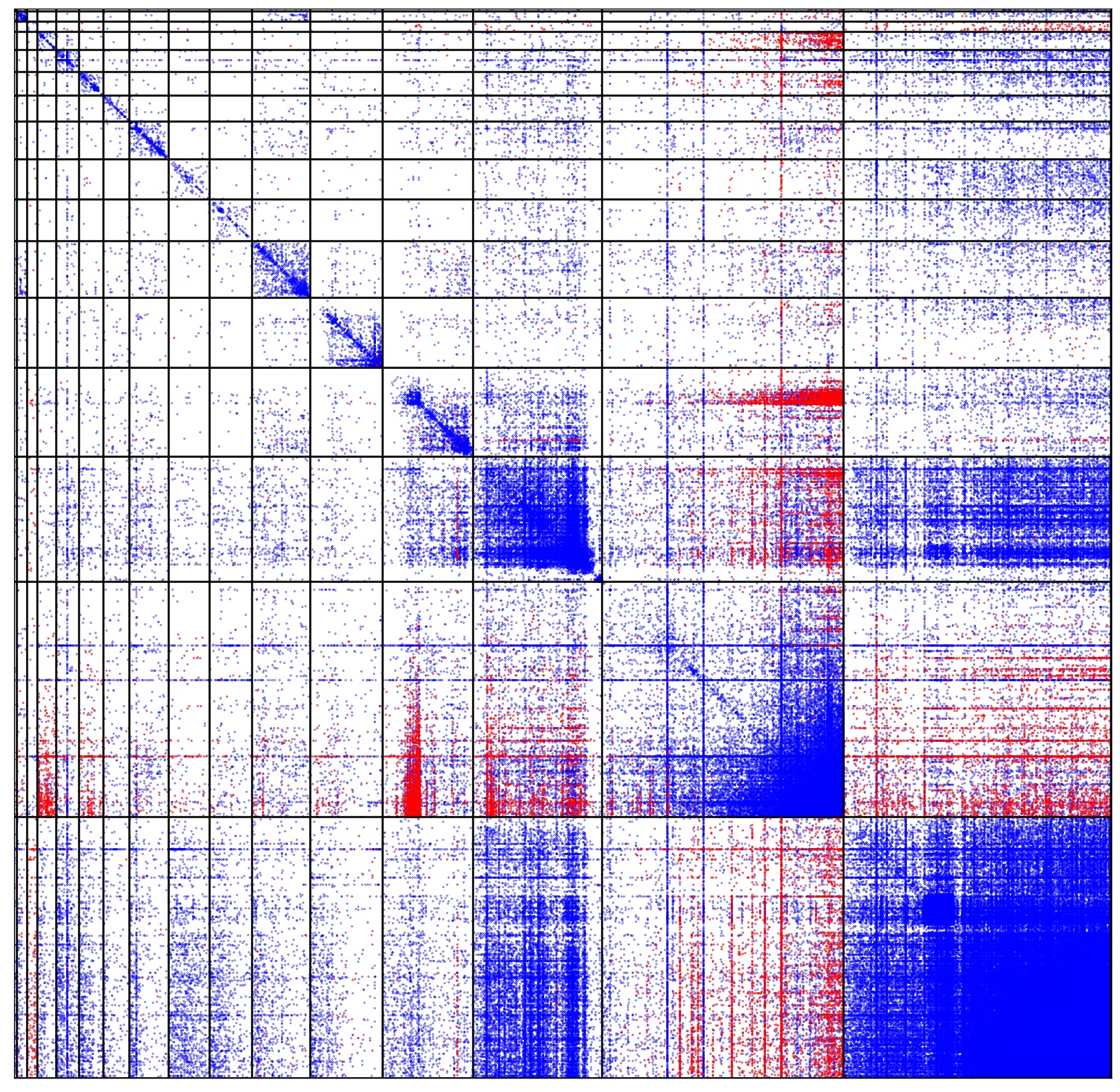}
    \caption{\textsc{OrA} $(D=16)$}
    
\end{subfigure}
\begin{subfigure}{0.24\textwidth}
    \includegraphics[width=0.9\textwidth]{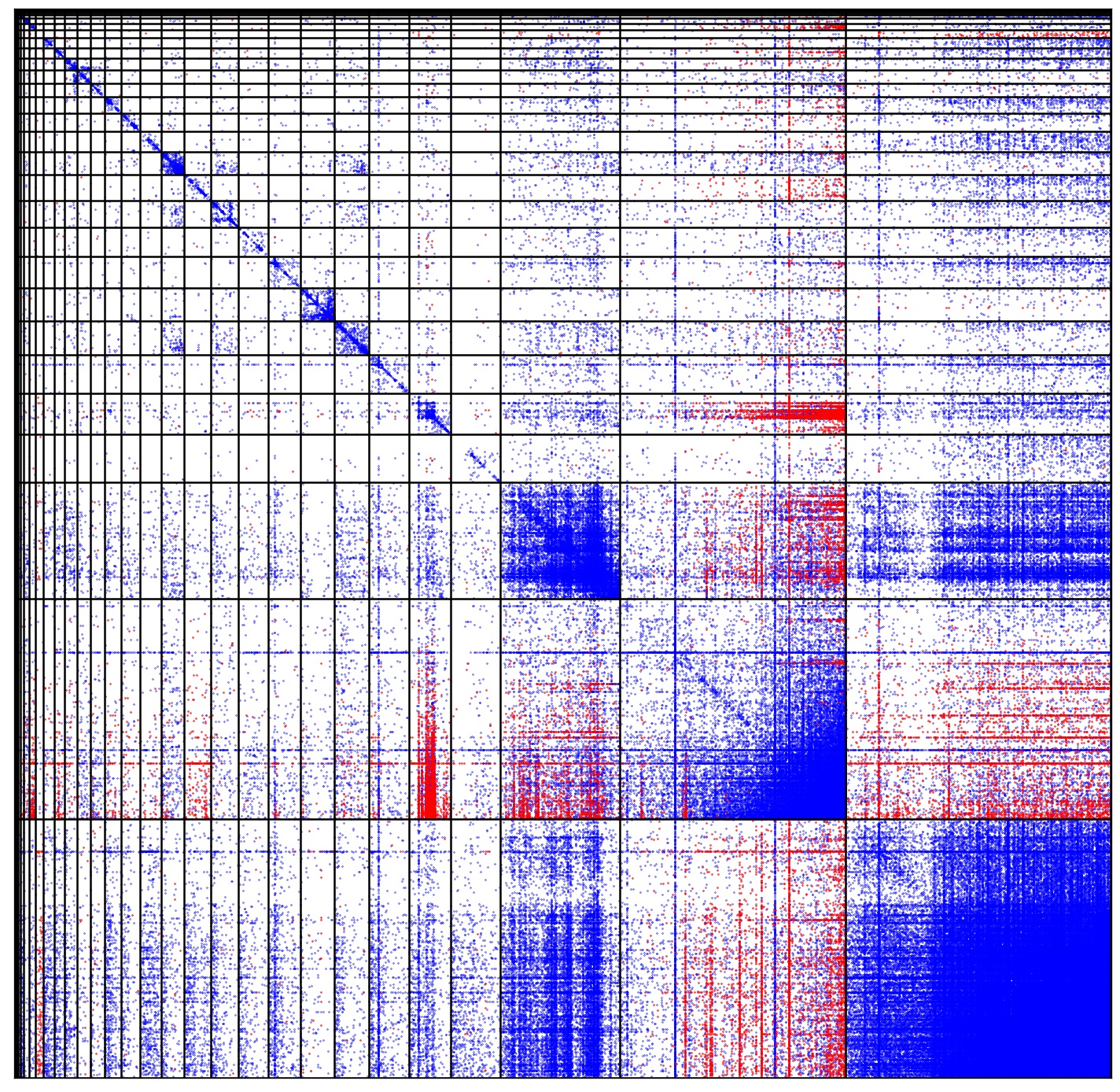}
    \caption{\textsc{OrA} $(D=32)$}
    
\end{subfigure}
\hfill
\begin{subfigure}{0.24\textwidth}
    \includegraphics[width=0.9\textwidth]{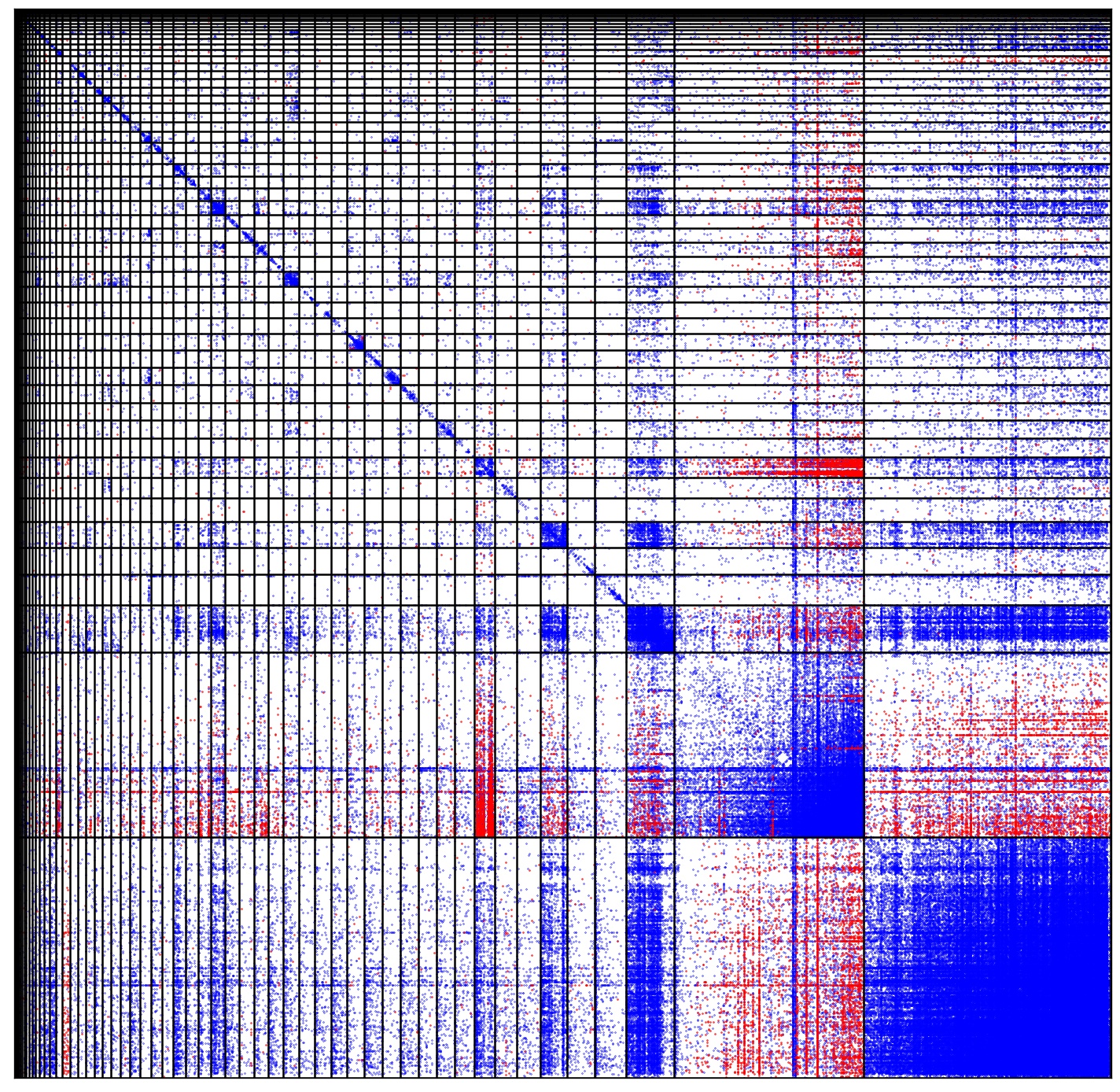}
    \caption{\textsc{OrA} $(D=64)$}
    
\end{subfigure}
\caption{\textbf{\textsc{sHM-LDM}(p=1)}: \textsl{Twitter} Network---Inferred simplex visualizations and ordered adjacency matrices for various dimensions $D$ and with simplex side lengths $\delta$ ensuring identifiability. The first row shows the latent space projection to the first two Principal Components---The second row provides a Negative Circular Plot (\textsc{NCP}) with red lines showcasing negative links between nodes---The third row shows a Positive Circular Plot (\textsc{PCP}) with the blue lines denoting positive links between node pairs---The fourth and final row shows the Ordered Adjacency (\textsc{OrA}) matrices sorted based on the memberships $\mathbf{w}_i$, in terms of maximum simplex corner responsibility, and internally according to the magnitude of the corresponding corner assignment for their reconstruction.}
\label{fig:soc_viz_p1}
\end{figure*}

\begin{figure*}
\centering
\begin{subfigure}{0.24\textwidth}
    \includegraphics[width=0.95\textwidth]{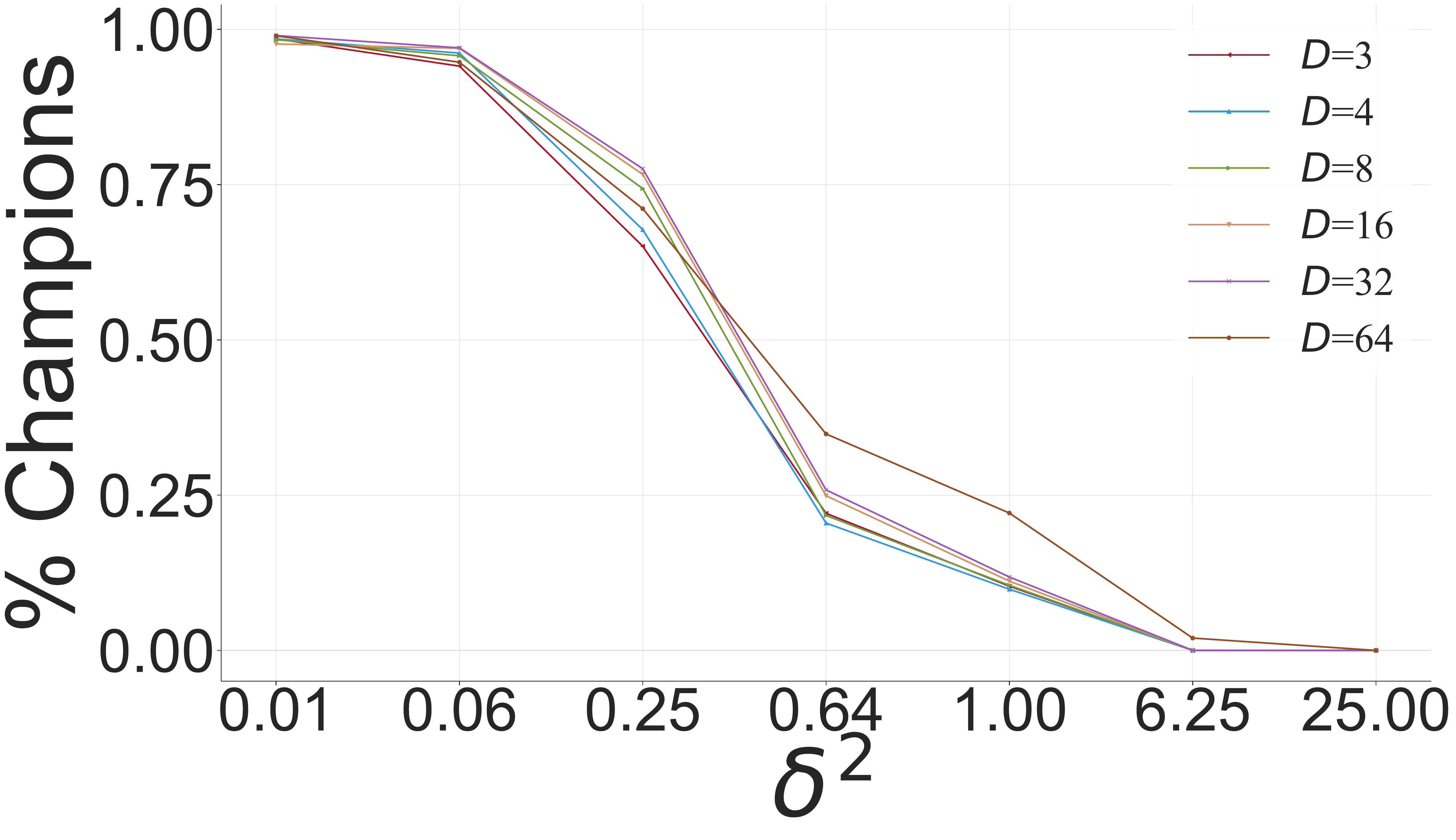}
    \caption{\% Champions}
    
\end{subfigure}
\hfill
\begin{subfigure}{0.24\textwidth}
    \includegraphics[width=0.95\textwidth]{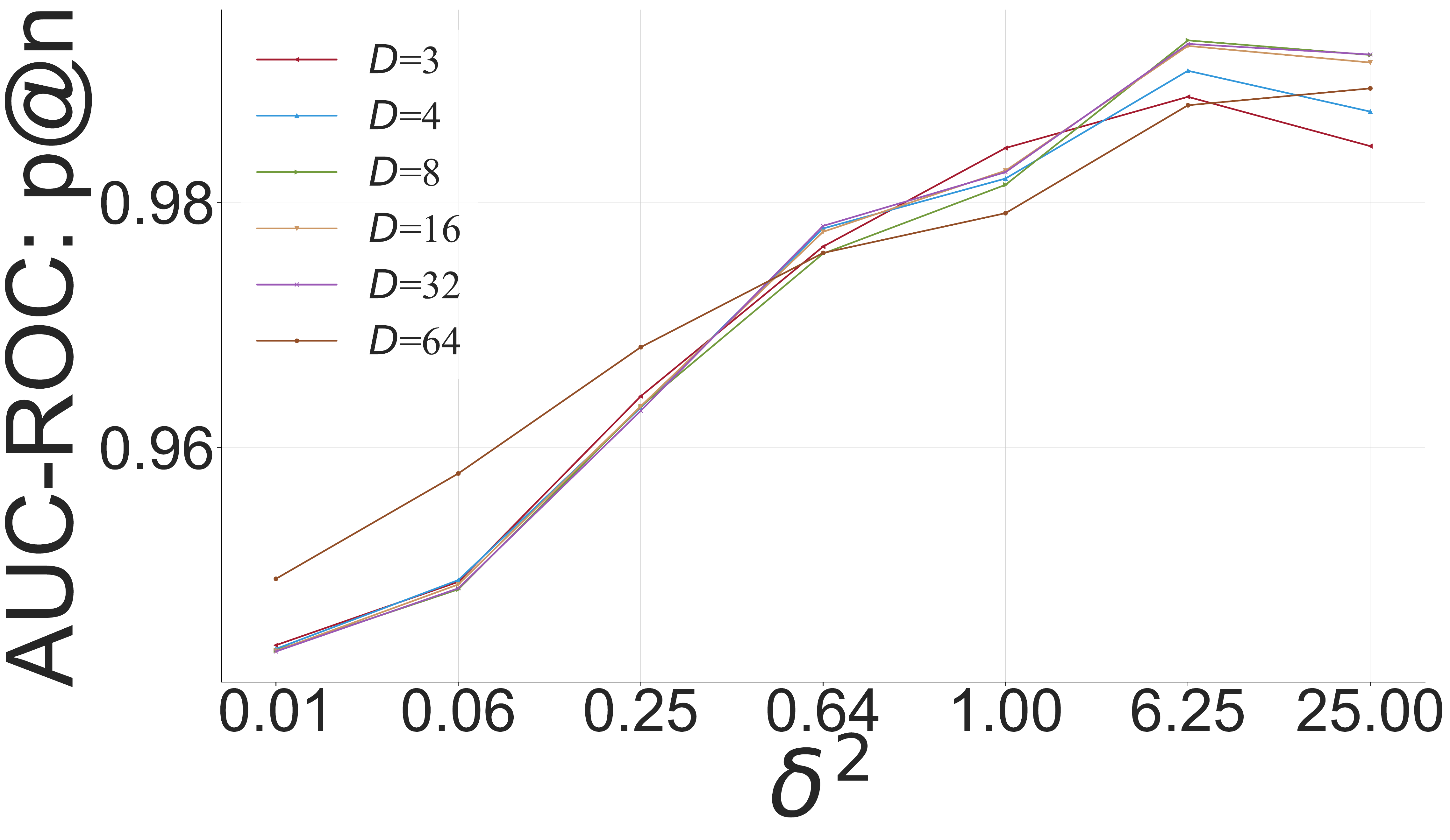}
    \caption{\textsc{LSP}: p@n}
    
\end{subfigure}
\hfill
\begin{subfigure}{0.24\textwidth}
    \includegraphics[width=0.95\textwidth]{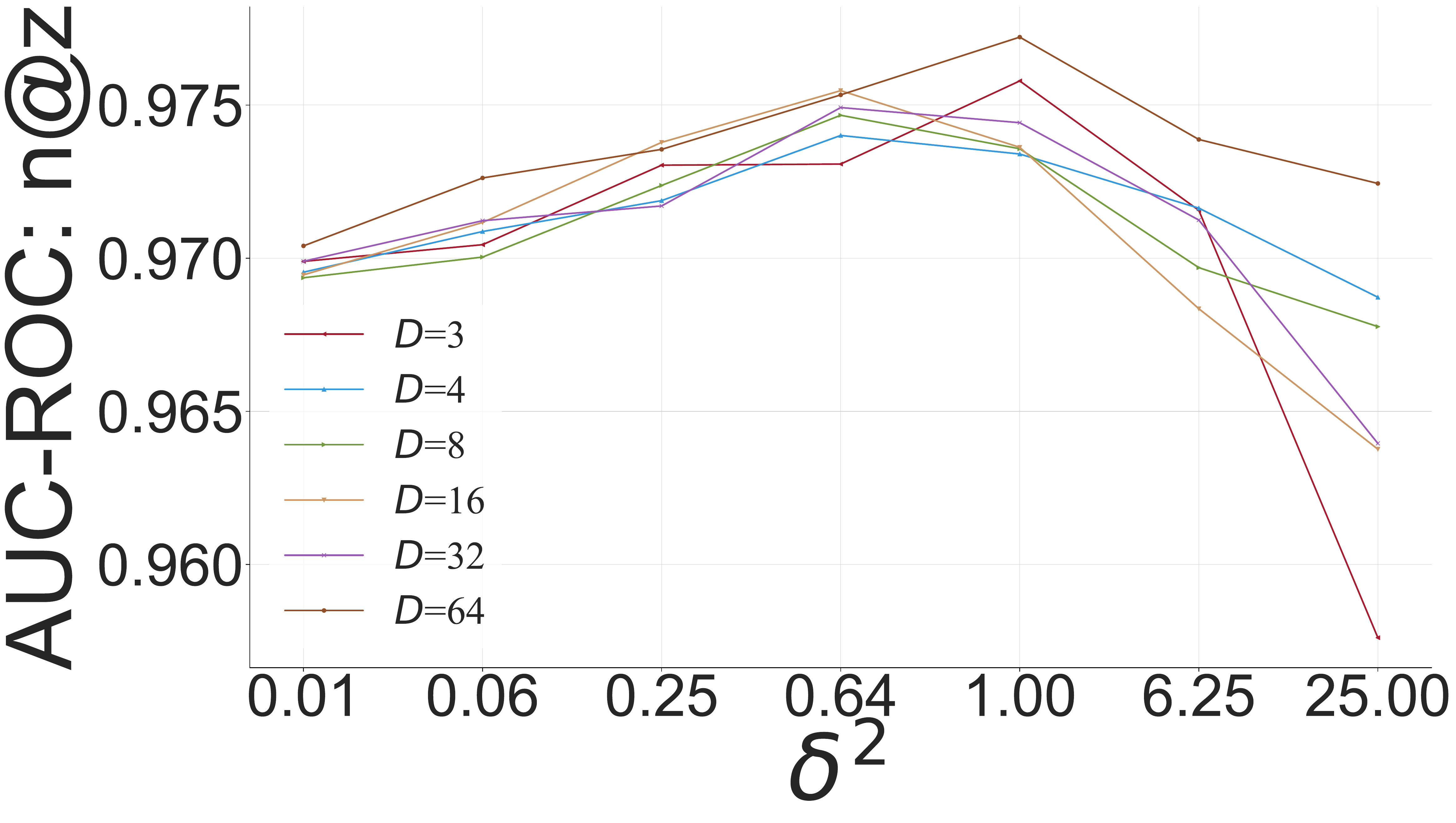}
    \caption{\textsc{SLP}: n@z}
    
\end{subfigure}
\begin{subfigure}{0.24\textwidth}
    \includegraphics[width=0.95\textwidth]{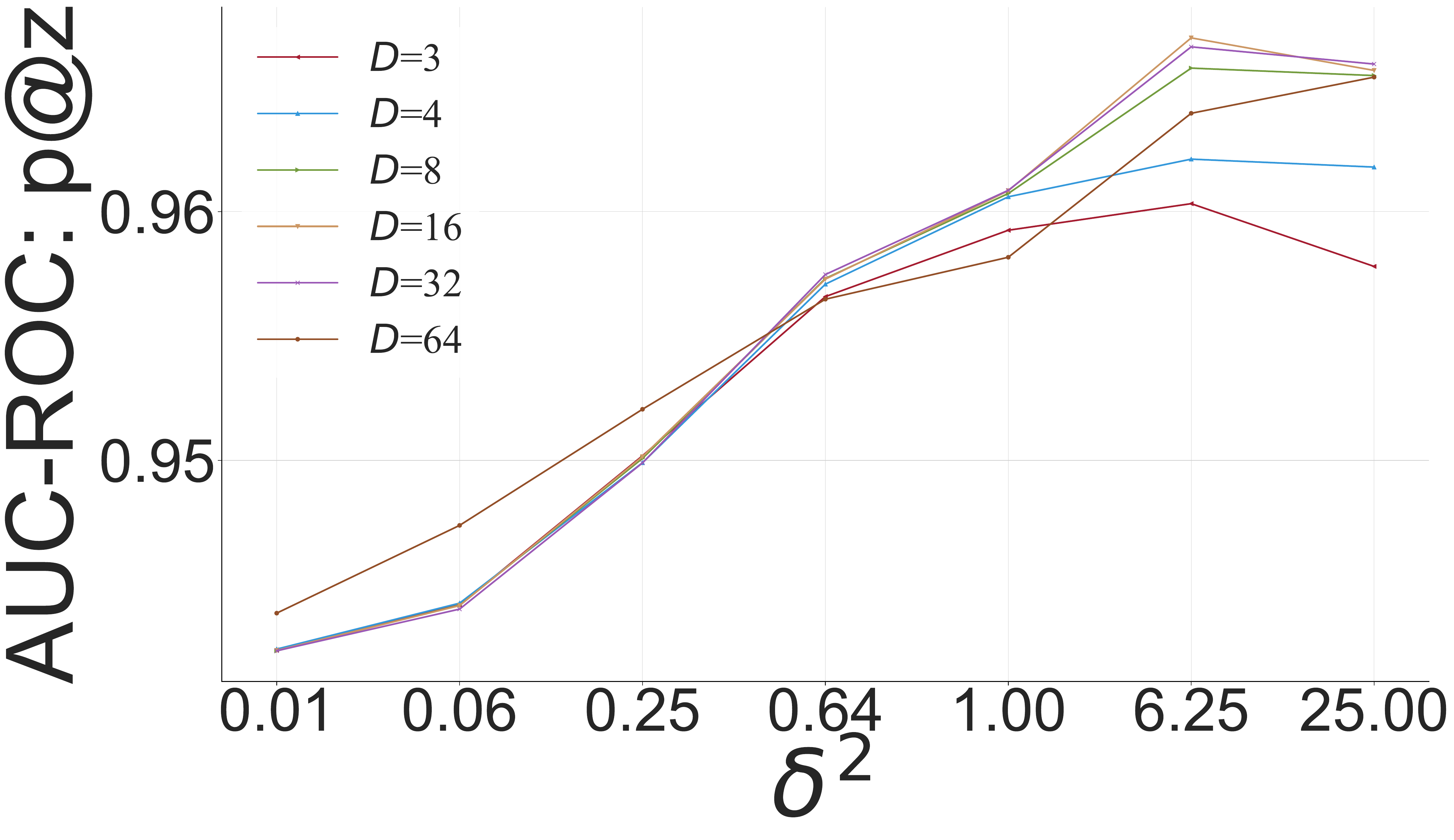}
    \caption{\textsc{SLP}: p@z}
    
\end{subfigure}
\hfill
\begin{subfigure}{0.24\textwidth}
    \includegraphics[width=0.95\textwidth]{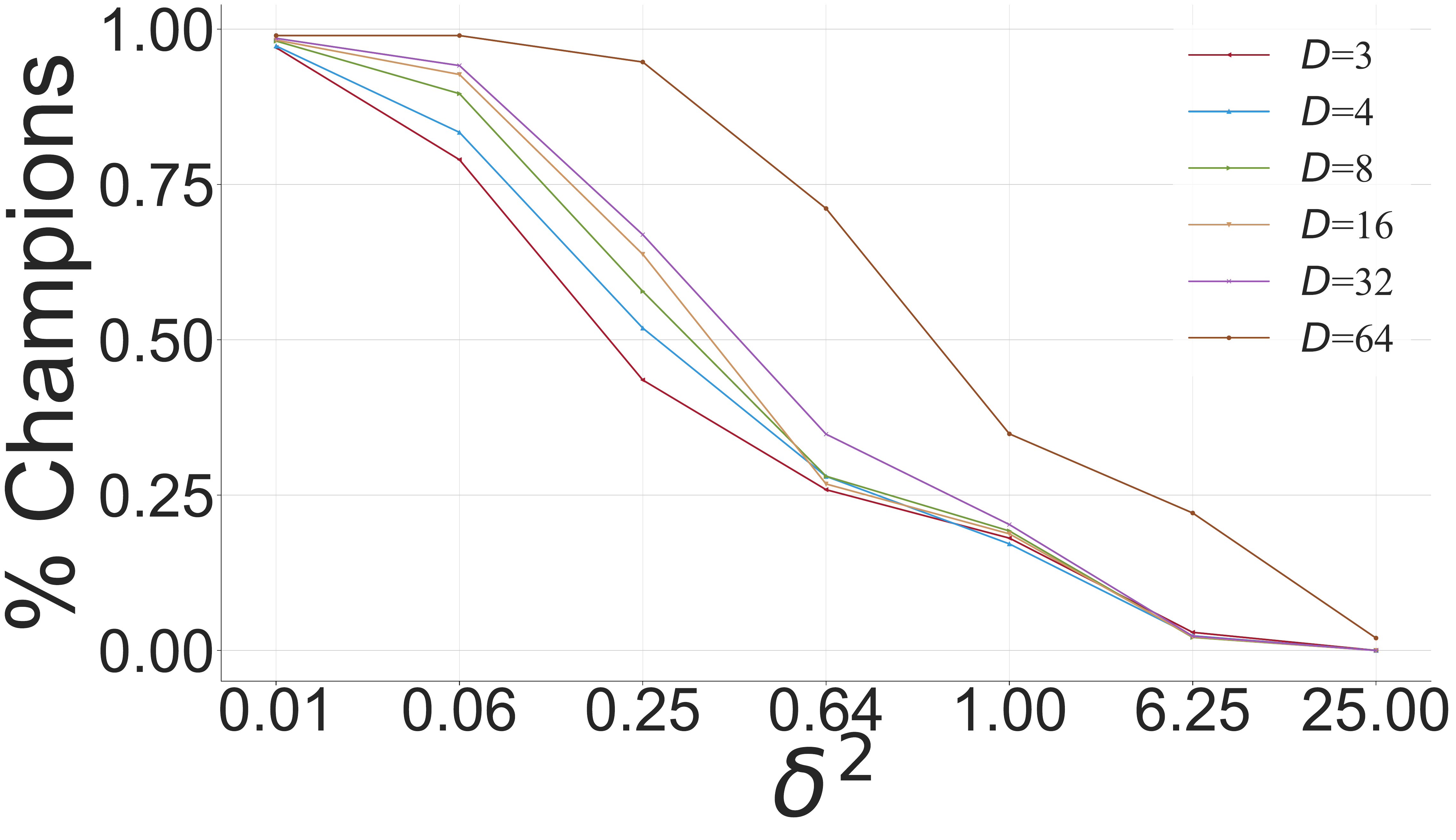}
    \caption{\% Champions}
    
\end{subfigure}
\hfill
\begin{subfigure}{0.24\textwidth}
    \includegraphics[width=0.95\textwidth]{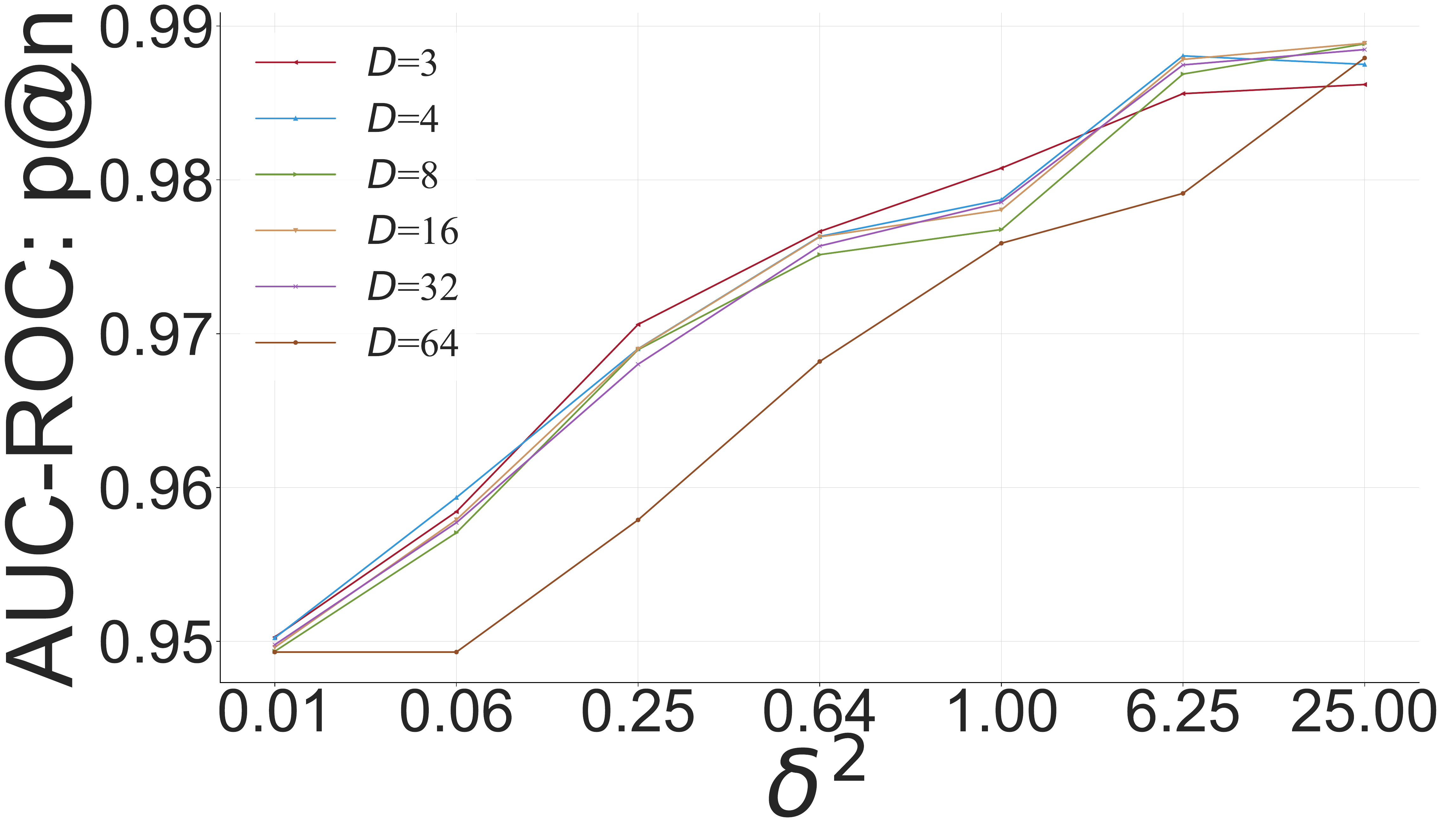}
    \caption{\textsc{LSP}: p@n}
    
\end{subfigure}
\hfill
\begin{subfigure}{0.24\textwidth}
    \includegraphics[width=0.95\textwidth]{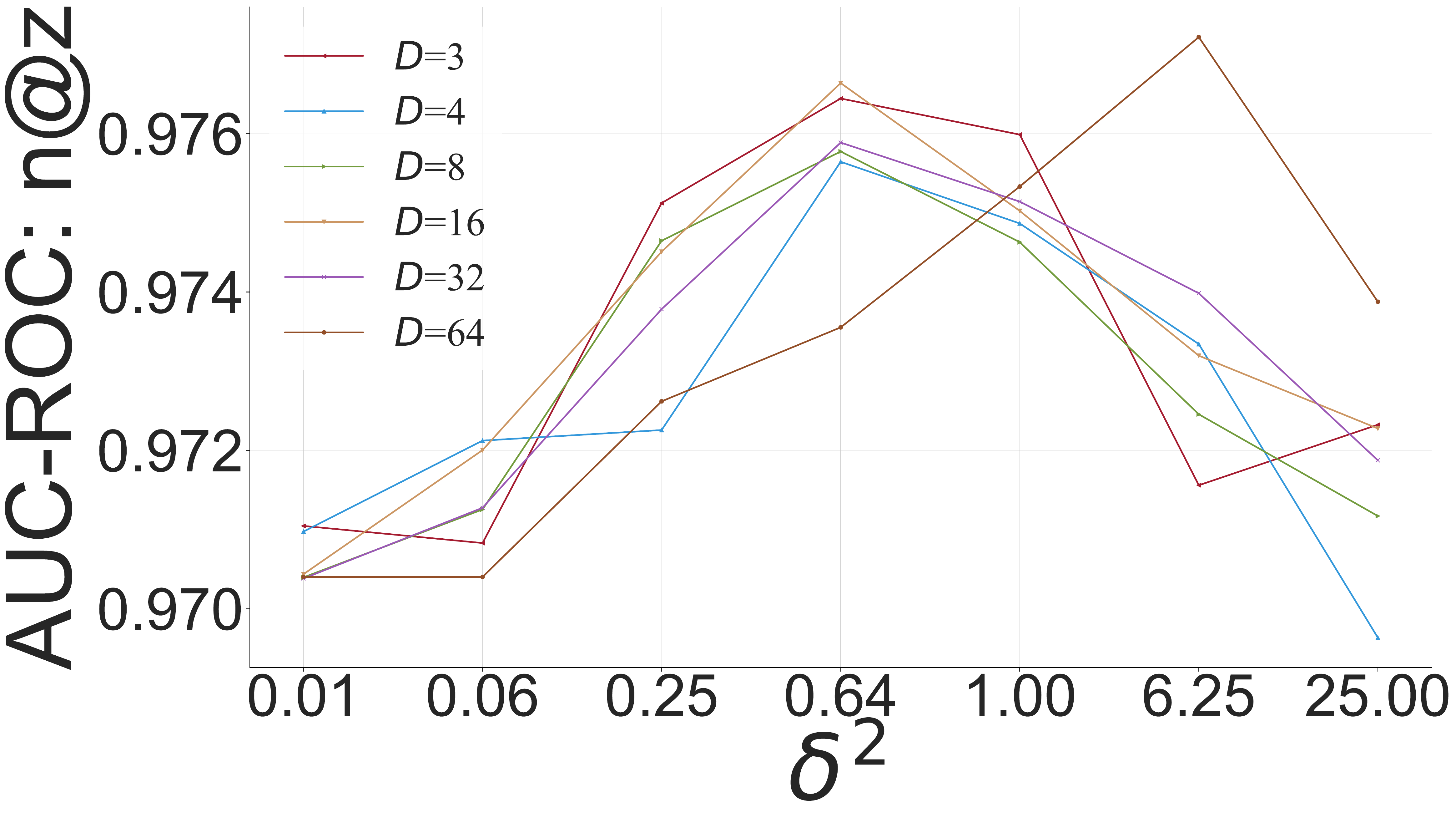}
    \caption{\textsc{SLP}: n@z}
    
\end{subfigure}
\begin{subfigure}{0.24\textwidth}
    \includegraphics[width=0.95\textwidth]{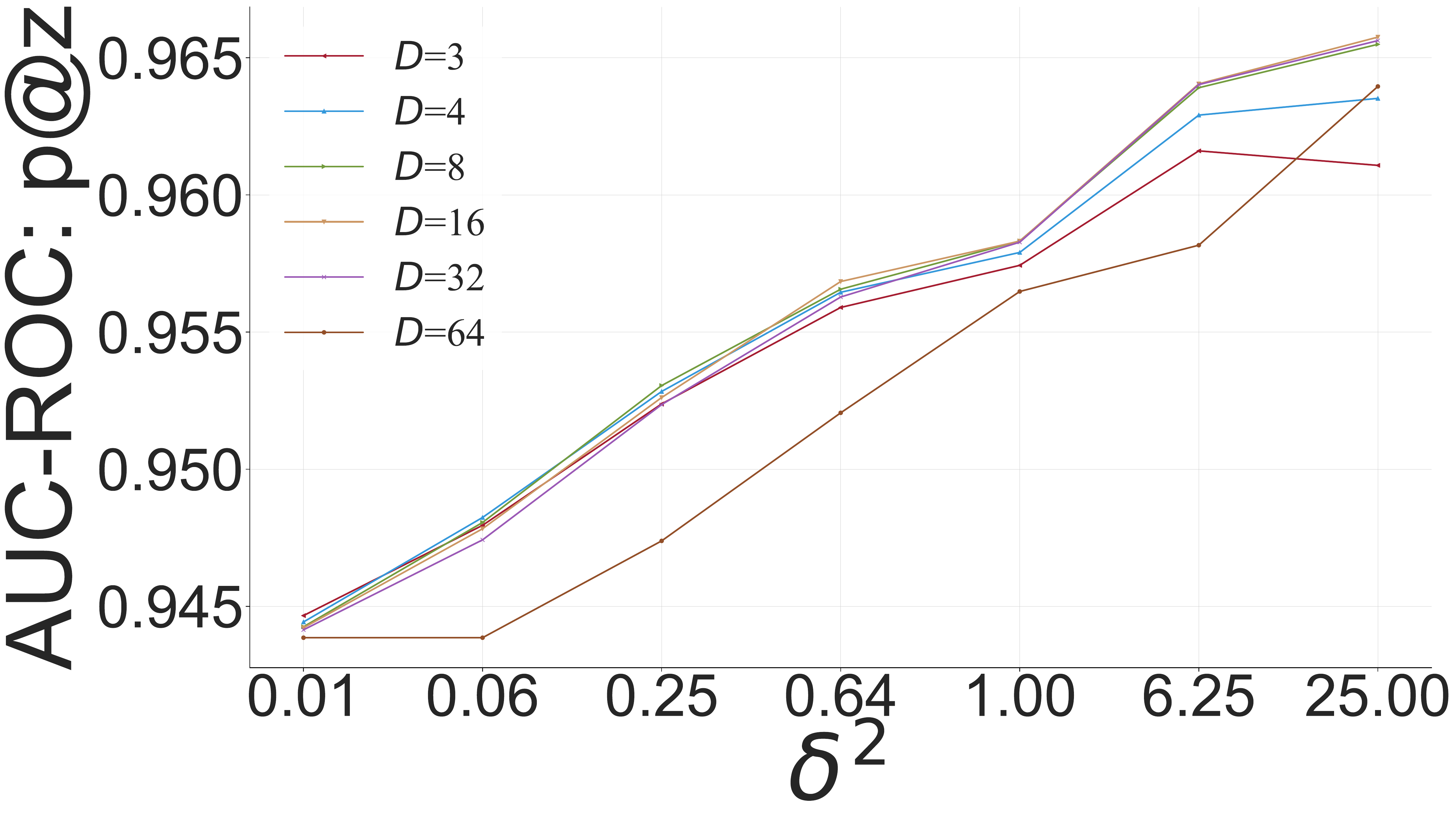}
    \caption{\textsc{SLP}: p@z}
    
\end{subfigure}
\caption{\textbf{\textsc{sHM-LDM}}: \textsl{Twitter} Network---Performance characteristics across different dimensions $D$ in terms of various values $\delta^2$ (simplex size). The first column shows the total community champions (\%)  across dimensions for \textsc{sHM-LDM}---The second column provides the Link Sign Prediction (\textsc{LSP}) performance for the task of inferring the sign of the test set links (p@n)---The third and fourth columns describe the performance for the Signed Link Prediction (\textsc{SLP}) tasks, distinguishing between negatively related and disconnected nodes (n@z), as well as, positively connected to disconnected nodes (p@z), respectively. Top row: $p = 2$ model specification. Bottom row $p = 1$ model specification.}
\label{fig:simplex_per}
\end{figure*}

\textbf{Visualizations:}
The inferred simplex of \textsc{sHM-LDM} extracts information about node memberships to distinct aspects of the network. Similar to \cite{slim}, we provide visualizations regarding the latent space as projected to the first two principal components and include circular plots describing the simplex and node memberships in two dimensions. Specifically, each corner of the simplex is positioned to the border of a circle, every $\text{rad}_k=\frac{2\pi}{D}$ radians, with $D$ being the number of the simplex corners. Furthermore, we provide the re-ordered adjacency matrices based on the inferred memberships for various dimensions. Visualizations for the \textsl{Twitter} are provided in Fig. \ref{fig:soc_viz_p2} and Fig. \ref{fig:soc_viz_p1}
for \textsc{sHM-LDM}($p=2$) and \textsc{sHM-LDM}($p=1$) models, respectively. For both models, visualizations are available for different dimensions while we see how the model successfully uncovers distinct aspects of the network when the simplex side length $\delta$ ensures identifiability. From the circular plots enriched with the corresponding negative (red lines) and positive (blue lines) links, we observe that the models always uncover simplex corners to act as dislike (high negative in-degree) and like (high positive in-degree) profiles of the network. We also observe controversial network profiles, sharing a high degree of both negative and positive connections. For the ordered adjacency matrices of the two models, we can observe successful structure extraction and discovery, and as we increase the dimensionality of the simplex structure it becomes finer and finer. Lastly, we also obtain simplex corners for the inferred simplex containing not-so-intensely connected nodes. This comes as a validation of stochastic equivalence presence that the \textsc{sHM-LDM} framework can express.

\textbf{Simplex size and performance evaluation:} In Fig. \ref{fig:simplex_per} we provide performance characteristics against various dimensions $D$ as a function of $\delta^2$ for \textsc{sHM-LDM}(p=2) and \textsc{sHM-LDM}(p=1) models, respectively. The first column shows the percentage of champion nodes as defined by the model whereas expected smaller simplex volumes lead to a higher percentage of hard-clustered nodes. In addition, it is clear that the dimensionality in \textsc{sHM-LDM}(p=1) has a bigger effect on the node champions than for the \textsc{sHM-LDM}(p=2) case. The last three columns showcase the performance across the p@z, n@z, and p@z tasks respectively. Comparing to the results of the\textsc{HM-LDM} we observe for the signed networks and \textsc{sHM-LDM} that the performance is not affected to the same degree by the shrinkage of the latent space (the maximum case is present in the p@n task accounting to just a $4\%$ decrease).

\begin{figure*}
\centering
\begin{subfigure}{0.32\textwidth}
    \includegraphics[width=0.7\textwidth]{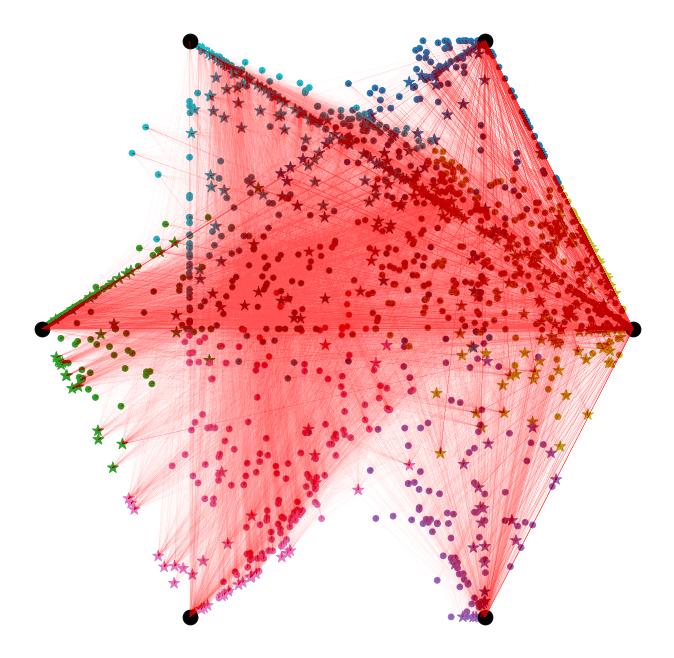}
    \caption{\textsc{NCP} \textsc{U.S.-Senate}}
    
\end{subfigure}
\hfill
\begin{subfigure}{0.32\textwidth}
    \includegraphics[width=0.7\textwidth]{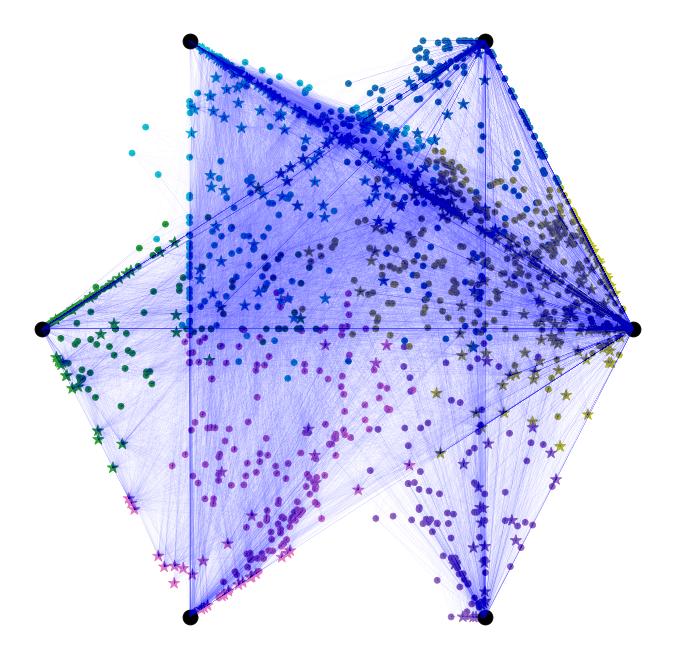}
    \caption{\textsc{PCP} \textsc{U.S.-Senate}}
    
\end{subfigure}
\hfill
\begin{subfigure}{0.32\textwidth}
    \includegraphics[width=0.7\textwidth]{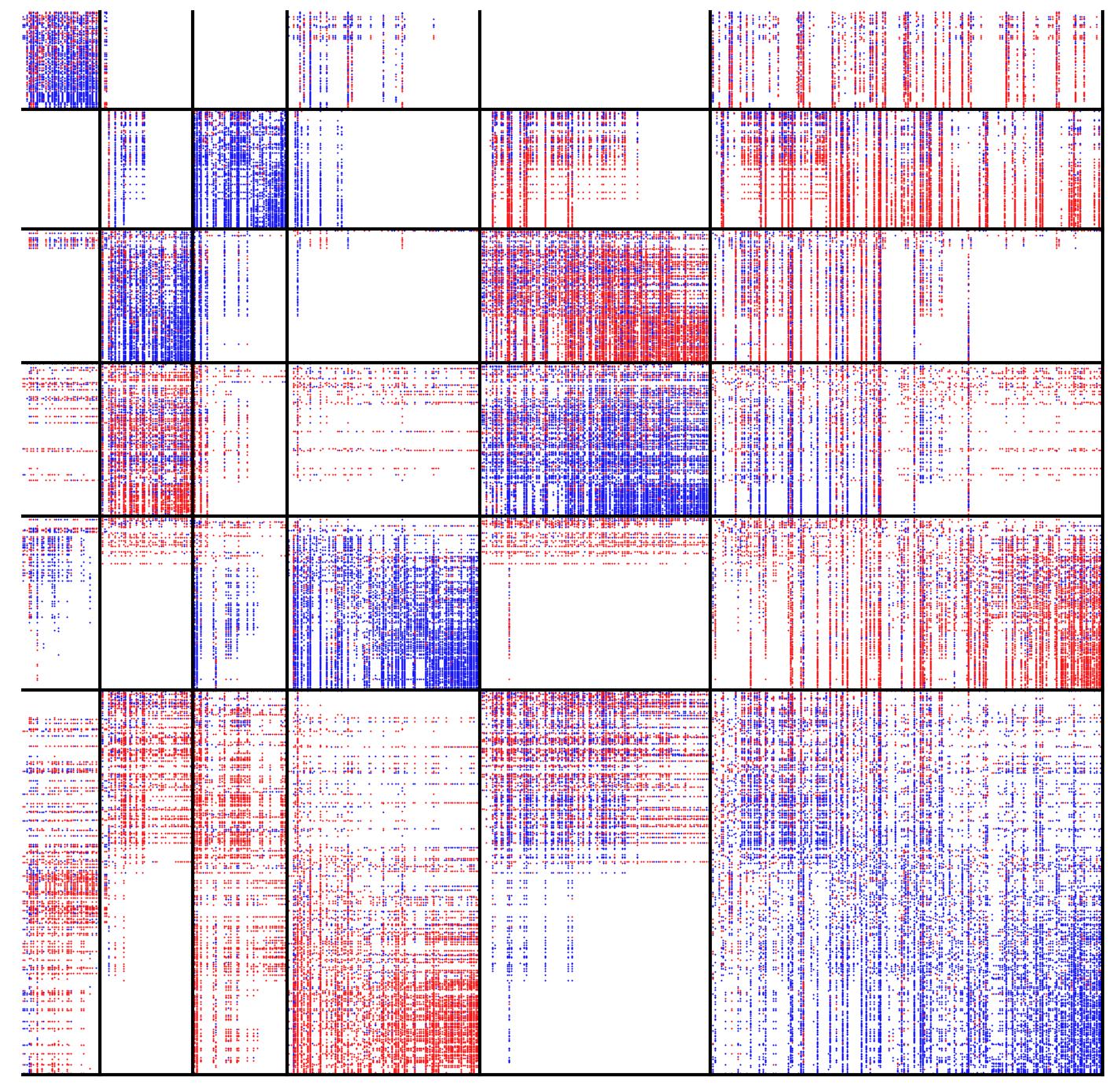}
    \caption{\textsc{OrA} \textsc{U.S.-Senate}}
    
\end{subfigure}
\begin{subfigure}{0.32\textwidth}
    \includegraphics[width=0.7\textwidth]{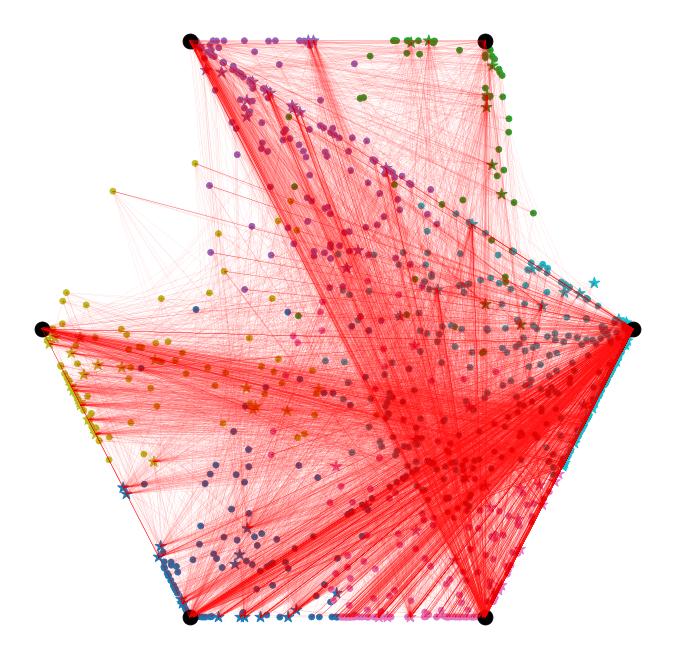}
    \caption{\textsc{PCP} \textsc{U.S.-House}}
    
\end{subfigure}
\hfill
\begin{subfigure}{0.32\textwidth}
    \includegraphics[width=0.7\textwidth]{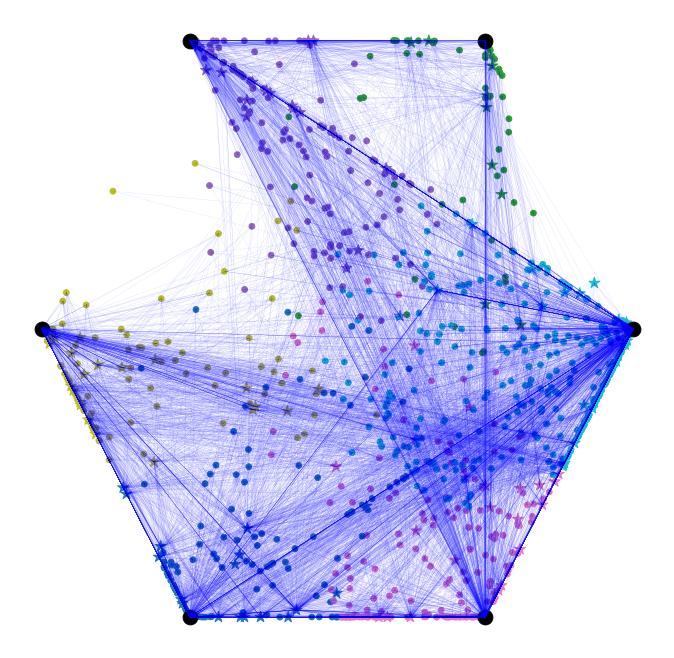}
    \caption{\textsc{PCP} \textsc{U.S.-House}}
    
\end{subfigure}
\hfill
\begin{subfigure}{0.32\textwidth}
    \includegraphics[width=0.7\textwidth]{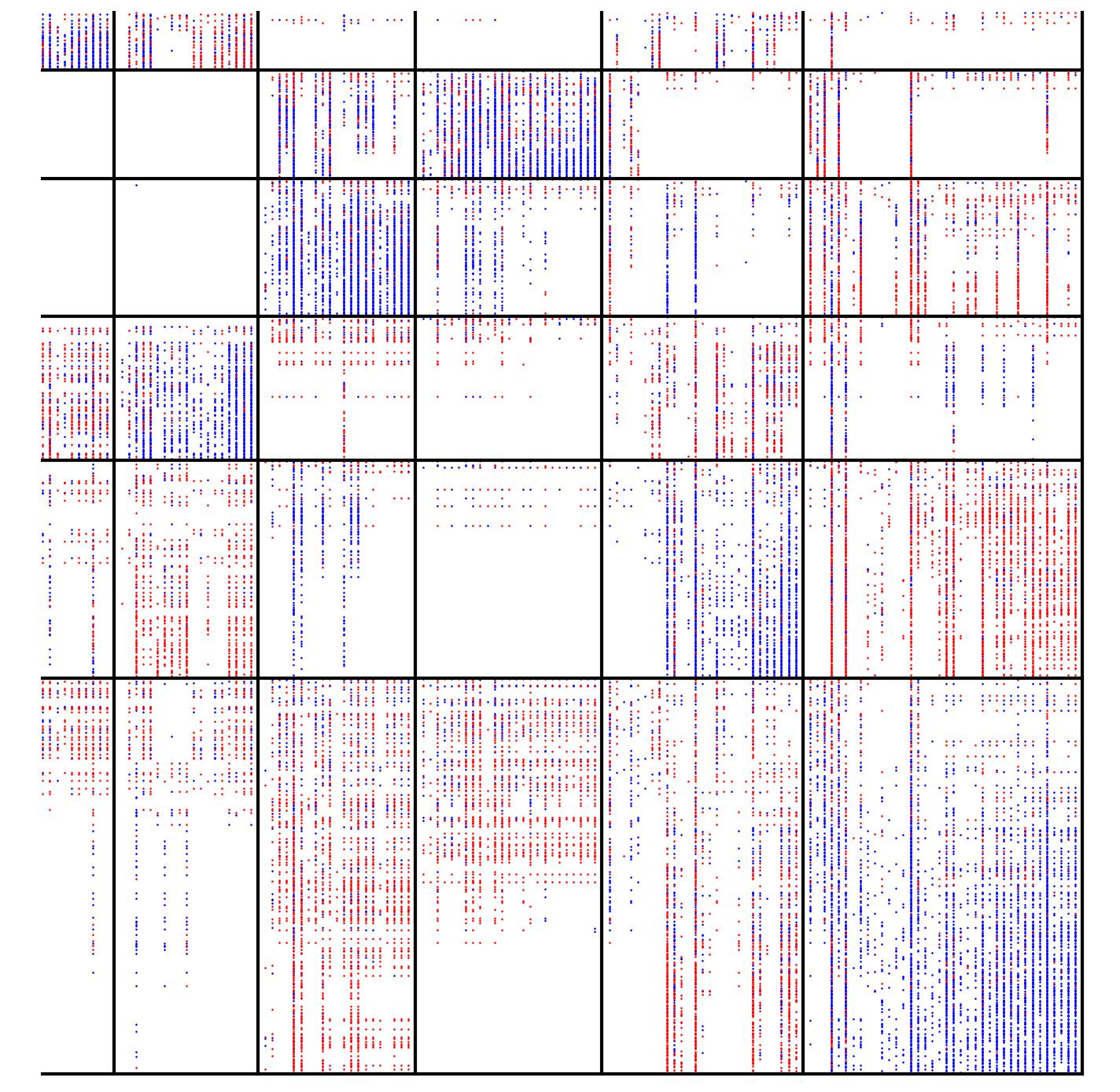}
    \caption{\textsc{OrA} \textsc{U.S.-House}}
    
\end{subfigure}
\caption{\textbf{\textsc{sHM-LDM}}(p=2): Inferred simplex visualizations and ordered adjacency matrices for a $D=6$ dimensional simplex with side lengths $\delta$ ensuring identifiability. The first column provides a Negative Circular Plot (\textsc{NCP}) with red lines showcasing negative links between nodes---The second column shows a Positive Circular Plot (\textsc{PCP}) with the blue lines denoting positive links between node pairs---The third and final column shows the Ordered Adjacency (\textsc{OrA}) matrices ordered based on the memberships, in terms of maximum simplex corner responsibility, and internally according to the magnitude of the corresponding corner assignment for their reconstruction. Top row: \textsc{U.S.-House}. Bottom row \textsc{U.S.-Senate}.}
\label{fig:bip_signed_p2}
\end{figure*}
\begin{figure*}
\centering
\begin{subfigure}{0.32\textwidth}
    \includegraphics[width=0.7\textwidth]{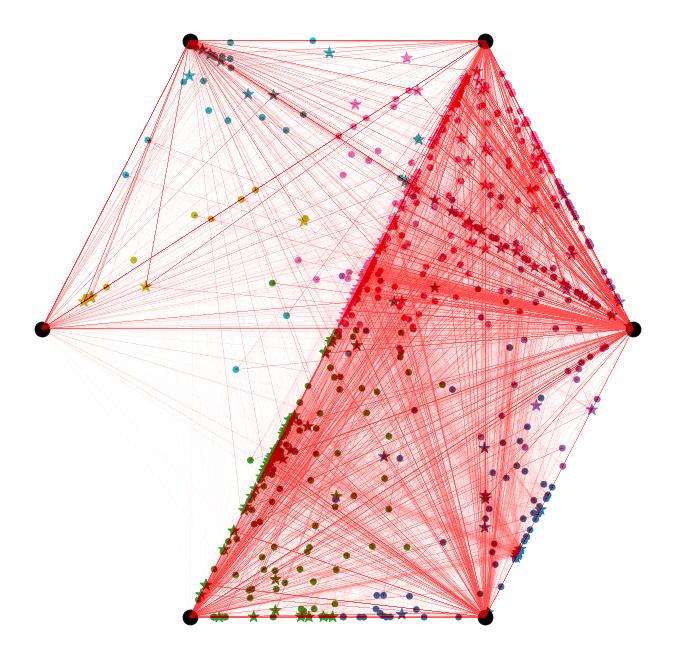}
    \caption{\textsc{NCP} \textsc{U.S.-Senate}}
    
\end{subfigure}
\hfill
\begin{subfigure}{0.32\textwidth}
    \includegraphics[width=0.7\textwidth]{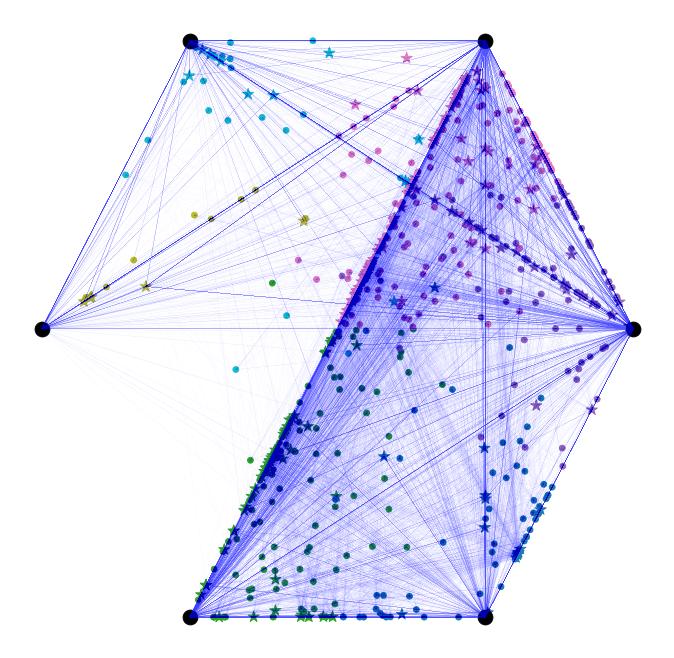}
    \caption{\textsc{PCP} \textsc{U.S.-Senate}}
    
\end{subfigure}
\hfill
\begin{subfigure}{0.32\textwidth}
    \includegraphics[width=0.7\textwidth]{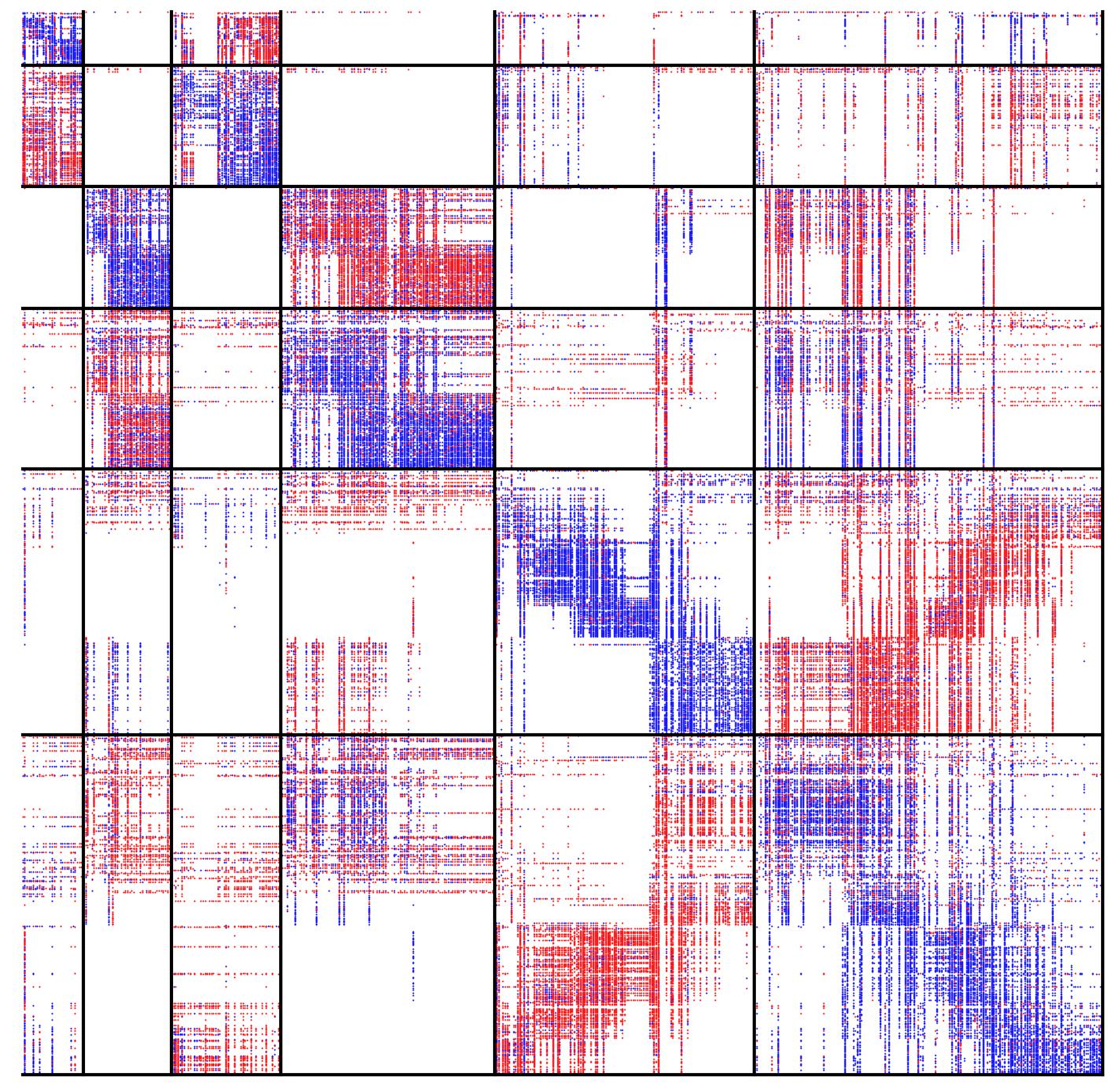}
    \caption{\textsc{OrA} \textsc{U.S.-Senate}}
    
\end{subfigure}
\begin{subfigure}{0.32\textwidth}
    \includegraphics[width=0.7\textwidth]{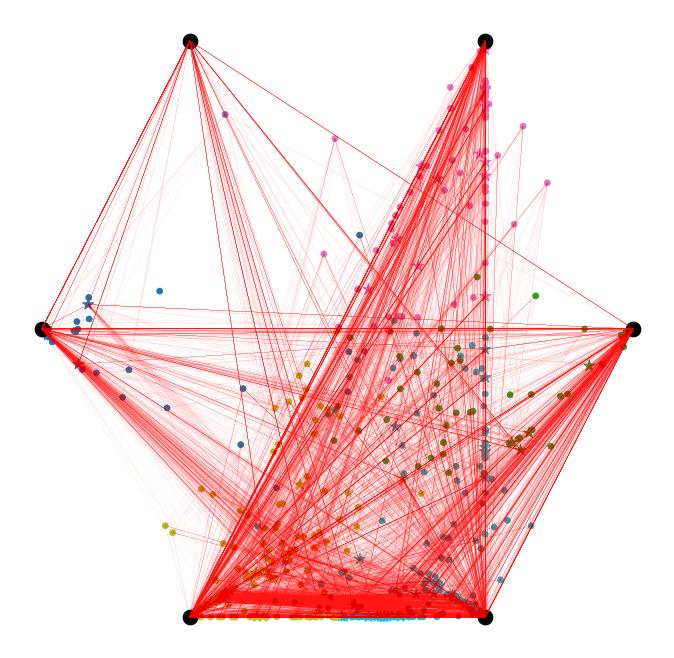}
    \caption{\textsc{PCP} \textsc{U.S.-House}}
    
\end{subfigure}
\hfill
\begin{subfigure}{0.32\textwidth}
    \includegraphics[width=0.7\textwidth]{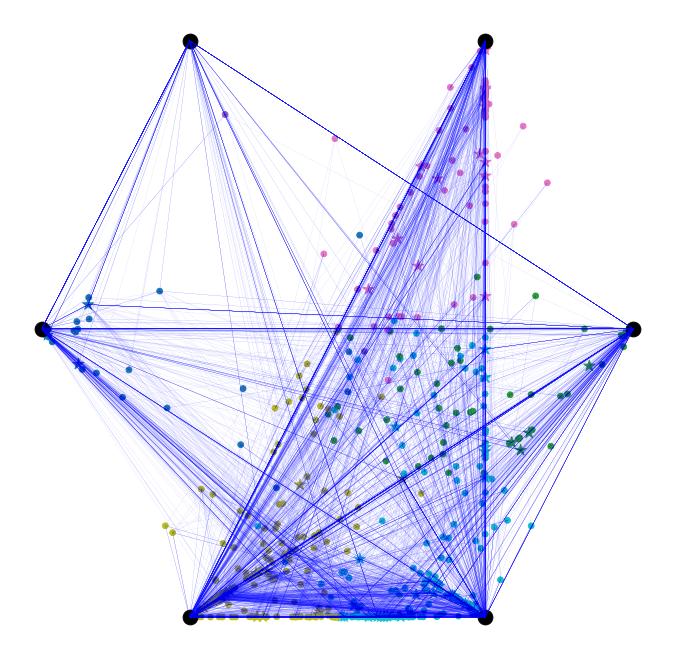}
    \caption{\textsc{PCP} \textsc{U.S.-House}}
    
\end{subfigure}
\hfill
\begin{subfigure}{0.32\textwidth}
    \includegraphics[width=0.7\textwidth]{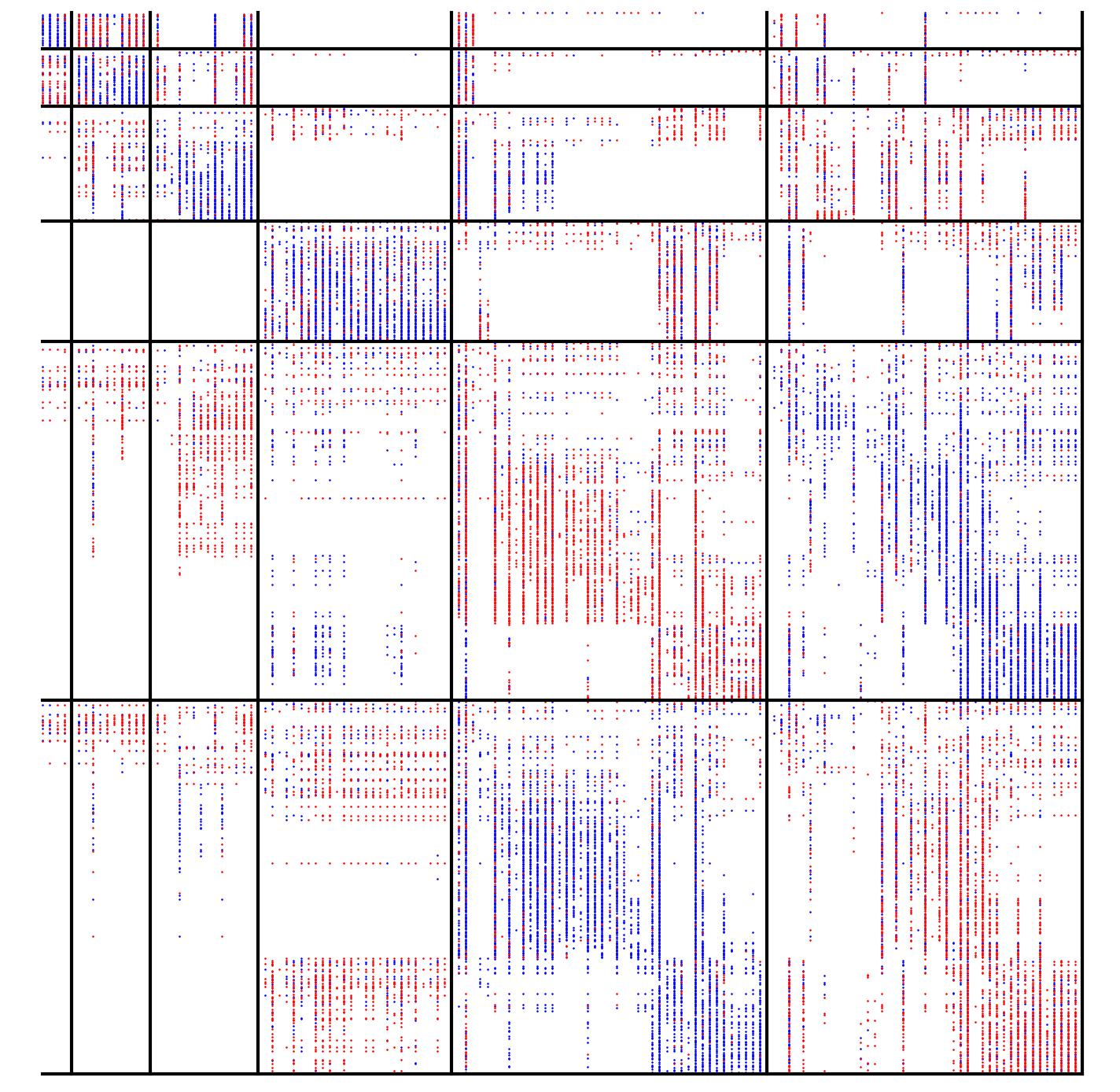}
    \caption{\textsc{OrA} \textsc{U.S.-House}}
    
\end{subfigure}
\caption{\textbf{\textsc{sHM-LDM}}(p=1): Inferred simplex visualizations and ordered adjacency matrices for a $D=6$ dimensional simplex with side lengths $\delta$ ensuring identifiability. The first column provides a Negative Circular Plot (\textsc{NCP}) with red lines showcasing negative links between nodes---The second column shows a Positive Circular Plot (\textsc{PCP}) with the blue lines denoting positive links between node pairs---The third and final column shows the Ordered Adjacency (\textsc{OrA}) matrices ordered based on the memberships, in terms of maximum simplex corner responsibility, and internally according to the magnitude of the corresponding corner assignment for their reconstruction. Top row: \textsc{U.S.-House}. Bottom row \textsc{U.S.-Senate}.}
\label{fig:bip_signed_p1}
\end{figure*}

\textbf{Extension to signed bipartite networks:} Here, similar to the unsigned network study, we extend the analysis to bipartite signed networks for \textsc{sHM-LDM}. The extension is again trivial by defining two sets of latent variables describing the two disjoint groups of nodes, as present in bipartite structures. In addition, we introduce four sets of random effects defining again node social and antisocial reach but now respecting target and source roles of the nodes in the corresponding networks links. We introduce two signed bipartite networks, \textsl{U.S.-House} \cite{house_senate} ($|\mathcal{V}|=1796$, $|\mathcal{E}^{+}|=61678$, $|\mathcal{E}^{-}|=52619$, Density=$0.1734$), and \textsl{U.S.-Senate} \cite{house_senate} ($|\mathcal{V}|=1201$, $|\mathcal{E}^{+}|=14964$, $|\mathcal{E}^{-}|=12096$, Density=$0.1769$), regarding voting records for proposed bills as made by the U.S. House of Representatives and the U.S. Senate, accordingly. For these networks, the first (rows) of the disjoint sets of nodes refer to the bills while the second (columns) to representatives or senators, accordingly. In Figs \ref{fig:bip_signed_p2} and \ref{fig:bip_signed_p1}, we provide the Positive Circular Plots \textsc{PCP}, Negative Circular Plots \textsc{NCP}, and Ordered Adjacency Matrices \textsc{OrA} for the corresponding networks and for both \textsc{sHM-LDM}(p=2) and \textsc{sHM-LDM}(p=1) frameworks, respectively. We witness how the  \textsc{sHM-LDM} framework generalizes to the study of bipartite networks, successfully uncovering distinct network aspects and profiles, that convey information about both homophily, as well as, animosity being present in the network.

\section{Complexity analysis} The proposed \textsc{HM-LDM} and its signed extension \textsc{sHM-LDM} belong to the family of latent distance models and thus require the calculation of the all-pairs distance matrix. This scales as $\mathcal{O}(N^2)$ in time and memory, making large-scale network analysis infeasible. To alleviate that problem we consider unbiased estimations of the log-likelihood through a random sampling approach. More specifically, in every model iteration, a set of network nodes, $S\subseteq\mathcal{V}$, is sampled (with replacement) and gradient steps are taken based on the log-likelihood of the block defined by the sampled node set. This effectively reduces the complexity of the models to $\mathcal{O}(S^2)$ both in time and memory. Another option is the case-control approach \cite{case_control} scaling by the number of network edges as $\mathcal{O}(E)$. Lastly, the Hierarchical Block Distance Model (\textsc{HBDM}) \cite{nakis2022hierarchical} is an attractive alternative option where gradient steps over the model parameters are based on a hierarchical approximation of the likelihood of the whole network. The \textsc{HBDM} model scales linearithmicly as $\mathcal{O}(N\log N)$ both in space and time while also offering hierarchical characterizations of structures at multiple scales.

\section{Conclusion and future work}\label{sec:conclusion}
In this study, we have presented the \textsc{\modelabbrv} reconciling graph representation learning and latent community detection. We extended the model to account for signed networks and showed that a minimum volume approach could uncover distinct profiles in social networks while ensuring model identifiability. Both presented frameworks were formulated to include a Euclidean as well as a squared Euclidean norm. For the latter, a direct relationship to an Eigenmodel in both the case of unsigned and signed networks was shown. Furthermore, by controlling the volume of the simplex by the magnitude of $\delta$, a sufficiently reduced simplex leads to unique representations. For unsigned networks, this resulted in the hard clustering of nodes to communities when the simplex was sufficiently contracted. Notably, the generalization to signed networks facilitated the extraction of distinct network profiles representing positive interactions and animosity. In regimes where \textsc{\modelabbrv} and \textsc{sHM-LDM} provide unique representations, we observed favorable link prediction performance and the ability to order the adjacency matrix based on prominent latent communities and distinct profiles. Notably, the proposed \textsc{\modelabbrv} combines network homophily and transitivity properties with latent community detection enabling explicit control of soft and hard assignment through the volume of the induced simplex. Importantly, the extended \textsc{sHM-LDM} merges homophily and heterophily properties to account for positive and negative ties as present in signed networks. To further evaluate the performance of \textsc{\modelabbrv} and \textsc{sHM-LDM}, future work should compare them against classical non-embedding methods such as the Degree Corrected Stochastic Block Model (\textsc{DC-SBM}) \cite{karrer2011stochastic} or the Mixed Membership Stochastic Block Model (\textsc{MM-SBM}) \cite{JMLR:v9:airoldi08a}, as well as, a Stochastic Block Model accounting for signed networks \cite{ssbm}.

\section*{Acknowledgements}
We would like to express sincere appreciation and thank the reviewers for their constructive feedback and their insightful comments. We gratefully acknowledge the Independent Research Fund Denmark for supporting this work [grant number: 0136-00315B].

\bibliographystyle{ws-acs}
\bibliography{ws-acs}

\begin{thebibliography}{10}
\providecommand{\urlprefix}{}
\expandafter\ifx\csname urlstyle\endcsname\relax
  \providecommand{\doi}[1]{doi:\discretionary{}{}{}#1}\else
  \providecommand{\doi}{doi:\discretionary{}{}{}\begingroup
  \urlstyle{rm}\Url}\fi

\bibitem{JMLR:v9:airoldi08a}
Airoldi, E.~M., Blei, D.~M., Fienberg, S.~E., and Xing, E.~P., Mixed membership
  stochastic blockmodels, \emph{J Mach Learn Res} \textbf{9} (2008) 1981--2014.

\bibitem{norm_lapl}
Atay, F. and Tunçel~Gölpek, H., On the spectrum of the normalized laplacian
  for signed graphs: Interlacing, contraction, and replication, \emph{Linear
  Algebra and its Applications} \textbf{442} (2014) 165–177.

\bibitem{nmf1}
Ball, B., Karrer, B., and Newman, M. E.~J., An efficient and principled method
  for detecting communities in networks, \emph{CoRR} \textbf{abs/1104.3590}
  (2011).

\bibitem{high_order1}
Beentjes, S.~V. and Khamseh, A., Higher-order interactions in statistical
  physics and machine learning: A model-independent solution to the inverse
  problem at equilibrium, \emph{Phys. Rev. E} \textbf{102} (2020) 053314.

\bibitem{louvainNE-wsdm20}
Bhowmick, A.~K., Meneni, K., Danisch, M., Guillaume, J.-L., and Mitra, B.,
  {LouvainNE}: Hierarchical louvain method for high quality and scalable
  network embedding, in \emph{WSDM} (2020), pp. 43--51.

\bibitem{HU}
Bioucas-Dias, J.~M., Plaza, A., Dobigeon, N., Parente, M., Du, Q., Gader, P.,
  and Chanussot, J., Hyperspectral unmixing overview: Geometrical, statistical,
  and sparse regression-based approaches, \emph{IEEE Journal of Selected Topics
  in Applied Earth Observations and Remote Sensing} \textbf{5} (2012) 354--379.

\bibitem{bueler2000exact}
B{\"u}eler, B., Enge, A., and Fukuda, K., Exact volume computation for
  polytopes: a practical study, in \emph{Polytopes—combinatorics and
  computation} (Springer, 2000), pp. 131--154.

\bibitem{expon_fam_emb}
{\c{C}}elikkanat, A. and Malliaros, F.~D., Exponential family graph embeddings,
  in \emph{{AAAI}} (2020), pp. 3357--3364.

\bibitem{com_metrics}
Chakraborty, T., Dalmia, A., Mukherjee, A., and Ganguly, N., Metrics for
  community analysis: A survey (2016).

\bibitem{cutler1994a}
Cutler, A. and Breiman, L., Archetypal analysis, \emph{Technometrics}
  \textbf{36} (1994) 338--347.

\bibitem{house_senate}
Derr, T., Johnson, C., Chang, Y., and Tang, J., Balance in signed bipartite
  networks, in \emph{Proceedings of the 28th ACM International Conference on
  Information and Knowledge Management}, CIKM '19 (Association for Computing
  Machinery, New York, NY, USA, 2019), ISBN 9781450369763, p. 1221–1230,
  \doi{10.1145/3357384.3358009},
  \urlprefix\url{https://doi.org/10.1145/3357384.3358009}.

\bibitem{node2vec-kdd16}
Grover, A. and Leskovec, J., {Node2Vec}: Scalable feature learning for
  networks, in \emph{KDD} (2016), pp. 855--864.

\bibitem{GRL_HAM}
Hamilton, W.~L., Graph representation learning, \emph{Synthesis Lectures on
  Artificial Intelligence and Machine Learning} \textbf{14} 1--159.

\bibitem{graphsage_hamilton}
Hamilton, W.~L., Ying, R., and Leskovec, J., Inductive representation learning
  on large graphs, in \emph{NIPS} (2017).

\bibitem{survey_hamilton_leskovec}
Hamilton, W.~L., Ying, R., and Leskovec, J., Representation learning on graphs:
  Methods and applications, \emph{{IEEE} Data Eng. Bull.} \textbf{40} (2017)
  52--74.

\bibitem{handcock2007model}
Handcock, M.~S., Raftery, A.~E., and Tantrum, J.~M., Model-based clustering for
  social networks, \emph{J R Stat Soc Ser A Stat Soc.} \textbf{170} (2007)
  301--354.

\bibitem{hart2015inferring}
Hart, Y., Sheftel, H., Hausser, J., Szekely, P., Ben-Moshe, N.~B., Korem, Y.,
  Tendler, A., Mayo, A.~E., and Alon, U., Inferring biological tasks using
  pareto analysis of high-dimensional data, \emph{Nature methods} \textbf{12}
  (2015) 233--235.

\bibitem{doi:10.1198/016214504000001015}
Hoff, P.~D., Bilinear mixed-effects models for dyadic data, \emph{JASA}
  \textbf{100} (2005) 286--295.

\bibitem{hoff2007modeling}
Hoff, P.~D., Modeling homophily and stochastic equivalence in symmetric
  relational data, in \emph{NIPS} (2007), p. 657–664.

\bibitem{exp1}
Hoff, P.~D., Raftery, A.~E., and Handcock, M.~S., Latent space approaches to
  social network analysis, \emph{JASA} \textbf{97} (2002) 1090--1098.

\bibitem{sigat}
Huang, J., Shen, H., Hou, L., and Cheng, X., Signed graph attention networks,
  in \emph{ICANN 2019: Workshop and Special Sessions} (2019), pp. 566--577.

\bibitem{SDGNN}
Huang, J., Shen, H., Hou, L., and Cheng, X., {SDGNN}: Learning node
  representation for signed directed networks, \emph{AAAI} \textbf{35} (2021)
  196--203.

\bibitem{nmf5}
Huang, K., Sidiropoulos, N.~D., and Swami, A., Non-negative matrix
  factorization revisited: Uniqueness and algorithm for symmetric
  decomposition, \emph{IEEE Trans. Signal Process} \textbf{62} (2014) 211--224.

\bibitem{pole}
Huang, Z., Silva, A., and Singh, A., {POLE}: Polarized embedding for signed
  networks, \emph{WSDM}  (2022) 390--400.

\bibitem{balance_theory}
Hummon, N.~P. and Doreian, P., Some dynamics of social balance processes:
  bringing heider back into balance theory, \emph{Social Networks} \textbf{25}
  (2003) 17--49.

\bibitem{signet}
Islam, M.~R., Aditya~Prakash, B., and Ramakrishnan, N., {SIGN}et: Scalable
  embeddings for signed networks, in \emph{Advances in Knowledge Discovery and
  Data Mining}, eds. Phung, D., Tseng, V.~S., Webb, G.~I., Ho, B., Ganji, M.,
  and Rashidi, L. (Springer International Publishing, Cham, 2018), pp.
  157--169.

\bibitem{868688}
{Jianbo Shi} and {Malik}, J., Normalized cuts and image segmentation,
  \emph{IEEE Transactions on Pattern Analysis and Machine Intelligence}
  \textbf{22} (2000) 888--905.

\bibitem{ssbm}
Jiang, J., Stochastic blockmodel and exploratory analysis in signed networks,
  \emph{Physical Review E} \textbf{91} (2015).

\bibitem{karrer2011stochastic}
Karrer, B. and Newman, M.~E., Stochastic blockmodels and community structure in
  networks, \emph{Physical review E} \textbf{83} (2011) 016107.

\bibitem{side}
Kim, J., Park, H., Lee, J.-E., and Kang, U., {SIDE}: Representation learning in
  signed directed networks, in \emph{Proceedings of the 2018 World Wide Web
  Conference} (International World Wide Web Conferences Steering Committee,
  2018), p. 509–518.

\bibitem{kingma2017adam}
Kingma, D.~P. and Ba, J., Adam: {A} method for stochastic optimization, in
  \emph{ICLR} (2015).

\bibitem{oversmoothing_gnn}
Kipf, T.~N. and Welling, M., Semi-supervised classification with graph
  convolutional networks, in \emph{ICLR} (2017).

\bibitem{KRIVITSKY2009204}
Krivitsky, P.~N., Handcock, M.~S., Raftery, A.~E., and Hoff, P.~D.,
  Representing degree distributions, clustering, and homophily in social
  networks with latent cluster random effects models, \emph{Social Networks}
  \textbf{31} (2009) 204 -- 213.

\bibitem{SymmNMF}
Kuang, D., Ding, C., and Park, H., Symmetric nonnegative matrix factorization
  for graph clustering, in \emph{SDM} (2012).

\bibitem{dataset_reddit}
Kumar, S., Hamilton, W.~L., Leskovec, J., and Jurafsky, D., Community
  interaction and conflict on the web, in \emph{WWW} (2018), pp. 933--943.

\bibitem{lee99}
Lee, D.~D. and Seung, H.~S., Learning the parts of objects by nonnegative
  matrix factorization, \emph{Nature} \textbf{401} (1999) 788--791.

\bibitem{dataset_wikielec}
Leskovec, J., Huttenlocher, D., and Kleinberg, J., Predicting positive and
  negative links in online social networks, in \emph{WWW} (2010), p. 641–650.

\bibitem{astroph_grqc_hepth}
Leskovec, J., Kleinberg, J., and Faloutsos, C., Graph evolution: Densification
  and shrinking diameters, \emph{TKDD} \textbf{1} (2007).

\bibitem{snapnets}
Leskovec, J. and Krevl, A., {SNAP Datasets}: {Stanford} large network dataset
  collection (2014).

\bibitem{facebook}
Leskovec, J. and Mcauley, J.~J., Learning to discover social circles in ego
  networks, \emph{NIPS}  (2012) 539--547.

\bibitem{nmf4}
Mao, X., Sarkar, P., and Chakrabarti, D., On mixed memberships and symmetric
  nonnegative matrix factorizations, in \emph{ICML}, Vol.~70 (2017).

\bibitem{MVCNMF}
Miao, L. and Qi, H., Endmember extraction from highly mixed data using minimum
  volume constrained nonnegative matrix factorization, \emph{IEEE Transactions
  on Geoscience and Remote Sensing} \textbf{45} (2007) 765--777.

\bibitem{fb_nets}
Mucha, P. and Porter, M., Social structure of facebook networks, \emph{Physica
  A: Statistical Mechanics and its Applications} \textbf{391} (2012)
  4165–4180.

\bibitem{high_order2}
Muolo, R., Gallo, L., Latora, V., Frasca, M., and Carletti, T., Turing patterns
  in systems with high-order interactions, \emph{Chaos, Solitons \& Fractals}
  \textbf{166} (2023) 112912.

\bibitem{5589222}
Mørup, M. and Kai~Hansen, L., Archetypal analysis for machine learning, in
  \emph{Workshop on Machine Learning for Signal Processing} (2010), pp.
  172--177.

\bibitem{hbdm}
Nakis, N., \c{C}elikkanat, A., and Mørup, M., Scalable hierarchical embeddings
  of complex networks (2022),
  \urlprefix\url{https://openreview.net/pdf?id=U-GB_gONqbo}.

\bibitem{hmldm}
Nakis, N., {\c{C}}elikkanat, A., and M{\o}rup, M., Hm-ldm: A hybrid-membership
  latent distance model, in \emph{Complex Networks and Their Applications XI},
  eds. Cherifi, H., Mantegna, R.~N., Rocha, L.~M., Cherifi, C., and
  Miccich{\`e}, S. (Springer International Publishing, Cham, 2023), ISBN
  978-3-031-21127-0, pp. 350--363.

\bibitem{slim}
Nakis, N., Çelikkanat, A., Boucherie, L., Djurhuus, C., Burmester, F.,
  Holmelund, D., Frolcová, M., and Mørup, M., Characterizing polarization in
  social networks using the signed relational latent distance model, in
  \emph{Proceedings of the 26th International Conference on Artificial
  Intelligence and Statistics} (2023).

\bibitem{nakis2022hierarchical}
Nakis, N., Çelikkanat, A., Jørgensen, S.~L., and Mørup, M., A hierarchical
  block distance model for ultra low-dimensional graph representations (2022).

\bibitem{newman}
Newman, M. E.~J., The structure and function of complex networks, \emph{SIAM
  Review} \textbf{45} (2003) 167--256.

\bibitem{10.5555/2980539.2980649}
Ng, A.~Y., Jordan, M.~I., and Weiss, Y., On spectral clustering: Analysis and
  an algorithm, in \emph{Proceedings of the 14th International Conference on
  Neural Information Processing Systems: Natural and Synthetic}, NIPS'01 (MIT
  Press, Cambridge, MA, USA, 2001), p. 849–856.

\bibitem{dataset_twitter}
Ordozgoiti, B., Matakos, A., and Gionis, A., Finding large balanced subgraphs
  in signed networks, in \emph{Proceedings of The Web Conference 2020} (2020),
  p. 1378–1388.

\bibitem{HOPE-kdd16}
Ou, M., Cui, P., Pei, J., Zhang, Z., and Zhu, W., Asymmetric transitivity
  preserving graph embedding, in \emph{KDD} (2016), pp. 1105--1114.

\bibitem{deepwalk-perozzi14}
Perozzi, B., Al-Rfou, R., and Skiena, S., Deepwalk: Online learning of social
  representations, in \emph{KDD} (2014), p. 701–710.

\bibitem{netsmf-www2019}
Qiu, J., Dong, Y., Ma, H., Li, J., Wang, C., Wang, K., and Tang, J., {NetSMF}:
  Large-scale network embedding as sparse matrix factorization, in \emph{WWW}
  (2019).

\bibitem{netmf-wsdm18}
Qiu, J., Dong, Y., Ma, H., Li, J., Wang, K., and Tang, J., Network embedding as
  matrix factorization: Unifying {DeepWalk}, {LINE}, {PTE}, and {Node2Vec}, in
  \emph{WSDM} (2018), pp. 459--467.

\bibitem{case_control}
Raftery, A.~E., Niu, X., Hoff, P.~D., and Yeung, K.~Y., Fast inference for the
  latent space network model using a case-control approximate likelihood,
  \emph{Journal of Computational and Graphical Statistics} \textbf{21} (2012)
  901--919.

\bibitem{ryan2017bayesian}
Ryan, C., Wyse, J., and Friel, N., Bayesian model selection for the latent
  position cluster model for social networks, \emph{Network Science} \textbf{5}
  (2017) 70--91.

\bibitem{skellam}
Skellam, J.~G., The frequency distribution of the difference between two
  poisson variates belonging to different populations., \emph{Journal of the
  Royal Statistical Society. Series A (General)} \textbf{109} (1946) 296--296.

\bibitem{NNSED}
Sun, B.-J., Shen, H., Gao, J., Ouyang, W., and Cheng, X., A non-negative
  symmetric encoder-decoder approach for community detection, in \emph{CIKM}
  (2017).

\bibitem{line}
Tang, J., Qu, M., Wang, M., Zhang, M., Yan, J., and Mei, Q., {LINE}:
  Large-scale information network embedding, in \emph{WWW} (2015), pp.
  1067--1077.

\bibitem{MNMF}
Wang, X., Cui, P., Wang, J., Pei, J., Zhu, W., and Yang, S., Community
  preserving network embedding, in \emph{AAAI} (2017).

\bibitem{dataset_wikirfa}
West, R., Paskov, H.~S., Leskovec, J., and Potts, C., Exploiting social network
  structure for person-to-person sentiment analysis, \emph{TACL} \textbf{2}
  (2014) 297--310.

\bibitem{nmf2}
Wind, D.~K. and Mørup, M., Link prediction in weighted networks, in
  \emph{Workshop on Machine Learning for Signal Processing} (2012), pp. 1--6.

\bibitem{slf}
Xu, P., Wu, J., Hu, W., and Du, B., Link prediction with signed latent factors
  in signed social networks, \emph{Proceedings of the Acm Sigkdd International
  Conference on Knowledge Discovery and Data Mining}  (2019) 1046--1054.

\bibitem{nmf3}
Yang, J. and Leskovec, J., Overlapping community detection at scale: A
  nonnegative matrix factorization approach, in \emph{WSDM} (2013).

\bibitem{GRL-survey-ieeebigdata20}
{Zhang}, D., {Yin}, J., {Zhu}, X., and {Zhang}, C., Network representation
  learning: A survey, \emph{IEEE Trans. Big Data} \textbf{6} (2020).

\bibitem{prone-ijai19}
Zhang, J., Dong, Y., Wang, Y., Tang, J., and Ding, M., Prone: Fast and scalable
  network representation learning, in \emph{IJCAI} (2019).

\bibitem{zhuang2019regularization}
Zhuang, L., Lin, C.-H., Figueiredo, M.~A., and Bioucas-Dias, J.~M.,
  Regularization parameter selection in minimum volume hyperspectral unmixing,
  \emph{IEEE Transactions on Geoscience and Remote Sensing} \textbf{57} (2019)
  9858--9877.

\bibitem{çelikkanat2022piecewisevelocity}
Çelikkanat, A., Nakis, N., and Mørup, M., Piecewise-velocity model for
  learning continuous-time dynamic node representations (2022).

\end{thebibliography}

\end{document}